\documentclass[british,titlepage]{ntnuthesis}

\title{First post-Newtonian correction to gravitational waves produced by compact binaries: \\ How to compute relativistic corrections to gravitational waves using Feynman diagrams}
\shorttitle{First PN correction to GWs produced by compact binaries}
\author{Vegard Undheim}
\shortauthor{V. Undheim}
\date{\today}

\newboolean{NaturalUnits}

\makeglossaries 

\newglossaryentry{latex}
{
        name=latex,
        description={Is a mark up language specially suited for
scientific documents}
}

\newglossaryentry{bibliography}
{
        name=bibliography,
        plural=bibliographies,
        description={A list of the books referred to in a scholarly work,
typically printed as an appendix}
}

\newglossaryentry{maths}
{
    name=mathematics,
    description={Mathematics is what mathematicians do}
}

\newglossaryentry{QFT}{
    name={q}uantum {f}ield {t}heory,
    description={The theory of fields endowed with quantum properties that can be used to describe forces and matter}
}

\newglossaryentry{BH}{
    name={b}lack {h}ole,
    description={A region of space-time curved to the point that no matter or radiation can escape. Usually taken to be a gravitationally collapsed star}
}

\newglossaryentry{GW150914}{
    name=GW150914,
    description={The gravitational wave event which occurred 14/09/2015. The first \acrshort{gw} event by \acrshort{ligo} \cite{FirstGW}}
}

\newglossaryentry{2bodyprob}{
    name=two body problem,
    description={Name of the physics problem of describing how a system consisting of two bodies (usually taken to be point particles) evolve in time, given they only interact with each other. For $r^{-1}$ potentials the two body problem generally has the solution of conic sections \cite{GoldsteinMech}}
}

\newglossaryentry{risco}{
    name=r$\ind{_{\text{ISCO}}}$,
    description={The \acrfull{isco} for a gravitational source, e.g. black hole. For a spherically symmetric, non-rotating, object, i.e. the Schwarzschild metric, $r\ind{_{\text{ISCO}}}=3R_s=6\frac{GM}{c^2}$}
}

\newglossaryentry{QCO}{
    name=quasi-stable circular orbit,
    description={Approximating the inspiral as circular orbits with gradually falling radii. The change in radius is negligible unless viewed over several periods}
}

\newglossaryentry{relativist}{
    name=relativist,
    description={Physicists using a geometrical interpretation of gravity, following in the footsteps of Einstein}
}

\newglossaryentry{field theorist}{
    name=field theorist,
    description={Physicists using fields on a static background space-time to model physical effects like forces and particles. In this thesis especially those who use fields to model gravity}
}

\newglossaryentry{Einstein-Hilbert action}{
    name=Einstein-Hilbert action,
    description={Einstein-Hilbert action is the action which when extremized generates the \gls{Einstein equations}, i.e. the action which governs \acrlong{gr} $S_\text{\acrshort{EH}}= \frac{c^4}{16\pi G}\int R \sqrt{-g}\dd[4]{x}$}
}

\newglossaryentry{Einstein equations}{
    name=Einstein's field equations,
    description={The equations of motion resulting from the \gls{Einstein-Hilbert action} which dictates the dynamics of space-time. Coupled to a matter source it reads $R\ind{_{\mu\nu}} - \frac{1}{2}R g\ind{_{\mu\nu}} = \frac{8\pi G}{c^4} T\ind{_{\mu\nu}}$}
}

\newglossaryentry{R_S}{
    name=Schwarzschild radius,
    description={The Schwarzschild radius $R_S$ is the radius associated with the event horizon of a non-rotating, static \acrlong{bh}. $R_S=\ifthenelse{\boolean{NaturalUnits}}{2M}{\frac{2GM}{c^2}}$}
}

\newacronym{phd}{PhD}{philosophiae doctor}
\newacronym{CoPCSE}{CoPCSE@NTNU}{Community of Practice in Computer ScienceEducation at NTNU}
\newacronym{gcd}{GCD}{Greatest Common Divisor}

\newacronym{gw}{GW}{gravitational wave}
\newacronym{gr}{GR}{general relativity}
\newacronym{sr}{SR}{special relativity}
\newacronym{qft}{QFT}{\gls{QFT}}
\newacronym{eom}{EoM}{equation of motion}
\newacronym{bh}{BH}{\gls{BH}}
\newacronym{ns}{NS}{neutron star}
\newacronym{ligo}{LIGO}{Laser Interferometer Gravitational-Wave Observatory}
\newacronym{isco}{ISCO}{innermost stable circular orbit}
\newacronym{pn}{PN}{post-Newtonian}
\newacronym{spa}{SPA}{stationary phase approximation}
\newacronym{lhs}{LHS}{left hand side}
\newacronym{rhs}{RHS}{right hand side}
\newacronym{fp}{FP}{Fierz-Pauli}
\newacronym{tt}{TT}{transverse-traceless}
\newacronym{em}{EM}{electromagnetic}
\newacronym{pp}{pp}{point particle}
\newacronym{gf}{gf}{gauge fixing term}
\newacronym{eft}{EFT}{effective field theory}
\newacronym{qcd}{QCD}{quantum chromodynamics}
\newacronym{qed}{QED}{quantum electrodynamics}
\newacronym{EIH}{EIH}{Einstein-Infeld-Hoffmann}
\newacronym{EH}{EH}{Einstein–Hilbert}
\newacronym{NASA}{NASA}{National Aeronautics and Space Administration}
\newacronym{stf}{STF}{symmetric trace free} 

\begin{document}
\includepdf[pages={2}]{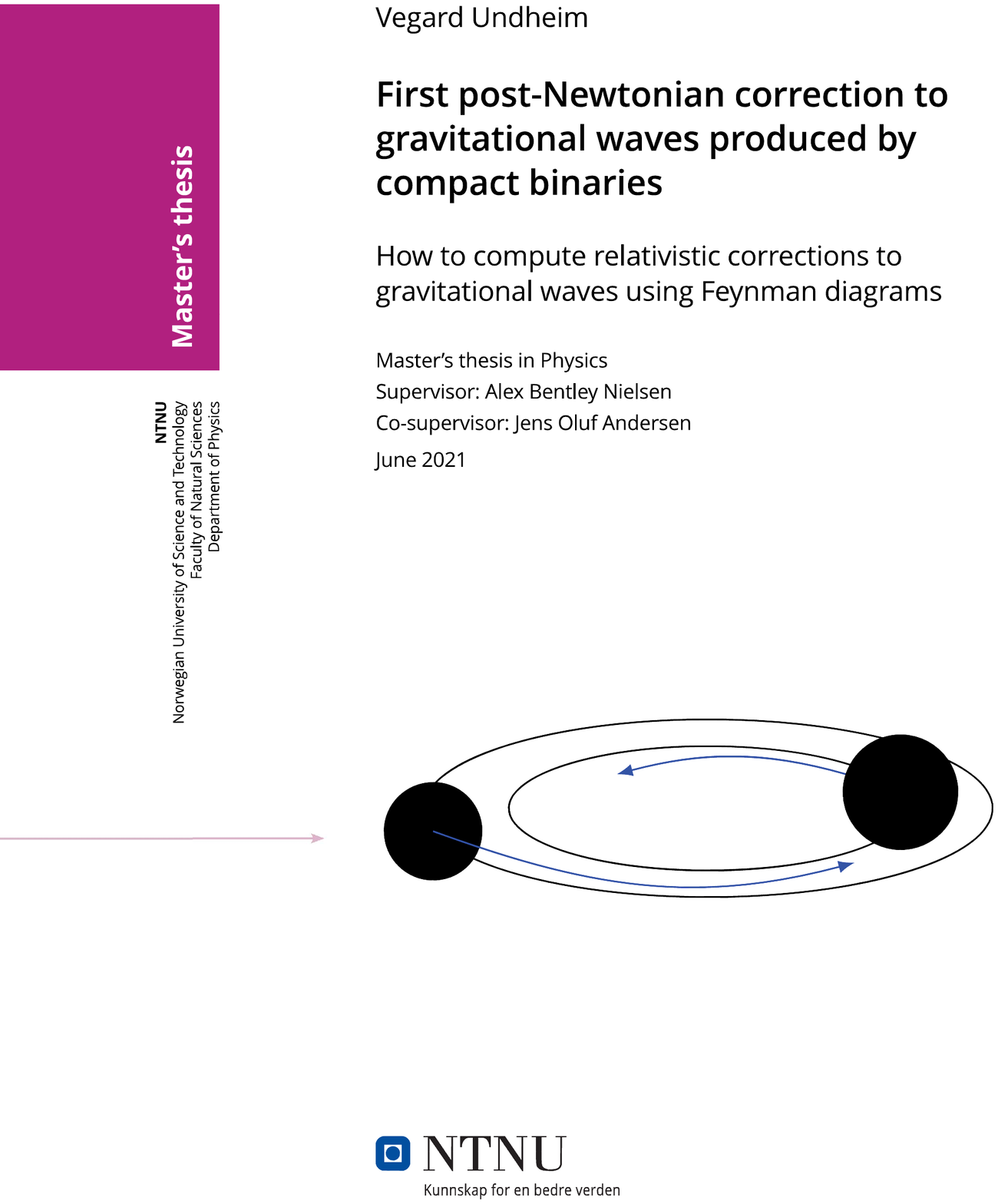}
\chapter*{Abstract}

    The purpose of this thesis is to calculate the relativistic correction to the \acrlong{gw}s produced by compact binaries in the inspiral phase. The correction is up to the next to leading order, the so-called first \acrlong{pn} order (1\acrshort{pn}), which are correctional terms proportional to $(v/c)^2$ compared to leading order, Newtonian, terms.

    These corrections are well known in the literature, even going beyond the first order corrections, so why is it computed again here? In later years, an alternative approach for computing these terms using \acrlong{eft} has emerged. This thesis investigates this approach by replicating it, and attempts to make this approach more accessible to those not familiar with effective field theories.

    It has been claimed that this approach greatly simplifies the complicated calculations of \acrlong{gw}forms, and even provides the required intuition for `physical understanding'. By this master student that was found not to be entirely correct. The calculations were made easier for those with a rich background in \acrlong{qft}, but for those who are not well acquainted with \acrlong{qft} this was not the case.

    It was, however, found to be a worthwhile method as a means for deepening one's understanding of gravity, and might provide a shorter route for some alternative theories of gravity to testable predictions.
\chapter*{Sammendrag}

    Hensikten med denne oppgaven er å beregne den relativistiske korreksjonen til gravitasjonsbølger som er produsert av kompakte binærsystemer i spiral-fall fasen. Korreksjonene er av den såkalte første post-Newtonske orden (1PN), som er korreksjonstermer proporsjonal med $ (v / c) ^ 2 $ sammenlignet med ledende, Newtonske, termerene.

    Disse korreksjonene er velkjente i litteraturen, og går til og med utover korreksjonene av første orden, så hvorfor blir de beregnet igjen her? I nyere tid har en alternativ tilnærming for å beregne disse størrelsene ved hjelp av effektiv feltteori dukket opp. Denne oppgaven undersøker tilnærmingen ved å reprodusere dem, og prøver å gjøre metoden mer tilgjengelig for de som ikke er kjent med effektive feltteorier.

    Det har blitt hevdet at beregningen av gravitasjonsbølgeformer kan gjøres mye enklere ved å bruke denne tilnærmingen, og til og med gir den nødvendige intuisjonen for `fysisk forståelse'. Ifølge denne masterstudenten er ikke dette helt riktig. Beregningene ble gjort enklere for de med en spesialisert bakgrunn i kvantefeltteori, og for de som er mindre kjent med kvantefeltteori var dette ikke tilfelle.

    Det ble imidlertid funnet å være en verdifull metode som et middel for å utdype forståelsen av tyngdekraften, og kan gi en kortere rute for noen alternative teorier for gravitasjon til testbare forutsigelser.
\chapter*{Acknowledgements}

    I would like to thank my supervisor \href{https://www.uis.no/nb/profile/447}{\textcolor{black}{Alex Bentley Nielsen}} for adhering to my wishes of working on \acrlong{gw}s, and as a consequence the numerous hours spent guiding me through this project. Of these hours, I am especially thankful for the time he spent discussing gravity, academia, and physics in general with me. I found these talks motivating and educational, and often the highlight of my week.

    I would also like to thank the \href{https://www.uis.no/en/mathematics-and-natural-science/department-mathematics-and-physics}{\textcolor{black}{Department of Mathematics and Physics of the University of Stavanger}}. During the COVID-19 pandemic, the department made the necessary arrangements to let me come visit them, for which I am grateful. The possibility to spend time physically with my supervisor was much appreciated. They welcomed me with open arms, and I thoroughly enjoyed my stay. A special thanks to \href{https://www.uis.no/nb/profile/631}{\textcolor{black}{Germano Nardini}} for conversations, coffee, and a scoop of ice cream during my visits to Stavanger.

    I also extend my thanks to my local supervisor, \href{https://www.ntnu.no/ansatte/jens.andersen}{\textcolor{black}{Jens Oluf Andersen}}, and \href{https://www.ntnu.edu/}{\textcolor{black}{NTNU}} for making the formal facilitations need to make this thesis. Especially for granting travel funds for me to visit Stavanger.

    Lastly, I thank Michelle Angell for proofreading the last draft of this thesis. There may still linger some typos in this document, but had it not been for her, it would have been many more.
\chapter*{Foreword}
    This document is for all intents and purposes my master's thesis, but it deviates to some degree from the document that was handed in for the examination. 

    The reason for this is that this document has been updated based on helpful comments from the examiner, \href{https://www.mn.uio.no/astro/english/people/aca/mota/}{\textcolor{black}{Professor David Fonseca Mota}}, with additional clarifications and purging of typos. This will hopefully make this document better suited than the original thesis for those who want to use it as an introduction to the use of field theories in \acrlong{gw} physics.

    The original thesis can be found at \href{https://www.ntnu.edu/}{\textcolor{black}{NTNU's}} digital thesis archive at \url{https://hdl.handle.net/11250/2785590}

\tableofcontents
\listoffigures

\printglossary[type=\acronymtype] 
\printglossary     

\setboolean{NaturalUnits}{false}    
\chapter{Introduction} \label{chap:introduction}
\section{Binary inspirals and gravitational waves}
    On the 14$\textsuperscript{th}$ of September 2015 the world was shocked, ever so slightly. So slightly in fact that the only reason we know about it is thanks to the effort of the \acrfull{ligo}, who measured this faint strain in their detectors. After careful testing and retesting, \acrshort{ligo} published their results on the 11$\textsuperscript{th}$ of February 2016 \cite{FirstGW}. They concluded that the event, called \gls{GW150914}, was a \acrfull{gw} produced by the merger of two \glspl{BH}, and was the first directly detected \acrlong{gw} event in human history.

    With the announcement of the historic detection of \gls{GW150914} came promises of a new era of astronomy, now equipped with a brand new type of data to constrain astronomical theories. Popular science lectures and books were given and written, and at the height of this hype I started my bachelor's degree in physics. Fascinated by these mysterious waves I wanted to learn more about them, and  when the time came to pick a topic for my master's thesis I requested to work on \acrlong{gw}s.

    My supervisor and I decided to work on relativistic corrections to the binary inspiral, using field theoretical methods. To date, all confirmed \acrshort{gw} events are thought to be produced by compact binaries. A compact object is a \acrfull{bh} or \acrfull{ns}, and a compact binary is a system consisting of two compact objects. When compact objects revolve around each other they produce so called \acrlong{gw}s which dissipate orbital energy from the system. As a result the compact objects fall toward each other, and in the end collide and merge together.

    The problem with compact binaries is that they are too heavy and fall too close to each other, making the gravitational interaction too strong to be adequately described by Newton's law of gravity. Although the \gls{2bodyprob} has a general solution in Newtonian mechanics, there is no known equivalent solution for the \gls{2bodyprob} in \acrlong{gr}, only the one body problem. To combat this issue, researchers have followed one of two approaches.\footnote{Or tried to find the actual, analytical, solution.}

\newpage
\begin{enumerate}
    \item Solve the full, non-linear, \gls{Einstein equations} numerically for the binary system in question.
    \item Use an approximate, analytical, solution and perturbatively expand it to account for relativistic corrections.
\end{enumerate}
    This thesis will focus on analytical approximations. With numerical simulations one obtains a picture of the dynamics at an arguably very high accuracy, but due to the complexity of \gls{Einstein equations} this is computationally costly, i.e. takes a lot of time and computing power. Furthermore, analytical expressions provide information about important quantities and intuition about the most important physical effects at play, that one simply does not gain from computer simulations.

\begin{figure}
    \centering
    \includegraphics[width=\textwidth]{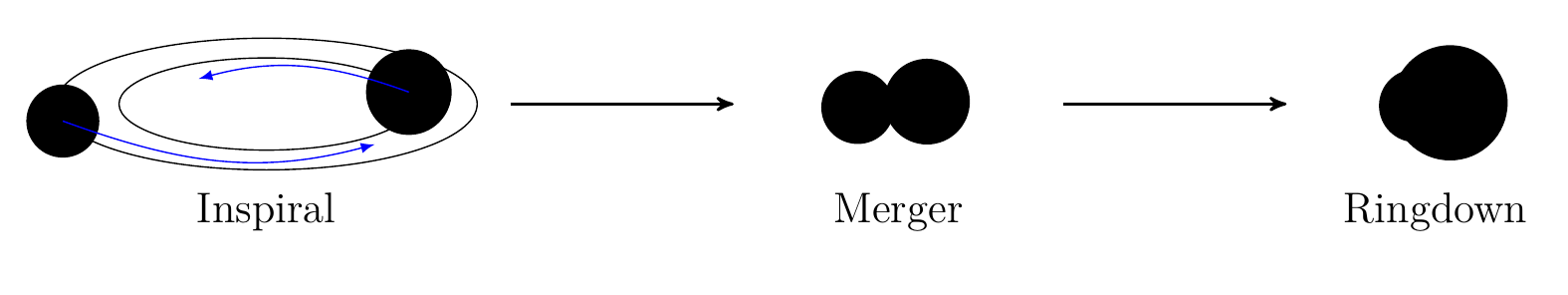}
    \caption[Phases of binary evolution.]{The evolution of compact binaries in three phases.}
    \label{fig:InspiralPhases}
\end{figure}

    In order to expand an analytical solution relativistically one first needs an approximate solution to expand. For this it is useful to divide the evolution of the compact binary into three phases, see \cref{fig:InspiralPhases}. The first phase is called the \textbf{inspiral phase}. Here the compact objects orbit each other at a distance, gradually falling closer together due to the emission of \acrlong{gw}s. Once the bodies are so close that a collision is imminent (typically when they `touch' or form a common event horizon) the system becomes highly non-linear, and enters the so-called \textbf{merger phase}. After the two objects have merged into one, the system enters the \textbf{ringdown phase}, in which the system can be described as a one body problem, but with remnant asymmetries from the merger. Typically, the merged object's asymmetries oscillate around the Kerr solution and gradually dampen down, hence the name ringdown.

    This is a useful division of the binary evolution as the different phases lend themselves to different approximations. The first phase, the inspiral, can be approximated as Keplerian orbits since the leading order term in the equations of motion is the Newtonian law of gravitation. The last phase can be approximated as a Schwarzschild or Kerr solution with perturbations. The merger phase is sandwiched between these two widely different approximations and is dominated by non-linear effects. Thus the merger phase has no good analytical approximation and must be simulated numerically.

    In this thesis I will work with the analytical approximation of the inspiral phase.

\newpage
\section{Structure of this thesis}
    As we will see in \cref{chap:fieldtheory} the frequency of the \acrlong{gw}s produced by compact binaries are directly dependent on the frequency at which the source oscillates, which is found to be inverse proportional to the total mass of the binary raised to the $5/8$ths power: $\omega_\text{source} \propto M^{-5/8}$ (see equation \eqref{eq:waveform:omegaOfTau}). Therefore, the waveform of \acrshort{gw}s measured here on Earth provides information about the dynamics of the binary which produced it, and can be compared with the predicted dynamics according to \acrlong{gr}. This is why \acrshort{gw} observation is a precise tool for constraining theories of gravity.

    To motivate these computations, \cref{chap:waveform} starts off by computing the waveform, using results from following chapters. Then in \cref{chap:fieldtheory} an alternative path to gravity is presented, that of a \emph{gauge field theory} on a static space-time background. It is demonstrated to recover the main results of standard \emph{linearized gravity}, which is the \gls{Einstein equations} expanded to linear order in metric perturbations over flat space-time. In \cref{chap:energy} and \ref{chap:flux} the main results needed to compute the waveform in \cref{chap:waveform} are derived, using the \acrfull{eft} based on the material presented in \cref{chap:fieldtheory}. Then the thesis ends with \cref{chap:conclusion}, which is concluding remarks on the \acrlong{eft} approach to gravitational waves.

    This is a form of \emph{top down} approach, starting with the final result (the waveform) and working back to the fundamental assumptions behind it. This structure has been chosen because of the large amount of laboursome calculations leading to the \acrlong{gw}form, and it will hopefully provide the overview needed to understand the motivation for each calculation as it appears.

\section{Why effective field theory?}
    In 2006, \textcite{Goldberger:EFT} wrote a paper showing how the \acrlong{gw}form could systematically be calculated to any \acrfull{pn} order using \acrshort{eft} formalism. Post-Newtonian expansion is ordering results like energy, the \acrlong{eom}, radiated energy flux, velocity, etc. as the Newtonian result plus relativistic corrections, usually expanded in factors of $v/c$.

    E.g.
    \begin{align}
        E = E_\text{Newt} \left[ 1 + \sum_{i=2}^\infty E_i \left( \frac{v}{c} \right)^i \right].
    \end{align}
    Here $E_\text{Newt} \cdot E_i \left( \frac{v}{c} \right)^i$ would be the $\frac{i}{2}$\acrshort{pn} term of the energy. This scaling as half the $v/c$ power is chosen to represent the \acrshort{pn} order such that the leading order correction is $1$\acrshort{pn}.

    In this thesis, working with fields on a non-dynamic, flat, space-time will be referred to as field theory, or the approach of \glspl{field theorist}, like Goldberger. This is supposed to be contrary to traditional geometrical theories of gravity, in the spirit of Einstein, which will be referred to as the approach of \glspl{relativist}. By any normal definition however, \acrlong{gr} and its interpretation by \glspl{relativist}, is a field theory. But they work with dynamical space-times, making it conceptually and mathematically quite differently formulated. Therefore, these constructed labels of \glspl{field theorist} and \glspl{relativist} will be employed in this thesis to emphasize the difference in approach.

    Formulating the computations in the language of \glspl{field theorist}, Goldberger and Rothstein unlocked all tools, tricks, and language usually reserved for \acrfull{qft}. Since then, this approach has been argued by \glspl{field theorist} to be easier and faster than the traditional \gls{relativist} approach, and has thus far produced results up to the 6th \acrshort{pn} order \cite{Blumlein:2021txj,Blumlein:2020znm} and a detailed description up to the 5th order \cite{Blumlein:2021txe,Blumlein:2020pyo,Blumlein:2020pog,Blumlein:2019bqq,Blumlein:2019zku}. One of these \glspl{field theorist}, R. Porto, has even claimed \cite{Porto2016}

    ``[...] that adopting an \acrshort{eft} framework, when possible, greatly simplifies the computations and provides the required intuition for `physical understanding'.''

    My supervisor, a self-proclaimed \gls{relativist}, got curious, and wondered just how easy the \acrlong{eft} approach would make the computation. Therefore, he asked me if I would try to go through these computations, to test if they made the computation manageable even for master's students. My comments on Porto's claim are given in the discussion of \cref{chap:conclusion}.

    With verifying or refuting Porto's claim as the ultimate goal of this thesis, it is mostly written as a \gls{relativist}'s guide to a \glspl{field theorist}' approach to \acrlong{gw}s. It should also be useful for those with a field theoretical background who wish to understand how Feynman diagrams can be used in classical gravity, and \acrlong{gw} physics.

\section{Notation}
    This thesis uses the mostly positive flat space-time metric
    \begin{align*}
       &\eta\ind{_{\mu\nu}} = \begin{pmatrix} -1 & 0 & 0 & 0 \\ 0 & 1 & 0 & 0 \\ 0 & 0 & 1 & 0 \\ 0 & 0 & 0 & 1 \end{pmatrix} & &\qq*{Flat metric}
    \end{align*}

    Four-vectors are written with Greek letter indices, and spatial vectors with Latin letter indices. The Einstein summation convention applies.
    \begin{align*}
        &x\ind{^\mu} = \MixVec{x\ind{^0}}{\tvec{x}} = \MixVec{\ifthenelse{\boolean{NaturalUnits}}{}{c} t}{ \tvec{x}} & &\qq*{Four-vector} \\
        &\partial\ind{_\mu} = \pdv{}{x\ind{^\mu}} = \MixVec{\ifthenelse{\boolean{NaturalUnits}}{}{\frac{1}{c}} \partial_t}{\nabla} & &\qq*{Four-gradienet} \\
        &\dd[4]{x} = \dd{x\ind{^0}} \dd[3]{x} = \ifthenelse{\boolean{NaturalUnits}}{}{c} \dd{t} \dd[3]{x} & &\qq*{Integration volume of space-time}
    \end{align*}

    Notably, the action is defined as
    \begin{align*}
        S = \int \dd{t} L = \int \ifthenelse{\boolean{NaturalUnits}}{\dd[4]{x}}{\frac{\dd[4]{x}}{c}} \L,
    \end{align*}
    with $L$ and $\L$ being the Lagrangian, and Lagrangian density, respectively.

    These tensor index notations are also used.
    \begin{align*}
    		&T\ind{_{[\mu\nu]}} \equiv \frac{1}{2!}\left( T\ind{_{\mu\nu}} - T\ind{_{\nu\mu}} \right) = A\ind{_{\mu\nu}} & &\qq*{Antisymmetrizing operation} \\
    		&T\ind{_{\{\mu\nu\}}} \equiv \frac{1}{2!}\left( T\ind{_{\mu\nu}} + T\ind{_{\nu\mu}} \right) = S\ind{_{\mu\nu}} & &\qq*{Symmetrizing operation} \\
	    	&T = T\ind{_\alpha^\alpha} = \eta\ind{_{\alpha\beta}} T\ind{^{\alpha\beta}} & &\qq*{Trace of tensor} \\
	    	&T\ind{_{\mu\nu,\alpha}} = \partial\ind{_\alpha}T\ind{_{\mu\nu}} & &\qq*{Partial derivative}\\
	    	&T\ind{_{\mu\nu,\alpha}^\alpha} = \partial\ind{_\alpha} \partial\ind{^\alpha} T\ind{_{\mu\nu}} = \dalembertian T\ind{_{\mu\nu}} & &\qq*{d'Alembertian operator}\\
    		&\bar{T}\ind{_{\mu\nu}} = \frac{1}{2} \left(T\ind{_{\mu\nu}} + T\ind{_{\nu\mu}} - T\ind{^\sigma_\sigma}\eta\ind{_{\mu\nu}}\right) & &\qq*{Bar operator}
    \end{align*}
    Colons will also appear in indices, but these have no mathematical meaning. Colons are simply used to separate pairs of indices that have distinct roles. E.g. could $T\ind{^{\mu\nu}} x\ind{^\lambda} \equiv S\ind{^{\mu\nu:\lambda}}$.

    Lastly, the Fourier transform, and inverse Fourier transform are defined by\footnote{Note that for most of this thesis, the tilde over the Fourier transformed function will be dropped, as the argument ($x$ or $k$) gives away whether it is a real-space or Fourier-space function.}
    \begin{align*}
        F(x) & = \int \frac{\dd[4]{k}}{(2\pi)^4} \Tilde{F}(k) e^{ik\ind{_\sigma} x\ind{^\sigma} },\\
        \Tilde{F}(k) & = \int \dd[4]{x} F(x) e^{-ik\ind{_\sigma} x\ind{^\sigma} }.
    \end{align*}
\chapter{The gravitational waveform} \label{chap:waveform}
    In this chapter the \acrlong{gw}form will be computed, both in the time domain \eqref{eq:waveform:Phi-tau} and in the frequency domain \eqref{eq:PsiSPA:f}.
    
    The computation follows standard methods, like presented in \textcite{Arun2008higherorder}.
    
\section{Setting up the equation for the gravitational waveform} \label{sec:SettingUpWaveform}
\subsection{What is a waveform?}
    As inferred by the name, \acrlong{gw}s are waves, which is to say they are solutions of the \emph{wave equation}.
    \begin{align}\label{eq:waveeq}
        \left( -\pdv[2]{}{\ifthenelse{\boolean{NaturalUnits}}{t}{(ct)}} + \nabla^2 \right) h\ind{_{\mu\nu}} = \partial\ind{_\alpha} \partial\ind{^\alpha} h\ind{_{\mu\nu}} \equiv \dalembertian h\ind{_{\mu\nu}} = 0.
    \end{align}
    Here the \emph{d'Alembert operator}, also called the \emph{d'Alembertian}, $\dalembertian$ has been defined, which is the operator of the wave equation.
    
    A simple solution to equation \eqref{eq:waveeq} is $h\ind{_{\mu\nu}} = \epsilon\ind{_{\mu\nu}} e^{-ik\ind{_\sigma} x\ind{^\sigma}}$, with $k\ind{_\mu}k\ind{^\mu} = -k_0^2 + \tvec{k}^2 = 0$, and where $\epsilon\ind{_{\mu\nu}}$ is some $x\ind{^\mu}$-independent tensor structure. The exponential is a plane wave solution, according to Euler's formula \eqref{eq:app:EulersFormula}. 
    
    Gravitational waves are rank two tensors, which means they have two indices and therefore $4\times 4=16$ components. It is also symmetric in these two indices: $h\ind{_{\mu\nu}}=h\ind{_{\nu\mu}}$, which means that only $10$ of these components are independent. The reason gravitational waves are rank two tensors follows in the \gls{relativist}s' approach because $h\ind{_{\mu\nu}}$ is a perturbation of the metric $g\ind{_{\mu\nu}}=\eta\ind{_{\mu\nu}} + h\ind{_{\mu\nu}}$, where $\eta\ind{_{\mu\nu}}$ is the flat space-time metric. In the \glspl{field theorist}' approach it is because gravity is the effect of a massless spin two field. $\epsilon\ind{_{\mu\nu}}$ is the polarization tensor of \acrshort{gw}s, and since it is a massless field it only has two independent polarizations. Gravitational waves are transverse, and thus $\epsilon\ind{_{ij}} k\ind{^j} =0$, i.e. the amplitude direction given by the polarization is orthogonal to the direction of propagation $\tvec{k}$.
    
    To solve the wave equation, the wave four-vector had to be null-like. This implies further that the wave itself must travel at the speed of light, $v=\pdv{\omega}{\abs{\tvec{k}}}=\pdv{\ifthenelse{\boolean{NaturalUnits}}{k_0}{ck_0}}{\abs{\tvec{k}}} = \ifthenelse{\boolean{NaturalUnits}}{1}{c}$. This is also a consequence of $h\ind{_{\mu\nu}}$ being a massless field.
    
    Because of the linearity of the wave operator, any sum of such exponential (or trigonometric) terms will also be a solution of the wave equation. The most general solution is thus, a sum over all null-like wave-vectors $k\ind{_\mu}$, and an expression which also leaves $h\ind{_{\mu\nu}}$ as a real function.\footnote{The exponential function with an imaginary argument is a great shorthand for trigonometric functions, but all observables must in the end be real valued.}
    \begin{align}\label{eq:wave:general}
        h\ind{_{\mu\nu}}(x\ind{^{\alpha}}) = \int \frac{\dd[3]{k}}{(2\pi)^3 \cdot 2\omega_k} \left\{ a\ind{_{\mu\nu}}(\tvec{k}) e^{-i k\ind{_\sigma} x\ind{^{\sigma}} } + a^\dagger_{\mu\nu}(\tvec{k}) e^{i k\ind{_\sigma} x\ind{^{\sigma}} } \right\}.
    \end{align}
    
    The expression above being real follows from the observation $h^\dagger_{\mu\nu}(x\ind{^{\alpha}}) = h\ind{_{\mu\nu}}(x\ind{^{\alpha}})$, which can only hold for real numbers. Here $\omega_k = \abs{\tvec{k}} = k_0$, which is to make the wave null-like, also known as `on shell'. For a derivation of this solution, see \cref{app:wave equation sol}.
    
    The coefficients $a\ind{_{\mu\nu}}(\tvec{k})$ are used to select particular solutions based on some initial condition, and are left to be determined.
    
    The frequency of the wave turns out to be integer multiples of the frequency at which the source binary orbits, which will be demonstrated in \cref{chap:flux}. Thus, it can be approximated as
    \begin{align} \label{eq:waveform:harmonicsDecomp}
        h\ind{_{ij}}(t) \simeq \epsilon\ind{_{ij}} \sum_{n=1}^\infty a_n(t) \cos(n\Phi(t)),
    \end{align}
    where $\Phi(t)$ is the phase of the source binary. The \emph{waveform} describes what kind of wave it is. $a_n(t)$ can be found, but the most important factor for detection of \acrlong{gw}s is contained in $\Phi(t)$. The reason for this is that \acrlong{gw} detectors receives faint signals with \emph{amplitudes} close to the amplitude of noise. However, the \emph{frequency} of \acrshort{gw}s is different from the major noise factors, and can thus be extracted using Fourier analysis. Therefore, in the rest of this chapter, and much of the literature, the word waveform will be used interchangeably about the phase, as it encodes information about the frequency spectrum.
    
    The orbital energy for circular, Newtonian motion is related to the frequency as \ifthenelse{\boolean{NaturalUnits}}{$E=-\frac{1}{2}\mu v^2 = -\frac{1}{2}\mu (M\omega)^{2/3} $}{ $E=-\frac{1}{2}\mu v^2 = -\frac{1}{2}\mu (GM\omega)^{2/3}$}, using $v=\omega r$ and Kepler's third law,
    \begin{align} \label{eq:KeplersThirdLaw}
        \omega^2 = \frac{\ifthenelse{\boolean{NaturalUnits}}{M}{GM}}{r^3},
    \end{align}
    to eliminate $r$ in favour of $\omega$.\footnote{How $v$, $\omega$, and $r$ are related follows from the \acrshort{eom}, which are presented in their 1\acrshort{pn} form in \eqref{eq:Relativistic: omega}-\eqref{eq:Relativistic: v2}.} As usual, $M=m_1+m_2$ is the total mass of the binary, $\mu=\frac{m_1m_2}{M}$ is the reduced mass, $G$ is Newton's gravitational constant, $r=\abs{\tvec{r}_2-\tvec{r}_1}$ is the spatial separation of the binary, and $v=\abs{\dot{\tvec{r}}}$ is the relative velocity. More details on Newtonian orbital mechanics and its associated masses and quantities can be found in \cref{app:equivOneBodyAndMassTerms}.
    
    The approximation of circular motion here might seem over idealized, but it turns out that the effect of \acrlong{gw} emission on elliptical orbits is to \emph{circularize} them.By the time the binary's frequency enters the detector range, near the time of coalescence, the orbits have become very circular, making circular orbits a sensible approximation.
    
    Noting that the energy was easier to handle with $v$ rather than $\omega$, as it has integer powers instead of fractional powers, one may use \ifthenelse{\boolean{NaturalUnits}}{$v = (M\omega)^{1/3}$}{$v = (GM\omega)^{1/3}$} as a proxy variable for the frequency. Note that as a Newtonian approximation this variable coincides with the relative velocity parameter, but this is no longer the case after relativistic corrections are accounted for.
    
    Then the phase of the orbit can be expressed as
    \ifthenelse{\boolean{NaturalUnits}}{
    \begin{align} \label{eq:waveform:partial:equation}
        \dv{\Phi}{t} = \omega = \frac{v^3}{M} \quad \Rightarrow \quad \dd{\Phi} = \frac{v^3}{M} \dd{t}.
    \end{align}
}{
    \begin{align} \label{eq:waveform:partial:equation}
        \dv{\Phi}{t} = \omega = \frac{v^3}{GM} \quad \Rightarrow \quad \dd{\Phi} = \frac{v^3}{GM} \dd{t}.
    \end{align}
}
    Sadly $v=v(t)$, which at this point is still an unknown function of time. However it is known that $v$ must evolve with time according to how the orbital energy evolves with time.
    
\subsection{Time evolution of orbital energy}
    The differential equation governing the dynamics of the orbital phase is
    \begin{equation}
		-\dv{E}{t} = \F, \label{eq:Energy-flux-relation}
	\end{equation}
    with $E$ the energy associated with conserved orbital motion, and $\F$ the total energy flux out of the system by means of \acrshort{gw}s. This is nothing but energy conservation for a gravitationally bound system.\footnote{It is not obvious that energy \emph{should} be conserved however. In full \acrshort{gr} there is no trivial argument why there should be a conserved energy quantity \cite{Szabados2009}, but in the \acrlong{pn} expansion the dynamics are expanded around the Newtonian problem, in which energy is conserved. Thus it it can be taken to be an artifact of the Newtonian background of which the solution is expanded in. Note however that energy conservation is not controversy free \cite{deharo2021noethers}.}
    
    Both $E$ and $\F$ can be analytically expanded in a relativistic parameter, like \ifthenelse{\boolean{NaturalUnits}}{$v$ ($v/c$ in non-natural units)}{$(v/c)$}. This requires a separation in scale, where on the short timescale the motion is conservative and has energy $E$, while on the long timescale the system loses energy to gravitational radiation at a rate $\F$, leading to an inspiral. This requires the inspiral to happen slowly compared to the orbital motion, so that at any one moment the motion can still adequately be described by Newtonian motion. Thus, it only works for relatively small values of $\F$, such that the objects do not fall down too rapidly.\footnote{Later in this chapter it will be shown that the requirement of slow infall can be fufilled by having $\dot{\omega}/\omega^2 \ll 1$ (see equation \eqref{eq:inspiral condition}), which is equivalent to having the orbital velocity $\omega r$ much greater than infall velocity $\dot{r}$.}
    
    Luckily, to leading order the flux term is suppressed by a factor of $c^{-5}$ compared to the leading order term of the energy. Thus, the approximation of so called \gls{QCO}s and post-Newtonian formalism holds surprisingly well, even when compared to numerical simulations of the full Einstein equations (see \textcite{CompOfPNandNR}).
    
    As will be demonstrated in \cref{chap:energy} and \ref{chap:flux}, the orbital energy \eqref{eq:1PN:Energy} and energy flux \eqref{eq:1PN:Flux} can be expanded in terms of \ifthenelse{\boolean{NaturalUnits}}{$v$}{$(v/c)$} as
    
    \ifthenelse{\boolean{NaturalUnits}}{
    \begin{align}
	\begin{split}\label{eq:Energy:Expansion:Gen}
		E & = E_\text{Newt} v^2 \left\{ 1 + \sum_{i=2}^\infty E_i v^i \right\} \\
		& = -\frac{\mu}{2} v^2 \left\{ 1+\left( -\frac{3}{4}-\frac{1}{12}\eta \right)v^2 + \order{v^3} \right\},
	\end{split}\\
	\begin{split}\label{eq:Flux:Expansion:Gen}
		\F & = F_\text{Newt} v^{10} \left\{ 1 + \sum_{i=2}^\infty F_i v^i  \right\} \\
		& = \frac{32}{5}\eta^2 v^{10} \left\{ 1+\left( -\frac{1247}{336}-\frac{35}{12}\eta \right)v^2 + \order{v^3} \right\}.
	\end{split}
	\end{align}
}{
    \begin{align}
	\begin{split}\label{eq:Energy:Expansion:Gen}
		E & = E_\text{Newt} v^2 \left\{ 1 + \sum_{i=2}^\infty E_i \left(\frac{v}{c}\right)^i \right\} \\
		& = -\frac{\mu}{2} v^2 \left\{ 1+\left( -\frac{3}{4}-\frac{1}{12}\eta \right)\frac{v^2}{c^2} + \order{\frac{v^3}{c^3}} \right\},
	\end{split}\\
	\begin{split}\label{eq:Flux:Expansion:Gen}
		\F & = F_\text{Newt} v^{10} \left\{ 1 + \sum_{i=2}^\infty F_i \left(\frac{v}{c}\right)^i  \right\} \\
		& = \frac{32}{5}\frac{\eta^2}{Gc^5} v^{10} \left\{ 1+\left( -\frac{1247}{336}-\frac{35}{12}\eta \right)\frac{v^2}{c^2} + \order{\frac{v^3}{c^3}} \right\}.
	\end{split}
	\end{align}
} 
    In the expansion \eqref{eq:Flux:Expansion:Gen} there is defined a \emph{Newtonian energy flux} $F_\text{Newt}= \frac{32 \eta^2}{5\ifthenelse{\boolean{NaturalUnits}}{}{Gc^5}}$ where $\eta\equiv \mu/M$ is the \emph{symmetric mass ratio} (see \cref{app:equivOneBodyAndMassTerms}, and especially equation \eqref{eq:masses:def}, for more details). This is strange since there are no \acrshort{gw}s in Newtonian theory, so what is this flux? It is merely a convention to call leading order terms Newtonian, and this is why it is referred to as `Newtonian'.
    
	Since $v$ is just a proxy for the frequency the expression \eqref{eq:Energy:Expansion:Gen}-\eqref{eq:Flux:Expansion:Gen} would be different expressed in terms of the \emph{actual} centre-of-mass frame relative velocity. This point will be revisited in \cref{chap:energy} and \ref{chap:flux}.
	
    Up to $(v/c)^2$ order corrections define the first \acrlong{pn} order, or 1\acrshort{pn} for short, and is the leading order correction. This has started the convention of calling terms $\sim \left(\ifthenelse{\boolean{NaturalUnits}}{v}{v/c}\right)^{2i}$ for $i$\acrshort{pn} order corrections, e.g. the leading order, Newtonian, term is 0\acrshort{pn} order. This has a somewhat awkward effect, since not all terms are even powers of $\ifthenelse{\boolean{NaturalUnits}}{v}{v/c}$, already the next order correction is $\sim \left(\ifthenelse{\boolean{NaturalUnits}}{v}{v/c}\right)^{3}$, and is thus of $1.5$\acrshort{pn} order. Higher order terms of both the energy and flux, and the final result of this chapter: The waveform, can be found in papers like \textcite{Arun2008higherorder}.
    
    Using equation \eqref{eq:Energy-flux-relation} the time evolution $\dd{t}$ can be expressed in terms of $v$ as
	\begin{empheq}[box=\widefbox]{align} \label{Eq:time-evolution-v} 
	    \dd{t} = -\frac{1}{\F}\dd{E} = -\frac{1}{\F}\dv{E}{v} \dd{v}.
	\end{empheq}
	Substituting \eqref{Eq:time-evolution-v} for $\dd{t}$ in \eqref{eq:waveform:partial:equation} results in the final expression for which the waveform can be derived (using \eqref{eq:Energy:Expansion:Gen}-\eqref{eq:Flux:Expansion:Gen})
	\ifthenelse{\boolean{NaturalUnits}}{
	\begin{empheq}[box=\widefbox]{align} \label{eq:waveform:diffEq}
	    \dd{\Phi} = -\frac{v^3}{M}\frac{1}{\F}\dv{E}{v} \dd{v}.
	\end{empheq}
	}{
	\begin{empheq}[box=\widefbox]{align} \label{eq:waveform:diffEq}
	    \dd{\Phi} = -\frac{v^3}{GM}\frac{1}{\F}\dv{E}{v} \dd{v}.
	\end{empheq}
	}
	
	Solving \eqref{Eq:time-evolution-v} will provide $v$ as a function of time. We proceed however by computing $\Phi$ as a function of $v$ directly rather than of time, as ultimately to be compared with experiments it is the waveform in the frequency domain (which will be called $\Psi$) which is needed. As already mentioned, this is because the signal is filtered in the frequency domain, and therefore the highest resolution is in the frequency spectrum.
\section{Computing the waveform} \label{sec:ComputingWaveform} \label{sect:CompWaveForm}
\subsection{Computing the waveform as a function of time}
    In order to compute $\Phi(t)$ it is convenient to first compute $\Phi(v)$ (equation \eqref{eq:Phi-v}), then $v(t)$ (equation \eqref{eq:waveform:vOfTau}), and lastly $\Phi(t)=\Phi(v(t))$ (equation \eqref{eq:waveform:Phi-tau}).
\subsubsection{Computing the waveform as a function of frequency} \label{sect:CompWaveForm:sub:waveformOfFrequency}
	Combining \eqref{eq:waveform:diffEq} with \eqref{eq:Energy:Expansion:Gen}-\eqref{eq:Flux:Expansion:Gen} yield up to 1\acrshort{pn} 
	\ifthenelse{\boolean{NaturalUnits}}{
	\begin{align}
		\nonumber \dd{\Phi} &= -\frac{v^3}{M} F_\text{Newt}^{-1} v^{-10} \left[ 1+\left( -\frac{1247}{336}-\frac{35}{12}\eta \right) v^2 \right]^{-1} \dv{v} \left[  E_\text{Newt} v^2 \left( 1+\left( -\frac{3}{4}-\frac{1}{12}\eta \right) v^2 \right) \right] \dd{v} \\
		\nonumber &= -\frac{1}{M}\frac{E_\text{Newt}}{F_\text{Newt}} \frac{1}{v^6} \frac{ 2 + \left( -3 -\frac{1}{3}\eta \right)v^2 }{ 1+\left( -\frac{1247}{336}-\frac{35}{12}\eta \right)v^2 } \dd{v} \\
		&\equiv -\frac{1}{M}\frac{E_\text{Newt}}{F_\text{Newt}} \frac{1}{v^6} \frac{ 2 + \alpha v^2 }{ 1 + \beta^2 v^2 } \dd{v}. \label{Eq:dPhi:Preliminary}
	\end{align}
	}{
	\begin{align}
		\nonumber \dd{\Phi} &= -\frac{v^3}{GM} F_\text{Newt}^{-1} v^{-10} \left[ 1+\left( -\frac{1247}{336}-\frac{35}{12}\eta \right)\frac{v^2}{c^2} \right]^{-1} \dv{v} \left[  E_\text{Newt} v^2 \left( 1+\left( -\frac{3}{4}-\frac{1}{12}\eta \right)\frac{v^2}{c^2} \right) \right] \dd{v} \\
		&= \frac{-2}{GM}\frac{E_\text{Newt}}{F_\text{Newt}} \frac{1}{v^6} \frac{ 1 + \left( -\frac{3}{2} -\frac{1}{6}\eta \right) v^2/c^2 }{ 1+\left( -\frac{1247}{336}-\frac{35}{12}\eta \right) v^2/c^2 } \dd{v} \equiv \frac{-2}{GM}\frac{E_\text{Newt}}{F_\text{Newt}} \frac{1}{v^6} \frac{ 1 +\alpha v^2/c^2 }{ 1 + \beta v^2/c^2 } \dd{v}. \label{Eq:dPhi:Preliminary}
	\end{align}
	}
	
	To evaluate this integral it would be advantageous to write the last fraction in an easier form. Utilising that \ifthenelse{\boolean{NaturalUnits}}{$v$ is small the last fraction can be Taylor expanded around $v=0$ up to 1\acrshort{pn}. Notice that this should only be done for the last fraction, as it contain all \acrshort{pn} corrections of the energy and flux, i.e. in none natural units only in the last fraction will $v \to v/c$.}{$v/c$ is small the last fraction can be Taylor expanded around $v/c=0$ up to 1\acrshort{pn}.}
	
	Performing the Taylor expansion results in
	\begin{align*}
	    \frac{1+\alpha x}{1+\beta x} \stackrel{ \text{for }x\sim 0 }{ \simeq } 1 + (\alpha-\beta)x + \beta(\beta-\alpha)x^2 + \dots
	\end{align*}
	This result inserted in \eqref{Eq:dPhi:Preliminary} yields the easily integratable 1\acrshort{pn} expression
	\ifthenelse{\boolean{NaturalUnits}}{
    \begin{align} \label{eq:dPhi:finalExpression}
		\dd{\Phi} &= -\frac{2}{M}\frac{E_\text{Newt}}{F_\text{Newt}} \frac{1}{v^6} \left\{ 1 + \left( \frac{743}{336} + \frac{11}{4}\eta \right) v^2 \right\} \dd{v}.
	\end{align}
	}{
	\begin{align} \label{eq:dPhi:finalExpression}
		\dd{\Phi} &= -\frac{2}{GM}\frac{E_\text{Newt}}{F_\text{Newt}} \frac{1}{v^6} \left\{ 1 + \left( \frac{743}{336} + \frac{11}{4}\eta \right) \frac{v^2}{c^2} \right\} \dd{v}.
	\end{align}
	}
	
	Integrating to obtain $\Phi(t)=\Phi(v(t))$ one must choose a reference point in time, usually referred to as $t_0$. For binary inspirals this reference point is canonically chosen to be the moment of coalescence $t_c$ (see \textcite{Maggiore:VolumeI} chapter 4), which for the duration of the inspiral is in the future. Therefore, the integration variables should go from $v(t)$ to $v_c=v(t_c)$, but a multiplication of $-1$ to both sides can flip this order. Performing the integral finally provides $\Phi(v)$
	\ifthenelse{\boolean{NaturalUnits}}{
	\begin{align}
	\begin{split}
        \Phi(v) & = \Phi_c - \frac{ 2E_\text{Newt} }{ MF_\text{Newt} } \int_{v_c}^v {v'}^{-6}\left\{ 1+ \left( \frac{743}{336} + \frac{11}{4}\eta \right) {v'}^2 \right\} \dd{v'} \\
        & = \Phi_c + \frac{ 2E_\text{Newt} }{ MF_\text{Newt} } \left[ \frac{1}{5}{v'}^{-5}\left\{ 1 + \frac{5}{3}\left( \frac{743}{336} + \frac{11}{4}\eta \right) {v'}^2 \right\}\right]_{v'=v_c}^{v'=v}
    \end{split}
	\end{align}
	}{
    \begin{align}
	\begin{split}
        \Phi(v) & = \Phi_c - \frac{ 2E_\text{Newt} }{ GMF_\text{Newt} } \int_{v_c}^v {v'}^{-6}\left\{ 1+ \left( \frac{743}{336} + \frac{11}{4}\eta \right) \frac{{v'}^2}{c^2} \right\} \dd{v'} \\
        & = \Phi_c + \frac{ 2E_\text{Newt} }{ GMF_\text{Newt} } \left[ \frac{1}{5}{v'}^{-5}\left\{ 1 + \frac{5}{3}\left( \frac{743}{336} + \frac{11}{4}\eta \right) \frac{{v'}^2}{c^2} \right\}\right]_{v'=v_c}^{v'=v}
    \end{split}
	\end{align}
	}
	
	Collecting all constant terms into one phase constant $\Phi_0$, writing out $E_\text{Newt}$ and $F_\text{Newt}$ from \eqref{eq:Energy:Expansion:Gen} and \eqref{eq:Flux:Expansion:Gen} respectively, results in the final result for the waveform as a function of $v$
	\ifthenelse{\boolean{NaturalUnits}}{
	\begin{empheq}[box=\widefbox]{align} \label{eq:Phi-v}
        \Phi(v) = \Phi_0 -\frac{ 1 }{ 2^5\eta } {v}^{-5}\left\{ 1 + \left( \frac{3715}{1008} + \frac{55}{12}\eta \right) {v}^2 + \order{v^3} \right\}.
	\end{empheq}
	}{
	\begin{empheq}[box=\widefbox]{align} \label{eq:Phi-v}
        \Phi(v) = \Phi_0 -\frac{ 1 }{ 2^5\eta } \frac{c^5}{v^5} \left\{ 1 + \left( \frac{3715}{1008} + \frac{55}{12}\eta \right) \frac{{v}^2}{c^2} + \order{\frac{{v}^3}{c^3}} \right\}.
	\end{empheq}
	
	The phase is dimensionless, as one should expect.\footnote{By definition the frequency measure $v=(GM\omega)^{1/3}$ has dimension of velocity, in accordance with the symbol used.}
	}
	To obtain the waveform as a direct function of time the frequency parameter $v$ must be given as a function of time.
\subsubsection{Computing the frequency as a function of time} \label{sect:CompWaveForm:sub:frequencyOfTime}
	The frequency parameter $v$ as a function of time can be obtained from the differential equation \eqref{Eq:time-evolution-v}, in an equivalent fashion to how \eqref{eq:Phi-v} was derived.
	\ifthenelse{\boolean{NaturalUnits}}{
	\begin{align}
        \dd{t} = -\frac{1}{\F}\dv{E}{v}\dd{v} & = \frac{M}{v^3} \dd{\Phi} \stackrel{\eqref{eq:dPhi:finalExpression}}{=} -\frac{2E_\text{Newt}}{F_\text{Newt}} v^{-9} \left\{ 1 + \left( \frac{743}{336} + \frac{11}{4}\eta \right) v^2 \right\} \dd{v}, \\
        \Rightarrow \quad \quad & t_c - t = \frac{5}{2^8} \frac{M }{\eta } v^{-8} \left\{ 1 + \frac{8}{6}\left( \frac{743}{336} + \frac{11}{4}\eta \right) v^2 \right\}. \label{eq:waveform:tOfv}
    \end{align}
	}{
	\begin{align}
        \dd{t} = -\frac{1}{\F}\dv{E}{v}\dd{v} & = \frac{GM}{v^3} \dd{\Phi} \stackrel{\eqref{eq:dPhi:finalExpression}}{=} -\frac{2E_\text{Newt}}{F_\text{Newt}} v^{-9} \left\{ 1 + \left( \frac{743}{336} + \frac{11}{4}\eta \right) \frac{v^2}{c^2} \right\} \dd{v}, \\
        \Rightarrow \quad \quad & t_c - t = \frac{5}{2^8} \frac{GM c^5}{\eta } v^{-8} \left\{ 1 + \frac{8}{6}\left( \frac{743}{336} + \frac{11}{4}\eta \right) \frac{v^2}{c^2} \right\}. \label{eq:waveform:tOfv}
    \end{align}
	}
	
	Notice that $t_c-t$ was chosen for the \acrfull{lhs} such that the expression on the \acrfull{rhs} becomes strictly positive. This is desired because both sides must be raised to the negative one 4$\textsuperscript{th}$ power, in order to produce a quadratic equation of $v^2$. Taking the square root of the resulting solution for $v^2$, and Taylor expanding it to the 1\acrshort{pn} order yields the expression for $v(t)$.
	
	Following the aforementioned procedure, and using $\tau=t_c-t$, the frequency can be determined to be
	\ifthenelse{\boolean{NaturalUnits}}{
	\begin{empheq}[box=\widefbox]{align} \label{eq:waveform:vOfTau}
	    v(\tau) = \frac{1}{2}\left( \frac{5 M}{\eta} \right)^{1/8} \tau^{-1/8} \left\{ 1 + \left( \frac{743}{8064} + \frac{11}{96}\eta \right) \left( \frac{5 M}{\eta} \right)^{1/4} \tau^{-1/4} + \order{\tau^{-1/2}} \right\}.
	\end{empheq}
	}{
	\begin{empheq}[box=\widefbox]{align} \label{eq:waveform:vOfTau}
	    v(\tau) = \frac{c}{2}\left( \frac{5 GM }{c^3\eta} \right)^{1/8} \tau^{-1/8} \left\{ 1 + \left( \frac{743}{8064} + \frac{11}{96}\eta \right) \left( \frac{5 GM }{c^3\eta} \right)^{1/4} \tau^{-1/4} \right\}.
	\end{empheq}
	}
	
	By the definition of $v$, the actual frequency $\omega(\tau)=v^3/GM$ can be computed as well, for completeness.
	\ifthenelse{\boolean{NaturalUnits}}{
	\begin{align} \label{eq:waveform:omegaOfTau}
	    \omega(\tau) = \frac{5}{8} \left( 5 \M \right)^{-5/8}  \tau^{-3/8} \left\{ 1 + \frac{8}{5}\left( \frac{3715}{8064}+\frac{55}{96}\eta \right) \frac{3}{8} \left( \frac{5 M }{\eta} \right)^{1/4} \tau^{-1/4} + \dots \right\}.
	\end{align}
	}{
	\begin{align} \label{eq:waveform:omegaOfTau}
	    \omega(\tau) = \frac{5}{8} \left( \frac{ 5 G \M }{ c^{3} } \right)^{-5/8}  \tau^{-3/8} \left\{ 1 + \frac{8}{5}\left( \frac{3715}{8064}+\frac{55}{96}\eta \right) \frac{3}{8} \left( \frac{5 GM }{c^3\eta} \right)^{1/4} \tau^{-1/4} \right\}.
	\end{align}
	}
	This result can be compared with e.g., \textcite{Maggiore:VolumeI} (equation (5.258) on p. 295). Note that he, and most of the rest of the literature, use dimensionless variables\footnote{These are commonly denoted $x=v^2/c^2=(GM\omega / c^3)^{2/3}$, $\gamma=GM/rc^2$, and $\Theta=(5GM/\eta c^3)^{-1}(t_c-t)$. Performing the substitutions for \eqref{eq:waveform:vOfTau} should be straightforward.}, but the 1\acrshort{pn} part of the expression is equivalent to \eqref{eq:waveform:vOfTau} and \eqref{eq:waveform:omegaOfTau}.
	
\subsubsection{Computing the waveform as a function of time} \label{sect:CompWaveForm:sub:waveformOfTime}
    Substituting \eqref{eq:waveform:vOfTau} for $v$ in \eqref{eq:Phi-v} yields
    \ifthenelse{\boolean{NaturalUnits}}{
    \begin{align} \label{eq:waveform:Phi-tau}
    \boxed{
        \Phi(\tau) = \Phi_0 - \left( 5\M \right)^{-5/8 } \tau^{5/8} \left\{ 1 + \left( \frac{3715}{8064} + \frac{55}{96}\eta \right) \left(\frac{5M}{ \eta}\right)^{1/4} \tau^{-1/4} \right\}.}
    \end{align}
    }{
    \begin{align} \label{eq:waveform:Phi-tau}
    \boxed{
        \quad \Phi(\tau) = \Phi_0 - \left( \frac{5G\M}{c^3} \right)^{-5/8 } \tau^{5/8} \left\{ 1 + \left( \frac{3715}{8064} + \frac{55}{96}\eta \right) \left(\frac{5GM}{c^3 \eta}\right)^{1/4} \tau^{-1/4} \right\}.\quad}
    \end{align}
    }
    
    \begin{figure}
        \centering
        \begin{subfigure}{0.5\textwidth}
            \centering
            \includegraphics[width=\textwidth]{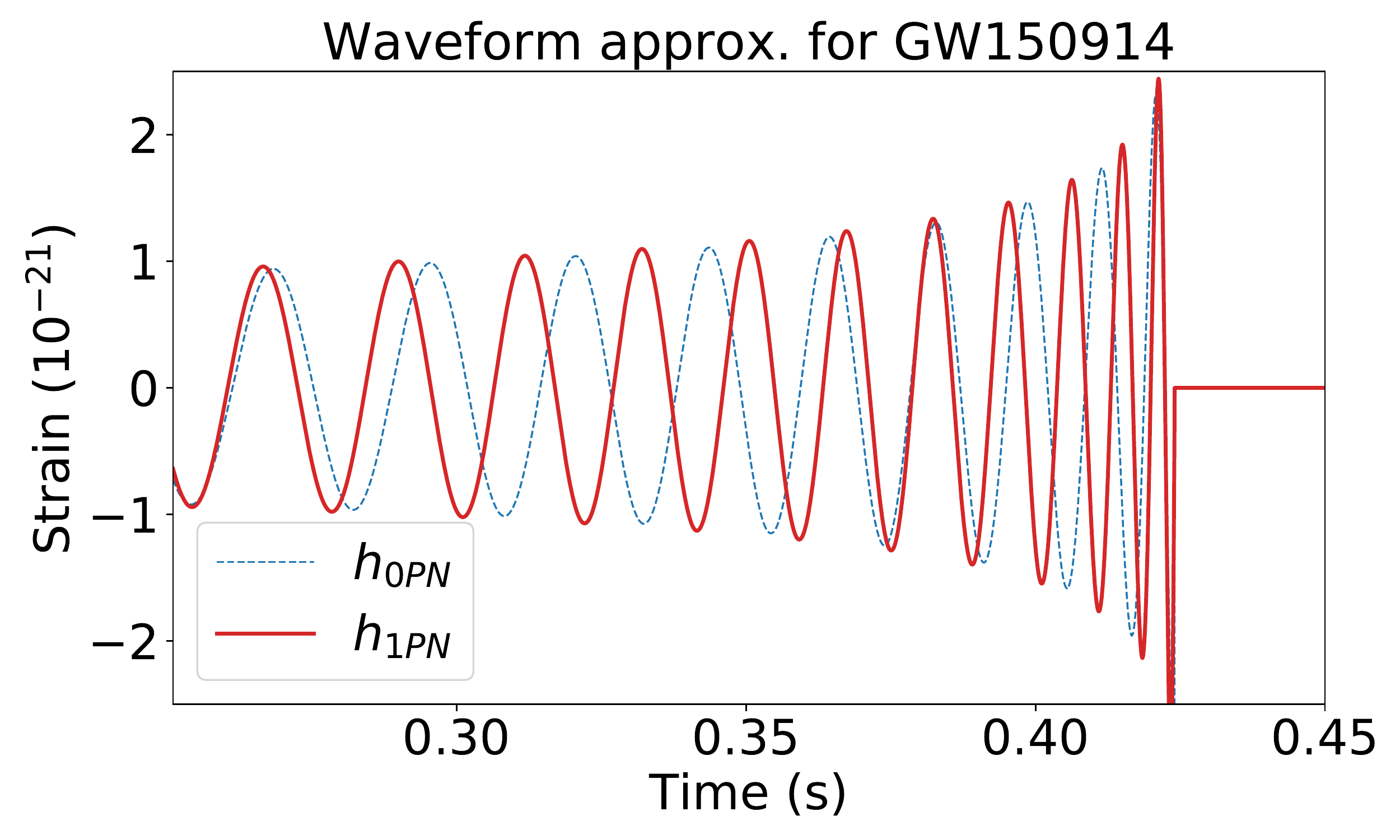}
            \caption{Waveform including (red) and excluding (blue) first order corrections. The waveform was plotted for values matching those of Table 1 of \textcite{FirstGW}. The plot depicts the waveform as seen in the detector frame, i.e. cosmologically redshifted.}
            \label{fig:MyWaveform}
        \end{subfigure}
        \begin{subfigure}{0.44\textwidth}
            \centering
            \includegraphics[width=\textwidth]{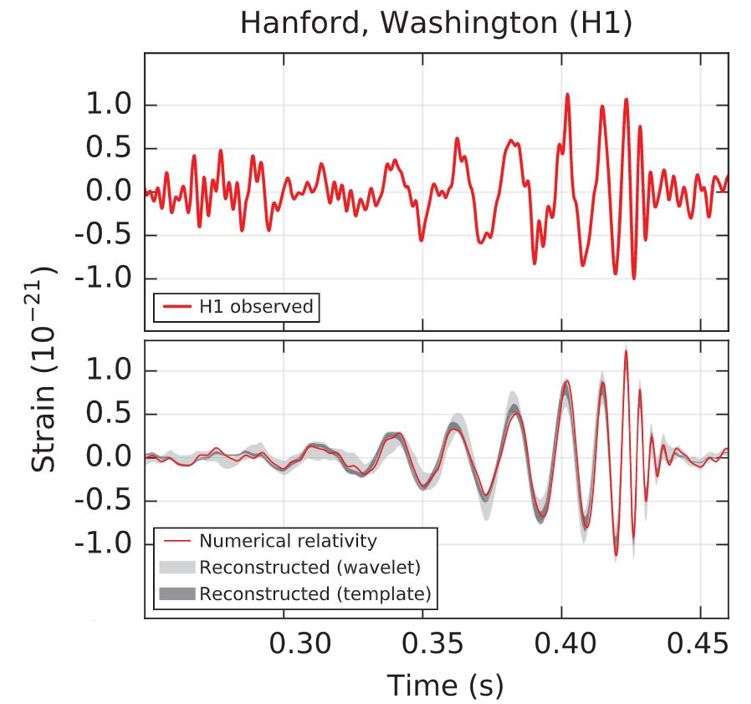}
            \caption{Data and model by \acrshort{ligo} for \gls{GW150914}, published in \textcite{FirstGW}.}
            \label{fig:LIGOData}
        \end{subfigure}
        
        \caption[Waveform of \acrshort{gw} produced by binary systems.]{Waveform based on computations of this thesis (\subref{fig:MyWaveform}) and \acrshort{ligo}'s data and model (\subref{fig:LIGOData}) for comparison. The plots share a similar structure, but it is clear by figure (\subref{fig:LIGOData}) that the signal-to-noise ratio is small, and much of the early-time structure is `washed out' by noise. That is to say, both the signal and the model in (\subref{fig:LIGOData}) has been filtered by frequency, making the plots somewhat jagged. Plot (\subref{fig:MyWaveform}) has not been filtered. In figure (\subref{fig:MyWaveform}) the amplitude diverges at $ t=0.425$s, while in the in (\subref{fig:LIGOData}) it does not. This is because the inspiral model breaks down here, and the merger phase takes over.
        }
    \end{figure}
    
    All that remains now to obtain the waveform is to find the amplitude of the different harmonics, and multiply them by $\cos(n\Phi(\tau))$. By using expression \eqref{eq:H:rad:fromMultipoles} from \cref{chap:flux} the amplitude of the $h_{1\text{\acrshort{pn}}}$ can be found to be
    \begin{align}
    \begin{split}
        h_{1\text{\acrshort{pn}}}(\tau) = -\frac{4G(1+z)\mu}{R c^2}  v^2_\text{1\acrshort{pn}}(\tau) \Biggl\{ &\left( 1 - \frac{23+27\eta}{42}  \frac{v^2_\text{0\acrshort{pn}}(\tau)}{c^2} \right) \cos\left( 2\Phi(\tau) \right) \\
        &+ \frac{77 \sqrt{1-4\eta}}{60} \frac{v_\text{0\acrshort{pn}}(\tau)}{c} \sin\left( 2\Phi(\tau) \right) \Biggr\}.
    \end{split}
    \end{align}
    Here $z$ is the cosmological redshift, see \textcite{Maggiore:VolumeI} for a derivation. 
    This function is plotted in \cref{fig:MyWaveform} using numbers from table 1 of \textcite{FirstGW} for $R$, $z$, $m_1$, and $m_2$. The functions $v(\tau)$ and $\Phi(\tau)$ are given by equation \eqref{eq:waveform:vOfTau} and \eqref{eq:waveform:Phi-tau} respectively. The 0\acrshort{pn} amplitude is obtained by using $v_\text{0\acrshort{pn}}(\tau)$ and neglecting the correctional terms inside the curly bracket that contain a factor of $c^{-1}$.
    
    In \cref{fig:MyWaveform} it is clear that 1\acrshort{pn} corrections does not affect the amplitude much \emph{directly}, but it has significant effect on the time evolution of the phase, and hence the frequency spectrum. It is however noticeable that the phase of \cref{fig:MyWaveform} does not match up with \cref{fig:LIGOData}. Either higher order corrections are required, or the model breaks down for such low values of $\tau$.
    
\subsection{Computing the Fourier transform of the waveform}	
    To obtain the high sensitivities in \acrshort{gw} detections the signal is Fourier transformed, in order to show which frequencies dominate the signal. This frequency spectrum can be compared to theoretical predictions to determine factors like the total mass at 0\acrshort{pn}, symmetric mass ratio at 1\acrshort{pn}, and more parameters at higher orders, e.g. spin at 1.5\acrshort{pn} \cite{Arun2008higherorder} and finite size effects like tidal deformation at 5\acrshort{pn} \cite{5PN}.
    
    In order to compare data with theoretical predictions these predictions must also be expressed in the frequency domain. Therefore, the desired waveform is $\Psi(f)$, which is the phase of the Fourier transformed waveform.
    
\subsubsection{The Fourier transform and \acrlong{spa}} \label{sect:CompWaveForm:sub:SPA}
    To compute the Fourier transformed $\tilde{B}(f)$ of some function $B(t)$ the \acrfull{spa} can be used, and it is commonly utilized to compute the Fourier transform of \eqref{eq:waveform:harmonicsDecomp}. Standardized in \acrshort{gw} physics by \textcite{SPAarticle} it approximates
    
    \begin{align}
        \nonumber \qq*{for} B(t) & = A(t)\cos(\Phi(t)), \\
        \begin{split}
            \Rightarrow \quad \tilde{B}(f) & \approx \frac{1}{2}A(t) \left(\dv{f}{t}\right)^{-1/2} \exp{ i \left( 2\pi f t - \Phi(f) - \pi/4 \right) }\\
            & \equiv \frac{1}{2}A(t) \left(\dv{f}{t}\right)^{-1/2} \exp{i \Psi(f)}, \label{eq:waveform:SPA:def}
        \end{split} \\
        \nonumber \qq*{provided} & \dv{\ln{A(t)}}{t} \ll \dv{\Phi}{t} \qq{and} \dv[2]{\Phi}{t}\ll\left(\dv{\Phi}{t}\right)^2.
    \end{align}
    
    This is exactly the type of expression which describes \acrshort{gw}s \eqref{eq:waveform:harmonicsDecomp}, and the conditions do indeed apply to the inspiral phase. 
    
    The leading order amplitude scales as $v^2 \sim \tau^{-1/4}$ \eqref{eq:waveform:vOfTau} (also, see equation \eqref{eq:Linearized:GWfromNewtonBinary:FinalExpression} in the next chapter for why the amplitude scales as $v^2$), while $\dd \Phi / \dd t = \omega(\tau) \sim \tau^{-3/8}$ (see \eqref{eq:waveform:omegaOfTau}). Thus, for large $\tau$, which is the time remaining till coalescence, $\dd{ \ln{a_n(t)}} / \dd t \sim \frac{1}{4}\tau^{-1} \ll  \omega(\tau) \sim \tau^{-3/8}$.
    
    As for the last prerequisite it can be shown to hold for \gls{QCO}s. Taking the time derivative of Kepler's third law \eqref{eq:KeplersThirdLaw} results in
    \ifthenelse{\boolean{NaturalUnits}}{
    \begin{align}
        \nonumber 2\omega \Dot{\omega} & = -3\Dot{r}\frac{M}{r^4} = -3\frac{\Dot{r}}{r} \omega^2, \\
        \Rightarrow \quad & \frac{-\Dot{r}}{\omega r} = \frac{2}{3} \frac{\Dot{\omega}}{\omega^2} \ll 1. \label{eq:inspiral condition}
    \end{align}
    }{
    \begin{align}
        \nonumber 2\omega \Dot{\omega} & = -3\Dot{r}\frac{GM}{r^4} = -3\frac{\Dot{r}}{r} \omega^2, \\
        \Rightarrow \quad & \frac{-\Dot{r}}{\omega r} = \frac{2}{3} \frac{\Dot{\omega}}{\omega^2} \ll 1. \label{eq:inspiral condition}
    \end{align}
    }
    For \gls{QCO}s the inspiral \emph{must} be slow compared to the orbital motion, and thus the radial velocity ($\Dot{r}$) must be small compared to the tangential velocity ($\omega r$), since for perfectly circular motion their fraction is identically zero. From Kepler's law this implies also that $\Dot{\omega}/\omega^2 \ll 1 \to \Ddot{\Phi} \ll \Dot{\Phi}^2$, which is exactly the condition required to use the \acrshort{spa}.
    
    This in hand also provides an estimate for the validity of this approximation, as $\omega$ is a known function of time \eqref{eq:waveform:omegaOfTau}
    \begin{align}
        \frac{2}{3} \frac{\Dot{\omega}}{\omega^2} \stackrel{\eqref{eq:waveform:omegaOfTau}}{\simeq} \frac{2}{5}\left(\ifthenelse{\boolean{NaturalUnits}}{5\M}{\frac{5G\M}{c^3}}\right)^{5/8} \tau^{-5/8} \ll 1.
    \end{align}
    This expression is indeed small for most values of $t<t_c$. 
    
    Since the most important part of the waveform for comparisons to experimental data is the frequency spectrum, the last computation of this chapter will be of the Fourier transformed phase $\Psi(f)$.
\subsubsection{Computing the \acrshort{spa} waveform}
	From equation \eqref{eq:waveform:SPA:def} the phase of the Fourier transformed waveform can be approximated as
	\ifthenelse{\boolean{NaturalUnits}}{
	\begin{equation}
		\Psi\ind{_{\text{\acrshort{spa}}}} = 2\pi f t(f) - \Phi(f) = \omega t(\omega) - \Phi(\omega) = \frac{v^3}{M} t(v) - \Phi(v).
	\end{equation}
	}{
	\begin{equation}
		\Psi\ind{_{\text{\acrshort{spa}}}} = 2\pi f t(f) - \Phi(f) = \omega t(\omega) - \Phi(\omega) = \frac{v^3}{GM} t(v) - \Phi(v).
	\end{equation}
	}
	$\Phi(v)$ being given by equation \eqref{eq:Phi-v}, and $t(v)$ by \eqref{eq:waveform:tOfv}, $\Psi(v)$ can easily be computed.
	
	\ifthenelse{\boolean{NaturalUnits}}{
	\begin{align}
		\frac{v^3}{M}t(v) & = \frac{v^3}{M}t_c - \frac{5}{2^8} \frac{1}{\eta} v^{-5} \left\{ 1 + \frac{8}{6}\left( \frac{743}{336} + \frac{11}{4}\eta \right) v^2 + \dots \right\},
	\end{align}
	\begin{empheq}[box=\widefbox]{align} \label{eq:PsiSPA:v}
		\Psi\ind{_{\text{\acrshort{spa}}}}(v) = \frac{v^3}{GM}t_c - \Phi_0 + \frac{3}{256}\frac{1}{\eta} v^{-5}\left\{ 1 + \left( \frac{3715}{756} + \frac{55}{9}\eta \right) v^2 + \dots \right\}.
	\end{empheq}
	}{
	\begin{align}
		\frac{v^3}{GM}t(v) & = \frac{v^3}{GM}t_c - \frac{5}{2^8} \frac{1}{\eta} \frac{c^5}{v^{5}} \left\{ 1 + \frac{8}{6}\left( \frac{743}{336} + \frac{11}{4}\eta \right) \frac{v^2}{c^2} + \dots \right\},
	\end{align}
	\begin{empheq}[box=\widefbox]{align} \label{eq:PsiSPA:v}
		\Psi\ind{_{\text{\acrshort{spa}}}}(v) = \frac{v^3}{GM}t_c - \Phi_0 + \frac{3}{256}\frac{1}{\eta} \frac{c^5}{v^{5}}\left\{ 1 + \left( \frac{3715}{756} + \frac{55}{9}\eta \right) \frac{v^2}{c^2} + \dots \right\}.
	\end{empheq}
	}
	
	Lastly the phase can be expressed in terms of the physical frequency by using  \ifthenelse{\boolean{NaturalUnits}}{
	$v= (M\omega)^{1/3}=(2\pi M f)^{1/3}$.
	\begin{empheq}[box=\widefbox]{align} \label{eq:PsiSPA:f}
		\Psi\ind{_{\text{\acrshort{spa}}}}(f)  = 2\pi f t_c - \Phi_0 + \frac{3}{256} \left( 2\pi \M f \right)^{-\frac{5}{3}} \left\{ 1 + \left( \frac{3715}{756} + \frac{55}{9}\eta \right) \left( 2\pi M f \right)^\frac{2}{3} \right\}.
	\end{empheq}
	}{$v= (GM\omega)^{1/3}=(2\pi GM f)^{1/3}$.
	\begin{empheq}[box=\widefbox]{align} \label{eq:PsiSPA:f}
	\begin{split}
		\Psi\ind{_{\text{\acrshort{spa}}}}(f)  = \hspace{3pt} & 2\pi f t_c - \Phi_0 + \frac{3}{256} \left( \frac{2\pi G \M f}{c^3} \right)^{-\frac{5}{3}} \\
		& \cdot \left\{ 1 + \left( \frac{3715}{756} + \frac{55}{9}\eta \right) \left( \frac{2\pi G M f}{c^3} \right)^\frac{2}{3} \right\}.
	\end{split}
	\end{empheq}
	}
	
	This is indeed equivalent to the expression found by \textcite{Arun2008higherorder} (equation (6.22), page 21) up to 1\acrshort{pn}, with some difference in notation.
	
	In order to compute this waveform all that is needed is the \acrshort{pn} expansion of the orbital energy, and the \acrshort{gw} energy flux, both associated with stable, energy conservative, motion. In \cite{Arun2008higherorder} these were provided with references to other papers. 
	
	In a sense, \eqref{eq:PsiSPA:f} is the final result of this thesis, computation wise, but it now remains to justify the expressions used for the \acrlong{pn} expansion of the orbital energy \eqref{eq:Energy:Expansion:Gen}, and energy flux expansion \eqref{eq:Flux:Expansion:Gen}, which will be derived in \cref{chap:energy} and \ref{chap:flux} respectively.
\chapter{Gravity as a gauge theory} \label{chap:fieldtheory}
    In this chapter the fundamental theory by which the orbital energy and energy flux will be calculated is derived. How can Einstein's general relativity be described as a classical field theory, and then recast into the language of \acrshort{eft}.
    
    The derivations presented in this chapter largely follows those presented in \textcite{Feynman:GravityLectures}, with supplements from \textcite{Maggiore:VolumeI} and \textcite{Porto2016}.
\section{Background} \label{sec:Background:ChapFieldTheory}
    The modern theory of gravity is partially split between two traditions. On the one hand there is the geometrical tradition following Einstein's approach by interpreting gravity as the effect of a curved space-time, which is curved according to the \gls{Einstein equations}. The followers of this tradition may be called \glspl{relativist}. On the other hand there is the tradition of using the formalism of Lorentz invariant fields on a static, Minkowskian, space-time, inspired by its monumental success for electrodynamics and quantum theory. The followers of this tradition may be called \glspl{field theorist}.
    
    Though these traditions are not entirely separated, the two different interpretations lend themselves to different \emph{natural} extensions of \acrlong{gr}. Thus the two traditions tend to separate \glspl{relativist} and \glspl{field theorist} by which theories they work on.
    
    In this thesis the 1\acrshort{pn} phase of \acrshort{gw} produced by compact binaries are computed using the formalism of field theory. Familiarity with basic \acrfull{qft} is expected, but the derivations are otherwise supposed to be elementary.
    
\subsubsection{Feynman and gravity}
    One of the most famous \glspl{field theorist}, R. P. Feynman had a ``gravity phase'' from 1954 to the late 1960s (\textcite{FeynmanReview}). After having worked on the foundations of \emph{quantum electrodynamics}, Feynman sought to uncover the quantum nature of gravity pursuing a similar method. He reckoned that gravity could, similarly to electromagnetism, be perturbatively expanded with respect to its coupling constant, and then quantized by quantizing the frequencies.
    
    Quantizing gravity turns out to be a little more complicated than that, but Feynman's approach to classical gravity as a massless, spin 2, gauge field has made a lasting impression on gravity physics, especially in the context of \acrshort{gw}s. This approach can be studied in the lecture notes from his lecture series of the 60's \cite{Feynman:GravityLectures}.
\section{Fierz-Pauli Lagrangian} \label{sec:FPLagrangian}
    To linearized order of the field ($h\ind{_{\mu\nu}}$) in the resulting \acrshort{eom}, the \gls{Einstein-Hilbert action} of \acrlong{gr} is equivalent to the massless \acrlong{fp} action from field theory \cite{Fierz:1939ix}.
\subsection{Deriving the graviton Lagrangian}
	When Feynman set out to study gravity, he took the mindset of a field theorist who until recently was unaware of gravitation, and just now have been presented with data suggesting that all masses attract other masses according to an inverse square law, proportional to the product of their masses,
	\begin{align*}
	    \tvec{F}\sim -\frac{m_1m_2\tvec{r}}{r^3}.
    \end{align*}
    
	Feynman envisioned this as the mindset of aliens on Venus who had just now acquired the technology to pierce through the atmosphere and measure the movement of the planets, but were still our equals in particle physics.
	
	Their first impulse would probably be to guess that this is an unknown effect of some known field. After finding no field that could replicate the solar system observation, their next guess would be that there exists a new kind of field which mediates this mysterious force. Calling this hypothetical field \emph{the graviton field}, and its associated quantum particle the graviton, the Venusians would next try to uncover its structure.
	
	To construct the Lagrangian for this new force of nature they would determine that it has to be of even spin, and thus an even tensor rank, for the resulting static force to be attractive for equal charges, where the charge for the graviton field would be mass. For the force to go as an inverse square the field must also be massless.
	
	Lastly, it must couple to all matter equally, but it must do so in a relativistic way. The natural suggestion is to somehow couple the field with the four-velocity of the source, like how the electric charge which the electric field couples to is promoted to the charge density four current $j^\mu = \gamma^{-1} \rho u^\mu$, and couples to the vector potential \ifthenelse{\boolean{NaturalUnits}}{$A_\mu = \MixVec{\phi}{A_i}$}{$A_\mu = \MixVec{\phi/c}{A_i}$}. See \textcite{SpesRel} or other textbooks on relativistic field theory.
	
	However, to let the graviton field couple to \emph{all} fields a natural candidate is the energy-momentum tensor $T\ind{^{\mu\nu}}$, induced by field invariance under space-time translations. Incidentally, for a \acrlong{pp} it is constructed by the four-velocity of the source: $T\ind{^{\mu\nu}} = \gamma^{-1} p\ind{^\mu} u\ind{^\nu} = \gamma^{-1} \rho u\ind{^\mu} u\ind{^\nu}$. Now for a scalar field it can be contracted to form a scalar, the trace, which is proportional to the mass density. Alternatively, a field of higher tensor rank can couple to the indices, also coupling the field to the mass density in the static frame $T\ind{^{\mu\nu}} = T\ind{^{00}} \delta\ind{^{\mu 0}} \delta\ind{^{\nu 0}}$.
	
	The spin zero / scalar field is a candidate for the graviton, but fails to couple to the electromagnetic energy-momentum tensor, as the electromagnetic energy-momentum tensor is traceless. It also fails to predict the perihelion procession of Mercury correctly \cite{SpesRel}.
	
	Thus, the Venusians would probably try a massless spin 2 field next. Since massless fields only have one (spin $s=0$) or two degrees of freedom (helicity $=\pm s$) the symmetric spin two field should be \emph{easiest} to work with, as it will have $10-2=8$ redundant degrees of freedom. The antisymmetric field by comparison only have $6-2=4$ redundant degrees of freedom.
	
	Thus demanding the Lagrangian to be composed of a massless, symmetric, rank two tensor field there are only four unique terms, containing only second / two derivatives, after considering partial integrations:
	\begin{enumerate}
		\item[] \hspace{-1cm} Two where the index of the tensor and the index of the derivative differ.
		\item $h\ind{_{\mu\nu,\rho}}h\ind{^{\mu\nu,\rho}}$
		\item \label{list:Divergenceless}$h\ind{_{\mu\nu,\rho}}h\ind{^{\mu\rho,\nu}}$
		\item[] \hspace{-1cm} And three where two of the indices contract for the individual $h\ind{_{\mu\nu}}$.
		\item \label{list:Divergnecefull} $h\ind{_{\mu\nu}^{,\nu}}h\ind{^{\mu\rho}_{,\rho}}$
		\item $h\ind{_{\mu\nu}^{,\nu}} h\ind{^{,\mu}} \qq{with} h\equiv h\ind{_\sigma^\sigma}$
		\item $h\ind{_{,\mu}}h\ind{^{,\mu}}$
	\end{enumerate}
	Why are these terms of quadratic order in $h\ind{_{\mu\nu}}$? Because it is action terms of quadratic order in a field which yields \acrshort{eom}s of linear order of that field.
	
	Note that term number \ref{list:Divergenceless}. and \ref{list:Divergnecefull}. are the same after two successive partial integrations $h\ind{_{\mu\nu,\rho}}h\ind{^{\mu\rho,\nu}} = - h\ind{_{\mu\nu}}h\ind{^{\mu\rho,\nu}_\rho} = h\ind{_{\mu\nu}^{,\nu}}h\ind{^{\mu\rho}_{,\rho}}$. Some texts use term \ref{list:Divergenceless}. (like \textcite{Maggiore:VolumeI}), but here term \ref{list:Divergnecefull}. will be employed (like in \textcite{Feynman:GravityLectures}). Thus, the free part of the Lagrangian\footnote{Terms $\sim h^1$ and $h^0$ only contribute constants to the \acrlong{eom}, and can thus be removed by field shifts. Terms proportional to $h^2$, but with no derivatives, determine the mass of the field $\sim m^2 h h$, and must therefore be zero for massless fields. Lastly, the Lagrangian must be a scalar in order to be Lorentz invariant. There are no contractions of only 1 derivative and two $h$'s that can produce a scalar. Therefore, to leading order in $h^n$, the Lagrangian must consist of terms proportional to $h^2$ with two derivatives. See e.g. \textcite{NoNonsenseQFT}, page 573-575, for a more detailed discussion.} must be of the form
	\begin{align}\label{eq:LagrangianH2coeffUndetermined}
		\L = a_1 h\ind{_{\mu\nu,\rho}}h\ind{^{\mu\nu,\rho}} + a_2 h\ind{_{\mu\nu}^{,\nu}}h\ind{^{\mu\rho}_{,\rho}} + a_3 h\ind{_{\mu\nu}^{,\nu}} h\ind{^{,\mu}} + a_4 h\ind{_{,\mu}}h\ind{^{,\mu}}.
	\end{align}
	It is possible to determine all the coefficients $a_{1-4}$ by imposing gauge invariance on the \acrfull{eom}. The \acrshort{eom} for fields is determined by the Euler-Lagrange equation \eqref{eq:EoM:FPL} (see e.g. \textcite{GoldsteinMech}, or \textcite{Kachelriess:2017cfe}), and for \eqref{eq:LagrangianH2coeffUndetermined} the \acrlong{eom} becomes \eqref{eq:Xi:def:MuNu}.
	\begin{subequations}
	\begin{align}
		\partial_\rho \pdv{\L}{h\ind{_{\mu\nu,\rho}}} - & \pdv{\L}{h\ind{_{\mu\nu}}} = 0 \label{eq:EoM:FPL}\\
		= \partial\ind{_\rho}\Bigl( 2a_1 h\ind{^{\mu\nu,\rho}} + a_2 \eta\ind{^{\nu\rho}} h\ind{^{\mu\sigma}_{,\sigma}} + a_2 \eta\ind{^{\mu\rho}} h\ind{^{\nu\sigma}_{,\sigma}} + a_3& \eta\ind{^{\nu\rho}}h\ind{^{,\mu}} + a_3 \eta\ind{^{\mu\nu}} h\ind{^{\rho\sigma}_{,\sigma}} + 2 a_4 \eta\ind{^{\mu\nu}}h\ind{^{,\rho}} \Bigr) \label{eq:Xi:Intermidiate:MuNuRho}  \\
		= 2 a_1 h\ind{^{\mu\nu,\rho}_\rho} + a_2 h\ind{^{\mu\rho,\nu}_\rho} + a_2 h\ind{^{\nu\rho,\mu}_\rho} + a_3 h\ind{^{,\mu\nu}} +& a_3 \eta\ind{^{\mu\nu}}h\ind{^{\rho\sigma}_{,\rho\sigma}} + 2a_4 \eta\ind{^{\mu\nu}} h\ind{^{,\rho}_\rho} \equiv \Xi\ind{^{\mu\nu}}. \label{eq:Xi:def:MuNu} 
	\end{align}
	\end{subequations}
	From the action of $\L+\L_\text{int} = \L + \frac{\lambda}{2} h\ind{_{\mu\nu}}T\ind{^{\mu\nu}}$ the inferred \acrshort{eom} should be
	\begin{subequations}
	\begin{align}
		\Xi\ind{^{\mu\nu}} &= -\frac{1}{2}\lambda T\ind{^{\mu\nu}},\\
		\label{eq:EoMLh2MustBeDivergenceless}T\ind{^{\mu\nu}_{,\nu}} = 0 \quad &\Rightarrow \quad \Xi\ind{^{\mu\nu}_{,\nu}} = 0.
	\end{align}
	\end{subequations}
	
	Equation \eqref{eq:EoMLh2MustBeDivergenceless} can be used to fix the coefficients of equation \eqref{eq:Xi:def:MuNu}, and thus also the Lagrangian.
	\begin{align}
		\nonumber \Xi\ind{^{\mu\nu}_{,\nu}} = \dalembertian h\ind{^{\mu\nu}_{,\nu}} (2a_1+a_2) + \dalembertian h\ind{^{,\mu}}(&a_3+2a_4) + h\ind{^{\rho\sigma,\mu}_{\rho\sigma}}(a_2+a_3) = 0, \\
		\Rightarrow \quad a_1 = -\frac{1}{2}a_2 &= \frac{1}{2}a_3 = -a_4.
	\end{align}
	
	Thus a Lagrangian of a symmetric, massless, rank 2 tensor field which couples to a divergenceless rank 2 tensor field (e.g. $\L_\text{int} = -\frac{\lambda}{2} h\ind{_{\mu\nu}}T\ind{^{\mu\nu}}$), consisting of only second derivatives, in a flat space-time, must to second power of $h$ take the form of the \emph{massless \acrlong{fp} Lagrangian} \cite{Fierz:1939ix}
	\begin{empheq}[box=\widefbox]{align}\label{eq:FierzPauliLagrangian}
		\L_\text{\acrshort{fp}} = -\frac{1}{2} h\ind{_{\mu\nu,\rho}}h\ind{^{\mu\nu,\rho}} + h\ind{_{\mu\nu}^{,\nu}}h\ind{^{\mu\rho}_{,\rho}} - h\ind{_{\mu\nu}^{,\nu}} h\ind{^{,\mu}} + \frac{1}{2} h\ind{_{,\mu}}h\ind{^{,\mu}}.
	\end{empheq}
	Here the overall factor has been set to $a_1=-1/2$.
\subsection{The equation of motion and gauge condition}
	
	The \acrshort{eom} for $\L_\text{\acrshort{fp}} + \frac{1}{2}\lambda h\ind{_{\mu\nu}}T\ind{^{\mu\nu}}$ follows directly from the Euler-Lagrange equation \eqref{eq:EoM:FPL} as
	\begin{align}
		-\dalembertian h\ind{_{\mu\nu}} + 2 h\ind{^{\alpha}_{\{\mu,\nu\}\alpha}} - h\ind{_{,\mu\nu}} - \eta\ind{_{\mu\nu}} \left( h\ind{_{\rho\sigma}^{,\rho\sigma}} -\dalembertian h \right) = \frac{\lambda}{2} T\ind{_{\mu\nu}}.
	\end{align}
	
	This is equivalent with the \acrlong{eom} found in the linear approximation of general relativity, for an appropriate choice of $\lambda$ (see e.g. equation (1.17) of \textcite{Maggiore:VolumeI}, or equation (9.16) of \textcite{Gron})
	
	Varying the \acrlong{fp} action directly should also provide the equations of motion
	\begin{align}
		\nonumber \var{\L_\text{\acrshort{fp}}} = \hspace{3pt} & \fdv{\L_\text{\acrshort{fp}}}{h\ind{_{\mu\nu,\rho}}} \var{h\ind{_{\mu\nu,\rho}}} + \fdv{\L_\text{\acrshort{fp}}}{h\ind{_{\mu\nu}}}\var{h\ind{_{\mu\nu}}} = \fdv{\L_\text{\acrshort{fp}}}{h\ind{_{\mu\nu,\rho}}} \partial\ind{_\rho} \var{h\ind{_{\mu\nu}}} + \fdv{\L_\text{\acrshort{fp}}}{h\ind{_{\mu\nu}}} \var{h\ind{_{\mu\nu}}}\\
		=& \left[ \fdv{\L_\text{\acrshort{fp}}}{h\ind{_{\mu\nu}}} - \partial\ind{_\rho} \fdv{\L_\text{\acrshort{fp}}}{h\ind{_{\mu\nu,\rho}}} \right]\var{h\ind{_{\mu\nu}}} = - \Xi\ind{^{\mu\nu}}\var{h\ind{_{\mu\nu}}} = 0. \label{eq:VariationOfFPLagrangian}
	\end{align}
	This automatically holds because of \eqref{eq:Xi:def:MuNu} ($\Xi\ind{^{\mu\nu}}=0$). But \eqref{eq:VariationOfFPLagrangian} can also be solved using the condition \eqref{eq:EoMLh2MustBeDivergenceless}, $\Xi\ind{^{\mu\nu}_{,\nu}}=0$. Letting $\var{h\ind{_{\mu\nu}}} = - \xi\ind{_{\mu,\nu}} - \xi\ind{_{\nu,\mu}}$ it is easy to show that the action stays invariant under this type of transformation, using partial integration.
	\begin{align}
		\var{\L_\text{\acrshort{fp}}} = \Xi\ind{^{\mu\nu}}\left( \xi\ind{_{\mu,\nu}} + \xi\ind{_{\nu,\mu}} \right) = - \Xi\ind{^{\mu\nu}_{,\nu}} \xi\ind{_{\mu}} -  \Xi\ind{^{\mu\nu}_{,\mu}} \xi\ind{_{\nu}} = -2 \Xi\ind{^{\mu\nu}_{,\nu}} \xi\ind{_{\mu}} \stackrel{\eqref{eq:EoMLh2MustBeDivergenceless}}{=} 0.
	\end{align}
	Thus the following transformation of the field leaves both the \acrshort{eom} and the gauge condition invariant.
	\begin{empheq}[box=\widefbox]{align}\label{eq:Gaugecondition:spin2field}
		h\ind{_{\mu\nu}}(x) \to h\ind{_{\mu\nu}}(x) + \var{h\ind{_{\mu\nu}}(x)} = h\ind{_{\mu\nu}}(x) - \xi\ind{_{\mu,\nu}}(x) - \xi\ind{_{\nu,\mu}}(x).
	\end{empheq}
	
	Again, this is equivalent to the gauge condition found in linear theory when linearizing metric invariance under change of coordinates (see equation (9.9) of \textcite{Gron}). 
	
	Also introducing the commonly used \emph{bar operator}, which symmetrize tensors and changes the sign of their trace,
	\begin{align}\label{eq:def:BarOperation}
	    \bar{S}\ind{_{\mu\nu}} \equiv \frac{1}{2} \left(S\ind{_{\mu\nu}} + S\ind{_{\nu\mu}} - S\ind{^\sigma_\sigma}\eta\ind{_{\mu\nu}}\right),
	\end{align}
	the gauge condition for the \emph{barred} $h$-field is obtained by transforming \eqref{eq:Gaugecondition:spin2field} as \eqref{eq:def:BarOperation}, resulting with
	\begin{align}\label{eq:transform law:hfield}
		\bar{h}\ind{_{\mu\nu}}(x) \to \bar{h}\ind{_{\mu\nu}}(x) - \xi\ind{_{\mu,\nu}}(x) - \xi\ind{_{\nu,\mu}}(x) + \eta\ind{_{\mu\nu}} \xi\ind{_\sigma^{,\sigma}}(x) \equiv \bar{h}\ind{_{\mu\nu}}(x) - \xi\ind{_{\mu\nu}}(x).
	\end{align}
	Thus the divergence of this barred field transforms as
	\begin{align}
		\bar{h}\ind{_{\mu\nu}^{,\nu}} \to \bar{h}\ind{_{\mu\nu}^{,\nu}} - \dalembertian\xi\ind{_\mu}.
	\end{align}
	As $\xi_\mu$ can be any vector without changing the \acrshort{eom}, it can be chosen such that $\dalembertian \xi\ind{_\mu} = \bar{h}\ind{_{\mu\nu}^{,\nu}}$, shifting the field such that $\bar{h'}\ind{_{\mu\nu}^{,\nu}}=0$, which is to impose the \emph{Lorenz gauge}\footnote{The divergenceless gauge can be refered to by many names, but the most common is to use the same name as in electro dynamics: Lorenz. Other names include Hilbert, De Donder and Harmonic gauge, though the latter two are more assosciated with curved backgrounds.}.
	\begin{empheq}[box=\widefbox]{align}
        \nonumber \text{Lorenz } & \text{gauge condition}: \\
        & \bar{h}\ind{_{\mu\nu}^{,\nu}} = 0. \label{eq:def:DeDonder gauge}
	\end{empheq}
	The exact expression for $\xi_\mu$ can be obtained by method of Green's functions, but this is unnecessary to compute. Simply keeping in mind that $\bar{h'}\ind{_{\mu\nu}^{,\nu}}=0$ shall suffice to simplify the Lagrangian \eqref{eq:FierzPauliLagrangian}. Notice that the following terms must be zero, using again \eqref{eq:def:BarOperation}, but in reverse.
	\begin{subequations}
		\begin{align}
	    	\label{eq:gaugefixingTermSingleh} \bar{h}\ind{_{\mu\nu}^{,\nu}} = \hspace{3pt} & 0 = h\ind{_{\mu\nu}^{,\nu}} - \frac{1}{2}\eta\ind{_{\mu\nu}}h\ind{^{,\nu}},\\
			\label{eq:gaugefixingTermDoubbleh} \bar{h}\ind{_{\mu\nu}^{,\nu}}\bar{h}\ind{^{\mu\rho}_{,\rho}} = \hspace{3pt} & 0^2 = h\ind{_{\mu\nu}^{,\nu}}h\ind{^{\mu\rho}_{,\rho}} - h\ind{_{\mu\nu}^{,\nu}}h\ind{^{,\mu}} + \frac{1}{4}h\ind{_{,\mu}}h\ind{^{,\mu}}.
		\end{align}
	\end{subequations}
	Adding and subtracting 0 from the Lagrangian should change nothing, and thus the following expression can be used as a \acrfull{gf}
	\begin{empheq}[box=\boxed]{align}
		&\L_\text{\acrshort{gf}} = - \bar{h}\ind{_{\mu\nu}^{,\nu}}\bar{h}\ind{^{\mu\rho}_{,\rho}} = \left[ -h\ind{_{\mu\nu}^{,\nu}}h\ind{^{\mu\rho}_{,\rho}} + h\ind{_{\mu\nu}^{,\nu}}h\ind{^{,\mu}} - \frac{1}{4}h\ind{_{,\mu}}h\ind{^{,\mu}} \right], \\
		\Rightarrow \quad \L_{(2)} + &\L_\text{int} \equiv \L_\text{\acrshort{fp}} + \L_\text{\acrshort{gf}} + \L_\text{int} = -\frac{1}{2} h\ind{_{\mu\nu,\rho}}h\ind{^{\mu\nu,\rho}} + \frac{1}{4} h\ind{_{,\mu}}h\ind{^{,\mu}} + \frac{\lambda}{2} h\ind{_{\mu\nu}}T\ind{^{\mu\nu}}.\quad \label{eq:h2Lagrangian:full}
	\end{empheq}
	with the subscript $(2)$ to signify that this is the action to quadratic order in $h$.
	
	The \acrshort{eom} of $\L_{(2)}+\L_\text{int}$ is the familiar
	\begin{empheq}[box=\widefbox]{align} \label{eq:EoM:FPlagrang:hbar}
		\dalembertian\left( h\ind{_{\mu\nu}} - \frac{1}{2}\eta\ind{_{\mu\nu}} h \right) = \dalembertian\bar{h}\ind{_{\mu\nu}} = - \frac{\lambda}{2} T\ind{_{\mu\nu}},
	\end{empheq}
	from linearized theory (see equation (1.24) of \textcite{Maggiore:VolumeI} or equation (9.22) of \textcite{Gron}).
	
    Comparing with the \acrshort{eom} of linearized \acrshort{gr} it is tempting to conclude that \ifthenelse{\boolean{NaturalUnits}}{$\lambda \equiv 4\kappa = 32\pi$}{$\lambda \equiv 4\kappa = 32\pi G / c^4$}, but then it is also common to make the \gls{Einstein-Hilbert action} dimensionless by scaling it with a factor of $(16\pi G/c^4)^{-1}$. Comparing \eqref{eq:h2Lagrangian:full} with the \gls{Einstein-Hilbert action} expanded to second order\footnote{Which is the necessary order needed to derive the linearized \gls{Einstein equations}.} the Lagrangian  \eqref{eq:h2Lagrangian:full} carries an additional factor of $2$, and is dimensionful. \Glspl{field theorist} usually fix the dimensionality of the action by rescaling their fields to become dimensionful. Doing this
    \begin{align} \label{eq:Rescaling of h}
        h_{\mu\nu}^\text{dim.ful} = \left( \frac{32\pi G}{c^4} \right)^{-1/2} h_{\mu\nu}^\text{dim.less},
    \end{align}
	$h$ adsorbs the dimensionful prefactor. To compare \eqref{eq:EoM:FPlagrang:hbar} with the linearized \gls{Einstein equations}, it must first be rescaled back to a dimensionless field according to \eqref{eq:Rescaling of h}, and then the coupling constant is revealed to be
	\begin{empheq}[box=\widefbox]{align}
	    \lambda \equiv \left(\frac{32\pi G}{c^4}\right)^{1/2} = \sqrt{4\kappa},
	\end{empheq}
	where $\kappa=\ifthenelse{\boolean{NaturalUnits}}{8\pi}{\frac{8\pi G}{c^4}}$ is the constant which appears in \gls{Einstein equations}.
	
	Some call this coupling constant $M_\text{Pl}^{-1}$ rather than $\lambda$ (\cite{Porto2016,Goldberger:EFT,Goldberger:LesHouches}). However it does not have dimesion of mass, nor is the Planck constant anywhere in the expression, so why do they do this? These articles use natural units $\hbar=c=1$, and $M_\text{Pl}^{-1} = \sqrt{G/\hbar c} = \lambda \cdot \sqrt{\hbar c^3/32\pi}$, which is just a numerical factor off from $\lambda$ (in natural units). 
	
	Furthermore, in natural units, legths $(L)$ are dimesionally equal to inverse mass $(M)^{-1}$ (using $[x]$ as dimesion of $x$: $L=[x]=[ct]=1 \cdot T$, and $E=[\hbar \omega] = 1 \cdot T^{-1} \stackrel{\text{also}}{=} [mc^2] = M\cdot 1^2, \quad \Rightarrow \quad M = T^{-1} = L^{-1} = E$), and thus the action has dimension of $[S] = [\int \dd[4]{x} \L] = L^4 [\L] = M^{-4}[\L] = 1$. The action must be dimesionless in \acrshort{qft}, since in the path integral approach it is exponated. Every field Lagrangian has a kinetic term $\sim \partial \phi \partial \phi$, which scale as $[\L] = M^4 = [\partial \phi]^2 = L^{-2} [\phi]^2 = M^2[\phi]^2 \quad \Rightarrow \quad [\phi]=M$. Thus for couplings $\L_\text{int} \sim \lambda h (\partial h)^2$ to have the same dimension as the kinetic term; $[\lambda] = M^{-1}$.
	
	Calling the coupling constant $M_\text{Pl}^{-1}$ might have the unfortunate consequense of making it look like a quantum theory, but make no mistake, this is all classical field theory. Therefore, it is simply labelled $\lambda$ in this thesis.
	\newpage
\section{Solutions of the graviton field} \label{sec:SolGraviton}
\subsection{Gravitational waves in vacuum, and their polarization}
    According to the \acrlong{eom} \eqref{eq:EoM:FPlagrang:hbar} the field will in a vacuum ($T\ind{_{\mu\nu}}=0$) behave as a relativistic wave
    \begin{align} \label{eq:wave equation}
        \dalembertian \bar{h}\ind{_{\mu\nu}} = 0,
    \end{align}
    which admits solutions of the form \eqref{eq:wave:general}. See \cref{app:wave equation sol} for a derivation of this solution. 
    
    Up until now the gauge has only been used to make sure the entire \acrshort{eom} \eqref{eq:EoM:FPlagrang:hbar} remains divergence free, just like the source term $T\ind{_{\mu\nu}^{,\nu}}=0$. In doing so it was determined that the field might only be shifted according to \eqref{eq:Gaugecondition:spin2field}. Furthermore, the divergence of the barred $h$-field could be eliminated only imposing further that $\dalembertian \xi_\mu = 0$.
    
    Keeping $\dalembertian \xi_\mu = 0$ still leaves
    \begin{align}
        \xi\ind{_{\mu\nu}}(x) \equiv \xi\ind{_{\mu,\nu}}(x) + \xi\ind{_{\nu,\mu}}(x) - \eta\ind{_{\mu\nu}} \xi\ind{_\sigma^{,\sigma}}(x)
    \end{align}
    with four degrees of freedom, as it is a function of the four independent parameters $\xi_\mu$, which satisfy $\dalembertian \xi_\mu = 0$. To make the graviton field divergence free imposes four additional conditions on $h\ind{_{\mu\nu}}$, by the four equations $h\ind{_{\mu\sigma}^{,\sigma}}=0$. This leaves $h\ind{_{\mu\nu}}$ with $10-4=6$ degrees of freedom. Subtracting further the four gauge freedoms reduces $h\ind{_{\mu\nu}}$ to only two effective degrees of freedom, as any massless spin two field should have. These four gauge freedoms $\xi_\mu$ can be used to impose four additional conditions on $\bar{h}\ind{_{\mu\nu}}$. $\xi_0$ can be used to set the trace $\bar{h}=0$, and since $\bar{h}\ind{_{\mu\nu}}$ is just $h\ind{_{\mu\nu}}$ with the reversed sign trace, in this gauge $h\ind{_{\mu\nu}}=\bar{h}\ind{_{\mu\nu}}$.
    
    The three remaining freedoms, $\xi_i$, can be used to set $h\ind{_{0i}}=0$ as well. Since $h\ind{_{\mu\nu}^{,\nu}}=0$ this implies $h\ind{_{0\nu}^{,\nu}}= \partial^0 h\ind{_{00}} + \partial^i h\ind{_{0i}} = 0 = \partial^0 h\ind{_{00}}$, making $h\ind{_{00}}$ a constant of time. A constant contribution to a \acrshort{gw} are uninteresting and for all intents and purposes it can be considered to be zero, making all $h\ind{_{0\mu}}=0$.
    
    This specific gauge is referred to as the \emph{\acrfull{tt} gauge}, and is defined by
    \begin{empheq}[box=\widefbox]{align}
        \nonumber &\text{\acrshort{tt} gauge condition:} \\
        &h\ind{_{0\mu}} = h\ind{_i^i} = h\ind{_{ij}^{,j}} = 0. \label{eq:TT gauge condition}
    \end{empheq}
    Note that this gauge can only be imposed in a vacuum, since the vacuum condition $\dalembertian \bar{h}\ind{_{\mu\nu}} = 0$ $\Rightarrow$ $\dalembertian \xi\ind{_{\mu\nu}}=0$ was used.
    
    Assuming $h\ind{_{\mu\nu}}(x^\alpha) = \epsilon\ind{_{\mu\nu}} h(x^\alpha)$ with $h(x^\alpha)$ as the scalar solution to the wave equation
    \begin{align}
        h(x^\alpha) = \int \frac{ \dd[3]{k} }{(2\pi)^3 \cdot 2\omega_k} & \left\{ a(\tvec{k}) e^{-ik\ind{_\sigma}x\ind{^\sigma}} + a^\dagger(\tvec{k}) e^{ik\ind{_\sigma}x\ind{^\sigma}} \right\},
    \end{align}
    which is \eqref{eq:scalar sol. wave eq.} from \cref{app:wave equation sol}. The \acrshort{tt} gauge condition then implies
    \begin{align*}
        h\ind{_{mn}^{,n}}(x^\alpha) \stackrel{\eqref{eq:TT gauge condition}}{=} \hspace{3pt} & 0 = \epsilon\ind{_{mn}} \partial^n h(x^\alpha) \\
        = \int \frac{ \dd[3]{k} }{(2\pi)^3 \cdot 2\omega_k} \epsilon\ind{_{mn}} ik^n & \left\{ -a(\tvec{k}) e^{-ik\ind{_\sigma}x\ind{^\sigma}} + a^\dagger(\tvec{k}) e^{ik\ind{_\sigma}x\ind{^\sigma}} \right\},
    \end{align*}
    hence
    \begin{align}
        k\ind{^j} \epsilon\ind{_{ij}} = 0.
    \end{align}
    For a wave travelling in the $z$-direction $k_\mu = \begin{pmatrix} k, & 0, & 0, & k \end{pmatrix}$, making $\epsilon\ind{_{\mu3}}=0$. With \eqref{eq:TT gauge condition} this is enough to determine the polarization down to the two essential degrees of freedom
    \begin{empheq}[box=\widefbox]{align}
        \epsilon\ind{_{\mu\nu}} = \begin{pmatrix}
        0 & 0 & 0 & 0 \\
        0 & \epsilon_+ & \epsilon_\times & 0 \\
        0 & \epsilon_\times & -\epsilon_+ & 0 \\
        0 & 0 & 0 & 0
        \end{pmatrix} = \begin{pmatrix}
            \epsilon_+ & \epsilon_\times \\
            \epsilon_\times & -\epsilon_+ \\
        \end{pmatrix}_\text{in plane $\perp$ to $\tvec{k}$}.
    \end{empheq}
\subsection{Source of gravitational waves}
    The general solution of the \acrlong{eom} \eqref{eq:EoM:FPlagrang:hbar} with sources can be obtained by method of Green's functions.
\subsubsection{Green's functions in general}
	The Green's function of a linear differential operator $\L\Op_x$ is defined as the function that satisfies
	\begin{align} \label{eq:Green's func. def}
		\L\Op_x \Delta(\fvec{x},\fvec{x'}) = \dirac{n}{\fvec{x}-\fvec{x'}},
	\end{align}
	where $\L\Op_x$ only acts on $x$.
	
	The differential equation in question
	\begin{align*}
	    \L\Op_x \psi(x) = f(x)
	\end{align*}
	admits solutions of the form
	\begin{align*}
	    \psi(x) = \int \dd[n]{x'} \Delta(x,x') f(x').
	\end{align*}
	
	This solution can easily be demonstrated to recover the original differential equation by using the definition of the Green's function,
	\begin{align*}
	    \L\Op_x \psi(x) & = \L\Op_x \int \dd[n]{x'} \Delta(x,x') f(x') = \int \dd[n]{x'} f(x') \L\Op_x \Delta(x,x')\\
	    & \stackrel{\eqref{eq:Green's func. def}}{=} \int \dd[n]{x'} f(x') \dirac{n}{x-x'} = f(x),
	\end{align*}
	which was the original differential equation.
\subsubsection{Green's function of the d'Alembert operator}
	In our case $\L\Op_x=\dalembertian_x$, which is invariant under translation. Thus $\Delta(x,x')=\Delta(x-x')$. In order to find the corresponding Green's function $\Delta(x-x')$ the easiest way is to go through Fourier space.
	\begin{align*}
	    \dalembertian\ind{_x} \Delta(x-x') & = \dalembertian\ind{_x} \int \frac{\dd[4]{k}}{(2\pi)^4} \tilde{\Delta}(k) e^{ ik\ind{_{\sigma}} (x\ind{^\sigma}-{x'}\ind{^\sigma}) } = \int \frac{\dd[4]{k}}{(2\pi)^4} \tilde{\Delta}(k) \left(i^2 k\ind{_\mu} k\ind{^\mu} \right) e^{ ik\ind{_{\sigma}} (x\ind{^\sigma}-{x'}\ind{^\sigma}) }\\
	    & = \dirac{4}{x-x'} = \int \frac{\dd[4]{k}}{(2\pi)^4} e^{ ik\ind{_{\sigma}} (x\ind{^\sigma}-{x'}\ind{^\sigma}) }.
	\end{align*}
	Matching the last equality of both lines indicates that the Fourier transform of $\Delta(x-x')$ must be\footnote{From here on out the tilde over $\tilde{\Delta}(k)$ will be dropped, and whether it is the Green's function in real or Fourier space will be expected to be understood by its argument.} \begin{empheq}[box=\widefbox]{align} \label{eq:Green's function fourier space}
	    \Delta(k) = \frac{-1}{k\ind{_\mu} k\ind{^\mu}}.
	\end{empheq}
	
	Obtaining the Green's function in real space is now just a matter of transforming \eqref{eq:Green's function fourier space}.
	
	One last remark about Green's functions is in the context of four dimensional space-time the solution in terms of Green's functions can be understood as counting up contributions from source terms, all over space, and across all time
	\begin{align*}
	    h(x^\alpha) = \int \dd[4]{x'} \Delta(x^\alpha-{x'}^\alpha) T({x'}^\alpha) = \int_{-\infty}^\infty \ifthenelse{\boolean{NaturalUnits}}{\dd{t'}}{\dd{ct'}} \int \dd[3]{x'} \Delta(x^\alpha-{x'}^\alpha) T({x'}^\alpha).
	\end{align*}
	Thus $\Delta(x-x')$ \emph{weighs the importance} of source contributions at different points in space and time. For physical solutions only contributions of source configurations from the \emph{past} contribute to $h(t)$. This is imposed by demanding $t\geq t'$.
	
	Back to deriving the real space Green's function. Performing the transform, and defining $r^\alpha = x^\alpha - {x'}^\alpha$
	\begin{align*}
	    \Delta(r^\alpha) & = \int \frac{\dd[4]{k}}{(2\pi)^4} \frac{-1}{k\ind{_\mu} k\ind{^\mu}} e^{ik\ind{_\sigma} r\ind{^\sigma}} = \int \frac{\dd{k\ind{_0}}}{2\pi} e^{-ik\ind{_0} r\ind{^0}} \int \frac{ \dd[3]{k} }{ (2\pi)^3 } \frac{1}{k_0^2-\tvec{k}^2} e^{i\tvec{k \cdot r}} \\
	    & = \int \frac{\dd{k\ind{_0}}}{2\pi} e^{-ik\ind{_0} r\ind{^0}} \int_0^\infty \int_{-1}^{1} \int_0^{2\pi} \frac{ \abs{\tvec{k}}^2 \dd{\abs{\tvec{k}}} \dd{\cos{\theta}} \dd{\phi} }{ (2\pi)^3 } \frac{1}{k_0^2-\tvec{k}^2} e^{i\abs{\tvec{k}} \abs{\tvec{r}} \cos{\theta} } \\
	    & = \int \frac{\dd{k\ind{_0}}}{2\pi} e^{-ik\ind{_0} r\ind{^0}} \int_0^\infty \frac{ \dd{\abs{\tvec{k}}} }{ (2\pi)^2 } \frac{ \abs{\tvec{k}}^2 }{ i\abs{\tvec{k}}\abs{\tvec{r}} } \frac{ e^{i\abs{\tvec{k}} \abs{\tvec{r}} } - e^{-i\abs{\tvec{k}} \abs{\tvec{r}}} }{k_0^2-\tvec{k}^2},
	\end{align*}
	where the spatial integral was performed in spherical coordinates. Relabelling $\abs{\tvec{k}}=k$ and $\abs{\tvec{r}}=r$, the expression can be worked further
	\begin{align*}
	    \Delta(r^\alpha) & = \int \frac{\dd{k\ind{_0}}}{2\pi} e^{-ik\ind{_0} r\ind{^0}} \int_{-\infty}^\infty \dd{k} \frac{ k }{ i(2\pi)^2r } \frac{ e^{ikr } }{k_0^2-k^2}.
	\end{align*}
	Extending to the complex plane, this integral can be evaluated using \emph{Cauchy's residue theorem}, from complex analysis. Shifting $k\ind{_0} \to k\ind{_0}+i\varepsilon$ is equivalent to imposing the \emph{retardation condition:} $t\geq t' \Leftrightarrow r\ind{^0}\geq 0$. Why this is the case should become apparent soon. According to the residue theorem
    \begin{align}
        \begin{split}
            &\oint \dd{z} f(z) = i2\pi \sum_k \text{Res}(f,a_k), \\
            \text{where $ a_k $ is } & \text{a pole of $ f(z) $ enclosed by the integral, and} \\
            & \text{Res}(f,a_k) = \lim_{z\to a_k} (z-a_k) \cdot f(z).
        \end{split}
    \end{align}
    
    Utilizing this theorem and integrating over the upper complex plane, the Green's function becomes
    \begin{align*}
        \Delta(r^\alpha) & = \lim_{\varepsilon\to 0} \int \frac{\dd{k\ind{_0}}}{2\pi} e^{-ik\ind{_0} r\ind{^0}} \frac{ -1 }{ i(2\pi)^2r } \oint \dd{k} k \frac{ e^{ikr } }{ (k-k_0-i\varepsilon)(k+k_0+i\varepsilon) } \\
        & = \lim_{\varepsilon\to 0} \int \frac{\dd{k\ind{_0}}}{2\pi} e^{-ik\ind{_0} r\ind{^0}} \frac{ -1 }{ i(2\pi)^2r } \frac{ i2\pi }{ 2 } e^{ ir(k\ind{_0} + i\varepsilon) } \\
        & = \frac{-1}{4\pi r} \int \frac{ \dd{k\ind{_0}} }{2\pi} e^{-ik\ind{_0} (r\ind{^0}-r)} = \frac{ -\dirac{}{ \ifthenelse{\boolean{NaturalUnits}}{t-r}{ct-r} } }{4\pi r}
    \end{align*}
    
	It is now apparent that this is the \emph{retarded solution}. Had $k_0$ rather been shifted by $-i\varepsilon$, the Dirac delta function would have been $\dirac{}{ \ifthenelse{\boolean{NaturalUnits}}{t+r}{ct+r}}$. The delta function picks out contributions on the light cone of $x^\alpha$, the retarded Green's function picks out on the past light cone, while the \emph{advanced} Green's function picks out on the future light cone.
	\begin{subequations}
	\begin{empheq}[box=\widefbox]{align}
	    \Delta_\text{ret}(r^\alpha) & = \frac{ -\dirac{}{ \ifthenelse{\boolean{NaturalUnits}}{\abs{t-\tvec{r}}}{ct-\abs{\tvec{r}}} } }{4\pi \abs{\tvec{r}}}. \label{eq:Retarded Green's func.}\\
	    \Delta_\text{adv}(r^\alpha) & = \frac{ -\dirac{}{ \ifthenelse{\boolean{NaturalUnits}}{\abs{t+\tvec{r}}}{ct+\abs{\tvec{r}}} } }{4\pi \abs{\tvec{r}}}.
	\end{empheq}
	\end{subequations}
	
	Beyond singling out contributions from the light cone, it is apparent that the importance of each contribution to the field is weighted by how far away it is from the point in question, according to the inverse power of the spatial distance.
\subsubsection{Solving the inhomogeneous \acrlong{eom}}
    Back to the \acrshort{eom} \eqref{eq:EoM:FPlagrang:hbar}, using the retarded Green's functions it admits solution of the form
	\begin{align}
	    \nonumber \dalembertian \bar{h}\ind{_{\mu\nu}}(x) & = -\frac{\lambda}{2} T\ind{_{\mu\nu}}(x) \\
		\nonumber \Rightarrow \quad \bar{h}\ind{_{\mu\nu}}(\fvec{x}) & = -\frac{\lambda}{2} \int\dd[4]{x'} \Delta_\text{ret}(\fvec{x}-\fvec{x'}) T\ind{_{\mu\nu}}(\fvec{x'}) \\
		\label{eq:Linearized:GW:SourceTerm}& = \frac{\lambda}{8\pi} \int\dd[3]{x'} \frac{T\ind{_{\mu\nu}}(t_\text{ret}, \tvec{x'})}{\abs{\tvec{x}-\tvec{x'}}}, \qq{where $t_\text{ret} \equiv t - \ifthenelse{\boolean{NaturalUnits}}{\abs{\tvec{x}-\tvec{x'}}}{\frac{\abs{\tvec{x}-\tvec{x'}}}{c}}$.}
	\end{align}
	
	Assuming the \acrshort{gw} is measured far away from the source compared to the size of the source system, then the approximation $ \abs{\tvec{x}-\tvec{x'}} \approx \abs{\tvec{x}} \equiv R $ holds. This simplifies the integral in equation \eqref{eq:Linearized:GW:SourceTerm} to only be dependent on the energy distribution of the source, and not where it is measured.
	
	Furthermore, waves measured in vacuum can be set into the \acrshort{tt} gauge, which implies that all physical information of the source can be captured by its spatial indices $\bar{h}\ind{_{ij}}$.
	\begin{align}
		\label{eq:Linearized:GWSource:rTerm}\bar{h}\ind{_{ij}}(t,r) = \frac{\lambda}{8\pi R} \int_{\V} T\ind{_{ij}}(t_\text{ret},\tvec{x'}) \dd[3]{x'}.
	\end{align}
	The volume $\V$ must cover the entirety of the spatial extension of the source, in order to be equivalent to the infinite integral of equation \eqref{eq:Linearized:GW:SourceTerm}.
	
	To simplify the expression of equation \eqref{eq:Linearized:GWSource:rTerm} further it is useful to note some properties of the energy-momentum tensor.
	
	\textbf{First:} it is divergenceless.
	\begin{align}
		T\ind{^{\mu\nu}_{,\nu}} = T\ind{^{\mu0}_{,0}} + T\ind{^{\mu i}_{,i}} = 0.
	\end{align}
	
	\textbf{Second:} use of the following integral will be made.
	\begin{align}
		\label{eq:Linearized:DivergenceIntergalOfT}\int_\V \left(T\ind{^{i k}}x\ind{^j}\right)\ind{_{,k}} \dd[3]{x} = \int_\V T\ind{^{i k}_{,k}}x\ind{^j} \dd[3]{x} + \int_\V T\ind{^{ij}} \dd[3]{x} = \oint_{\partial \V} T\ind{^{ik}}x\ind{^j} \dd{A}\ind{_k} .
	\end{align}
	In the last line the divergence theorem has been used.
	
	If the integration boundary is taken to encapsulate the entire source, $\eval{T\ind{^{ik}}}_{\partial V} = 0 $, equation \eqref{eq:Linearized:DivergenceIntergalOfT} becomes equal to 0. It then follows
	\begin{align}
	\begin{split}
		\int_\V T\ind{^{ij}} \dd[3]{x} & = -\int_\V T\ind{^{k\{i}_{,k}}x\ind{^{j\}}} \dd[3]{x} = \int_\V T\ind{^{0\{i}_{,0}}x\ind{^{j\}}} \dd[3]{x} = \dv{\ifthenelse{\boolean{NaturalUnits}}{t}{ct}} \int_\V T\ind{^{0\{i}}x\ind{^{j\}}} \dd[3]{x} \\
		& = \frac{1}{2} \dv{\ifthenelse{\boolean{NaturalUnits}}{t}{ct}} \int_\V \left( T\ind{^{i0}}x\ind{^j} + T\ind{^{j0}}x\ind{^i} \right) \dd[3]{x}.
	\end{split}
	\end{align}
	In the last line the symmetry $T\ind{^{\{ij\}}}=\frac{1}{2}(T\ind{^{ij}} +T\ind{^{ji}})$ was written out explicitly.
	
	So far not much has been accomplished, but notice how this procedure can be repeated to eliminate \textit{all} dependence of the spatial components of $T\ind{^{\mu\nu}}$.
	\begin{align}
		\begin{split}
			\int_\V \left( T\ind{^{k0}}x\ind{^i}x\ind{^j} \right)\ind{_{,k}} \dd[3]{x} & = \int_\V T\ind{^{k0}_{,k}}x\ind{^i}x\ind{^j} \dd[3]{x} + \int_\V \left( T\ind{^{i0}}x\ind{^j} + T\ind{^{j0}}x\ind{^i} \right) \dd[3]{x} \\
			& = \oint_{\partial \V} T\ind{^{k0}}x\ind{^i}x\ind{^j}\dd{A}\ind{_k} = 0
		\end{split}\\
		\Rightarrow \quad \int_\V \left( T\ind{^{i0}}x\ind{^j} + T\ind{^{j0}}x\ind{^i} \right)\dd[3]{x} & = - \int_\V T\ind{^{0k}_{,k}}x\ind{^i}x\ind{^j} \dd[3]{x} = \int_\V T\ind{^{00}_{,0}}x\ind{^i}x\ind{^j} \dd[3]{x}
	\end{align}
	
	Using that $T\ind{^{00}}=\ifthenelse{\boolean{NaturalUnits}}{\rho}{\rho c^2}$ is the mass-energy density, the integral in equation \eqref{eq:Linearized:GWSource:rTerm} takes a simple form of the second time derivative of the so-called \textit{quadrupole moment}
	\begin{subequations}
	\begin{empheq}[box=\widefbox]{align} \label{eq:QuadrupoleMoment:def}
	    Q\ind{_{ij}}(t) & \equiv \int_\V \rho(t,\tvec{x}) x\ind{_i}x\ind{_j} \dd[3]{x}, \\
		\ddot{Q}\ind{_{ij}}(t) & = 2\int_\V T\ind{_{ij}}(t,\tvec{x}) \dd[3]{x}.
	\end{empheq}
	\end{subequations}
	
	Finally, the source of linearized \acrshort{gw}s is provided as
	\begin{empheq}[box=\widefbox]{align} \label{eq:linearized:GW:QuadPole}
	    \bar{h}\ind{_{ij}}(t,\tvec{R}) & = \frac{\lambda}{16\pi R} \ddot{Q}\ind{_{ij}}(t_\text{ret}).
	\end{empheq}
	
	Utilizing this result the \acrshort{gw}s generated from any source, with a non-vanishing energy-momentum tensor, can be calculated at distances sufficiently far away from the source. 
	
	Note that this does \emph{not} mean that the energy-momentum tensor of a source is converted into \acrshort{gw}s. This formula is simply a result of the interaction term of the graviton action to source energy-momentum tensors, e.g. from the energy-momentum tensor of stars or \acrlong{bh}s. However, in section \ref{sec:Energy-Momentum tensor of GW} it will be demonstrated that these \acrshort{gw}s \emph{do} carry energy, and since this energy must come from somewhere it will be assumed to come from the source. This amendment of the theory is the topic of section \ref{sec:GravitonBeyondQuadratic}.
\section{Gravity from gravitons} \label{sec:GravityFromGravitons}
    Equipped with a field Lagrangian it is now desirable to check that it in fact reproduces the Newtonian law of universal gravity for non-relativistic sources.
    
    In special relativity the action of \acrlong{pp}s are the \emph{geodesics}\footnote{Extrema, e.g. shortest, path between two points in space-time. In Euclidian space it is the straigh line connecting the two points.} of Minkowski space-time. The geodesic of some non-trivial space-time, with metric $g\ind{_{\mu\nu}}=\eta\ind{_{\mu\nu}} + \lambda h\ind{_{\mu\nu}}$ is
    \begin{subequations} \label{eq:pp action expansion}
    \begin{align}
        S_{\acrshort{pp}} & = -m\ifthenelse{\boolean{NaturalUnits}}{}{c} \int \dd{s} = -m\ifthenelse{\boolean{NaturalUnits}}{}{c} \int \sqrt{-g\ind{_{\mu\nu}}\dd{x}^\mu\dd{x}^\nu}\\
        & = -m\ifthenelse{\boolean{NaturalUnits}}{}{c} \int \sqrt{-(\eta\ind{_{\mu\nu}} + \lambda h\ind{_{\mu\nu}}) \dd{x}^\mu \dd{x}^\nu} \\
        & = -m\ifthenelse{\boolean{NaturalUnits}}{}{c} \int \sqrt{-\dd{\tau}^2(\eta\ind{_{\mu\nu}} + \lambda h\ind{_{\mu\nu}}) \dot{x}^\mu \dot{x}^\nu}\\
        & = -m\ifthenelse{\boolean{NaturalUnits}}{}{c} \int \dd{\tau} \sqrt{\ifthenelse{\boolean{NaturalUnits}}{1}{c^2} - \lambda h\ind{_{\mu\nu}}\dot{x}^\mu\dot{x}^\nu}\\
        & = -m\ifthenelse{\boolean{NaturalUnits}}{}{c^2} \int \dd{\tau} \sqrt{1 - \lambda h\ind{_{\mu\nu}} \ifthenelse{\boolean{NaturalUnits}}{\dot{x}^\mu \dot{x}^\nu}{\frac{\dot{x}^\mu}{c}\frac{\dot{x}^\nu}{c}}}\\
        & \approx -m\ifthenelse{\boolean{NaturalUnits}}{}{c^2} \int \dd{t} \gamma^{-1} + \frac{\lambda}{2} \int\dd{t} h\ind{_{\mu\nu}} m \gamma^{-1} \dot{x}^\mu \dot{x}^\nu + \dots\\
        & \approx \int \left[ \left( -m\ifthenelse{\boolean{NaturalUnits}}{}{c^2} +\frac{1}{2}mv^2 + \dots \right) + \frac{\lambda}{2}h\ind{_{\mu\nu}} T^{\mu\nu}_{\acrshort{pp}} \right] \dd{t}.
    \end{align}
    \end{subequations}
    
    Why compute the geodesic for a perturbed Minkowski space? In the \glspl{relativist}' approach \acrshort{gw}s is such a perturbation of the metric (at least for vacuum solutions), thus this is the action of \acrlong{pp}s to linearized order in $h\ind{_{\mu\nu}}$, according to \glspl{relativist}. It is included here to motivate the interaction term $\L_\text{int}=\frac{\lambda}{2}h\ind{_{\mu\nu}} T\ind{^{\mu\nu}}$, with the \acrlong{pp} energy momentum-tensor\footnote{The factor of $\gamma^{-1}_a$ is a result of rewriting the proper time integral into a generic time integral $\dv{t}{\tau_a}=\gamma_a$, common for all particles. Writing out the four-velocities should result in an overall factor of $\gamma_a^1$. For more info on the \acrshort{pp} energy-momentum tensor see e.g. \textcite{SpesRel}.}
    \begin{subequations} \label{eq:pp EnergyMomentumTensor}
    \begin{align}
        &T_{\acrshort{pp}}^{\mu\nu}(x) = \sum_a \gamma^{-1}_a m_a \dot{x}\ind{^\mu} \dot{x}\ind{^\nu} \dirac{3}{\tvec{x} - \tvec{x}_a(t)}, \\
        &\qq*{with} \dot{x}^\mu_a = \gamma_a \MixVec{\ifthenelse{\boolean{NaturalUnits}}{1}{c}}{\tvec{v}} = \left(1-\ifthenelse{\boolean{NaturalUnits}}{\abs{\tvec{v}}^2}{\frac{\abs{\tvec{v}}^2}{c^2}} \right)^{-\frac{1}{2}} \MixVec{\ifthenelse{\boolean{NaturalUnits}}{1}{c}}{\tvec{v}}.
    \end{align}
    \end{subequations}
    Otherwise, it just shows that the free \acrlong{pp} Lagrangian (the kinetic part) is just $L_{\text{free }\acrshort{pp}} \simeq \frac{1}{2}m_av^2_a = \int \frac{1}{2}m_a\dot{x}^2\dirac{3}{\tvec{x}-\tvec{x}_a} \dd[3]{x} $. The constant term $-m\ifthenelse{\boolean{NaturalUnits}}{}{c^2}$ does not contribute to the \acrshort{eom}, and can therefore be neglected.
    
    Thus, the total Lagrangian of two \acrlong{pp}s interacting only via the graviton field, up to the 0\acrshort{pn} order, is
    \begin{align}
        \L_{\acrshort{pp}} = \sum_{a=1}^2 \left[ \frac{1}{2}m_a\dot{\tvec{x}}^2\dirac{3}{\tvec{x}-\tvec{x}_a} + \frac{\lambda}{2}h\ind{_{\mu\nu}} m_a \dot{x}\ind{^\mu} \dot{x}\ind{^\nu} \dirac{3}{\tvec{x}-\tvec{x}_a} \right] + \L_\text{\acrshort{fp}} + \L_\text{\acrshort{gf}}
    \end{align}
    
    The \acrshort{eom} for the $h$ field is known from \eqref{eq:EoM:FPlagrang:hbar}. Since the bar operator is its own inverse operator when used on symmetric tensors, $\bar{\bar{S}}\ind{_{\mu\nu}}=S\ind{_{\mu\nu}}$, equation \eqref{eq:EoM:FPlagrang:hbar} may also be written as
    \begin{subequations}
    \begin{align}
        \dalembertian h \hspace{3pt} & = -\frac{\lambda}{2}\bar{T}\ind{_{\mu\nu}}, \\
    \begin{split} \label{eq:EoM of h by Tbar}
        \Rightarrow \quad h\ind{_{\mu\nu}}(x) & = - \frac{\lambda}{2} \int \dd[4]{y} \Delta_{\text{ret}}(x-y) \bar{T}\ind{_{\mu\nu}}(y) \\
        & = - \frac{\lambda}{2} \int \dd[4]{y} \Delta_{\text{ret}}(x-y) P\ind{_{\mu\nu:\alpha\beta}} T\ind{^{\alpha\beta}}(y),
    \end{split}\\
        \qq*{with} & P\ind{_{\mu\nu:\alpha\beta}} = \frac{1}{2} \left( \eta\ind{_{\mu\alpha}}\eta\ind{_{\nu\beta}} + \eta\ind{_{\mu\beta}}\eta\ind{_{\nu\alpha}} - \eta\ind{_{\mu\nu}} \eta\ind{_{\alpha\beta}} \right). \label{eq:Pmunu:ab:def}
    \end{align}
    \end{subequations}
    
    Substituting \eqref{eq:EoM of h by Tbar} for $h\ind{_{\mu\nu}}$ in the interaction term results in the following action for the \acrlong{pp}s.
    \begin{align}
        \nonumber S_{\acrshort{pp}} = & \sum_{a=1}^2 \int \ifthenelse{\boolean{NaturalUnits}}{\dd[4]{x}}{\frac{\dd[4]{x}}{c}} \Biggl\{ \frac{1}{2}m_a\dot{x}^2\dirac{3}{\tvec{x}-\tvec{x}_a} \\
        \nonumber &- \frac{\lambda}{2} \left[ \int \dd[4]{y} \frac{\lambda}{2} \Delta_{\text{ret}}(x-y) P\ind{_{\mu\nu:\alpha\beta}} T\ind{_{\alpha\beta}}(y) \right] m_a \dot{x}\ind{^\mu} \dot{x}\ind{^\nu} \dirac{3}{\tvec{x}-\tvec{x}_a} \Biggr\} \\
        \begin{split} \label{eq:Newtonian Action Prop}
            = & \sum_{a=1}^2 \int \frac{1}{2}m_a\dot{x}^2\dirac{3}{\tvec{x}-\tvec{x}_a} \ifthenelse{\boolean{NaturalUnits}}{\dd[4]{x}}{\frac{\dd[4]{x}}{c}} + \sum_{b > a} \frac{\lambda^2}{4} \iint \left[ m_a \dot{x}\ind{^\mu} \dot{x}\ind{^\nu} \dirac{3}{\tvec{x}-\tvec{x}_a} \right] \\
            & \cdot \left\{ P\ind{_{\mu\nu:\alpha\beta}} \frac{ \dirac{}{ x\ind{^{0}}-{y}\ind{^{0}} } }{4\pi \abs{\tvec{x}-\tvec{y}}} \right\} \left[m_b {\dot{y}}^\alpha {\dot{y}}^\beta \dirac{3}{\tvec{y}-\tvec{x}_b} \right] \ifthenelse{\boolean{NaturalUnits}}{\dd[4]{x}}{\frac{\dd[4]{x}}{c}} \dd[4]{y}
        \end{split}
    \end{align}
    The next trick is to first approximate the space-time distance to mostly be in time, for slow moving, not too far separated \acrlong{pp}s. Then the middle Dirac delta is approximately $\dirac{}{x\ind{^0}-y\ind{^0}}$. After using that approximation to eliminate the $y\ind{^0}$ integral, the spatial integrals are next, which due to the remaining Dirac deltas just makes all $\tvec{x}\to \tvec{x}_a(t)$, $\tvec{y}\to \tvec{x}_b(t)$, and $r= \abs{\tvec{x}_1-\tvec{x}_2}$. Lastly, to leading order in powers of $c$ the only contributing factor of $\dot{x}\ind{^\mu}$ is the time component $\dot{x}_a^0=\gamma_a\ifthenelse{\boolean{NaturalUnits}}{}{c^2}\approx \ifthenelse{\boolean{NaturalUnits}}{1}{c^2}$. The result is
    \begin{empheq}[box=\widefbox]{align}
        \nonumber S_{\text{Newt }\acrshort{pp}} & = \int \left( \frac{1}{2}\left(m_1v_1^2 + m_2v_2^2\right) + \frac{\lambda^2}{8} \frac{m_1 m_2 \ifthenelse{\boolean{NaturalUnits}}{}{c^4}}{4\pi r}  \right) \dd{t}\\
        & = \int \left( \frac{1}{2}\left(m_1v_1^2 + m_2v_2^2\right) + \frac{\ifthenelse{\boolean{NaturalUnits}}{}{G} m_1 m_2}{r}  \right) \dd{t}, \label{eq:NewtonianAction}
    \end{empheq}
    which is exactly the action of Newtonian theory for two \acrlong{pp}s with mass.
    
    Notice in equation \eqref{eq:Newtonian Action Prop}, the potential can be understood as the energy-momentum tensor $T_a^{\mu\nu}(x)$ connected to the energy momentum tensor $T_b^{\alpha\beta}(y)$ by some sort of \emph{propagator}, defined by the contents of the curly brackets. This can be expressed graphically, as in \cref{fig:Feynman:H-diagram:detailed}, and will be further developed in \cref{chap:energy}.
    
    Once the action is known, the energy can simply be derived by finding the corresponding Hamiltonian
    \begin{empheq}[box=\widefbox]{align} \label{eq:Hamiltonian:LegendreTransform:def}
        H \equiv \sum_i \dot{q}^i \pdv{L}{\dot{q}^i} - L = \dot{q}^i p_i -L,
    \end{empheq}
    where $p_i$ is the canonical momentum, and $q^i$ are generalized coordinates. This is nothing but a Legendre transformation of the Lagrangian for $\dot{q}^i \to p_i$.
    
    Unsurprisingly, for the Newtonian action it reads
    \begin{align}
        H_\text{Newt} = \sum_{i=1}^2 \tvec{v}_i \tvec{ \cdot } m_i\tvec{v}_i - L_\text{Newt} = \frac{1}{2}m_1 v_1^2 + \frac{1}{2}m_2 v_2^2 - \frac{Gm_1m_2}{r} = E_\text{Newt}.
    \end{align}
    
    In relative coordinates it reads (see \cref{app:equivOneBodyAndMassTerms} for derivations and tricks)
    \begin{empheq}[box=\widefbox]{align}
        E_\text{Newt} = \frac{1}{2}\mu v^2 - \frac{GM\mu}{r} \stackrel{\eqref{eq:KeplersThirdLaw}}{=} \frac{1}{2}\mu v^2 - r^2 \omega^2 \mu = -\frac{1}{2}\mu v^2, \label{eq:NewtonianEnergy pp}
    \end{empheq}
    which is exactly the first order term of the energy expansion \eqref{eq:Energy:Expansion:Gen}.
\section{The energy-momentum tensor of gravitational waves} \label{sec:Energy-Momentum tensor of GW}
	Equipped with the Lagrangian for linearized theory \eqref{eq:h2Lagrangian:full} it is straight forward to utilize Noether's theorem to obtain the energy-momentum tensor of the graviton field.
	\begin{subequations}
	\begin{align}
		&\qq*{For coordinate transformations} x^\nu \rightarrow x^\nu + \epsilon^a A^\nu_a(x) \\
		&\qq*{And field transformations} \phi_i(x) \rightarrow \phi_i(x) + \epsilon^a F_{i,a}(\phi,\partial\phi) \\
		\begin{split}
		    &\Rightarrow \quad \partial_\nu j^\nu_a = 0, \qq{where} \\
		    &j^\nu_a = \left[ \pdv{ \L }{ \phi_{i,\nu} } \phi_{i,\rho} - \delta^\nu_\rho \L \right]A^\rho_a(x) - \pdv{ \L }{ \phi_{i,\nu} } F_{i,a}(\phi,\partial \phi).
        \end{split}
	\end{align}
	\end{subequations}
	Noether's theorem is a central result of modern physics, and for a detailed derivation consult any field theory book, e.g. \textcite{Kachelriess:2017cfe}, \textcite{Maggiore:VolumeI} or \textcite{GoldsteinMech}.
	
	For pure translations $A_\alpha^\beta = \delta_\alpha^\beta$, $F_{i,\alpha} = 0$, and thus
	\begin{align}
		j\ind{_\alpha^\beta} \equiv -t\ind{_\alpha^\beta} = \L_{(2)} \delta\ind{_\alpha^\beta} -\pdv{\L_{(2)}}{h\ind{_{\mu\nu,\beta}}} h\ind{_{\mu\nu,\alpha}}.
	\end{align}
	The terms in this equation is known, the first from \eqref{eq:Xi:Intermidiate:MuNuRho},\footnote{To match this expression to \eqref{eq:h2Lagrangian:full} set $a_1=-2a_4=-1/2$ and $a_2=a_3=0$, which is to impose the Lorenz gauge. This gauge can only be imposed \emph{after} evaluating the $\pdv*{\L}{h\ind{_{\mu\nu,\beta}}}$ term.} and the second from \eqref{eq:h2Lagrangian:full}.
	\begin{align} \label{eq:GWt:nonaveraged}
		t^{\mu\nu} = h\ind{_{\sigma\rho}^{,\mu}}h\ind{^{\sigma\rho,\nu}} - \frac{1}{2}h\ind{^{,\mu}}h\ind{^{,\nu}} + \L_{(2)}\eta\ind{^{\mu\nu}}.
	\end{align}
	
	For a wave-packet centred around a reduced wavelength $\lambdabar \equiv \lambda/2\pi$, the total 4-momentum flux is obtained by integrating over a volume $\V\sim L^3$ where $L\gg\lambdabar$, such that $t\ind{^{\mu\nu}}$ is zero on the boundary $\partial\V$. Then the effective energy-momentum tensor is\footnote{The Noehter current need not be physical \emph{in itself}, it is the volume integral of $j_a^\nu$, where $j_a^\nu$ goes sufficently fast to zero on the boundary, which is conserved, and thus physical. In field theory one defines the \emph{effective} energy-momentum tensor, which is the average value over the volum integral, i.e. the spatial average, as the physical energy-momentum tensor. All terms lost under spatial averaging would not contribute to physical effects anyway. This is why \eqref{eq:t:for:h2} is averaged over space.
	
	An interesting lesson here is that the energy of \acrshort{gw}s can not be isolated to one space-time point, which can be understood by them being non-localisable.}
	\begin{align}
		\label{eq:t:for:h2}& t\ind{^{\mu\nu}} = \expval{ \L_{(2)} } \eta\ind{^{\mu\nu}} - \expval{ \pdv{\L_{(2)}}{h\ind{_{\sigma\rho,\nu}}} h\ind{_{\sigma\rho,\mu}} } = \expval{ h\ind{_{\sigma\rho}^{,\mu}}h\ind{^{\sigma\rho,\nu}} } - \frac{1}{2}\expval{ h\ind{^{,\mu}}h\ind{^{,\nu}} }
	\end{align}
	Here the $\expval{x}$ is to be understood as spatial averaging. Also, for radiation detectable far away from the source the \acrshort{eom} can be considered to be for a vacuum, and thus $\dalembertian\bar{h}\ind{_{\mu\nu}}=0$. Setting the vacuum as the zero solution of the $h$-field the Lagrangian averages to zero $\expval{\L_{(2)}}=0$, and is thus dropped from \eqref{eq:t:for:h2}. 
	
	Additionally, in vacuum the Lorenz gauge can be promoted to the \acrshort{tt} gauge, further imposing that $h=h\ind{_{0\mu}}=h\ind{_{ij}^{,j}}=0$ \eqref{eq:TT gauge condition}, results in the final expression for the energy-momentum (pseudo-)\footnote{Note that this is the energy-momentum tensor in the \acrshort{tt} gauge, which is to say it is gauge dependent! This might sound strange, but as \textcite{Maggiore:VolumeI} points out this is also the case for electromagnetism. 
	
	In the geometric approach the gauge dependence of $t_{\mu\nu}$ is a coordinate dependence, since the gauge transformation is a coordinate transform in the geometrical picture. Thus $t_{\mu\nu}$ is not a real tensor, but rather a \emph{pseudotensor}. Also in the geometrical picture, $t_{\mu\nu}$ is averaged over several wavelengths to extract the \acrshort{gw} contribution to the background metric, making it equivalent to the energy-momentum tensor obtained by \glspl{field theorist}.} tensor
	\begin{subequations}
	\begin{empheq}[box=\widefbox]{align}
		& t\ind{^{\mu\nu}} = \expval{ \partial\ind{^\mu}h^\text{\acrshort{tt}}_{ij} \partial\ind{^\nu} h^{ij}_\text{\acrshort{tt}} },\\
		P\ind{^\mu} = \ifthenelse{\boolean{NaturalUnits}}{}{\frac{1}{c}} \int_\V & \dd[3]{x} t\ind{^{\mu0}} = \ifthenelse{\boolean{NaturalUnits}}{}{\frac{1}{c}} \int_\V \dd[3]{x} \expval{ \partial\ind{^\mu}h^\text{\acrshort{tt}}_{ij} \partial\ind{^0} h^{ij}_\text{\acrshort{tt}}}.
	\end{empheq}
	\end{subequations}
\subsection{Total radiated energy flux}
	The total radiated energy flux $\F$ can be obtained from $P\ind{^0}=\ifthenelse{\boolean{NaturalUnits}}{E}{E/c}$ and the divergencelessness of the energy-momentum tensor, which follows from Noether's theorem 
	\begin{align}
	    \nonumber t\ind{^{\mu0}_{,\mu}} = 0, \\
	    \Rightarrow \quad \pdv{P\ind{^{0}}}{\ifthenelse{\boolean{NaturalUnits}}{t}{ct}} = - P\ind{^{i}_{,i}} = - \frac{1}{c}\int_\V & \dd[3]{x} \partial\ind{_i} t\ind{^{i0}} = -\frac{1}{c} \int_{\partial\V} t\ind{^{r0}} \dd{A},
	\end{align}
	where the last equality used the divergence theorem, expressed in spherical coordinates. For the next step notice that $t\ind{^{r0}} = \expval{ \partial\ind{^r} h^\text{\acrshort{tt}}_{mn} \partial\ind{^0} h^{mn}_\text{\acrshort{tt}} }$, and recall that $h$'s sufficiently far from the source can be written as $h^\text{\acrshort{tt}}_{mn}=\frac{1}{r}f\ind{_{mn}}(t_\text{ret})$ with $t_\text{ret}=t-\ifthenelse{\boolean{NaturalUnits}}{r}{r/c}$ and $f\ind{_{mn}}$ some function. Then observe that $\pdv{}{r}f\ind{_{mn}}(t_\text{ret})=-\pdv{}{\ifthenelse{\boolean{NaturalUnits}}{t}{ct}}f\ind{_{mn}}(t_\text{ret})$, which makes
	\begin{align}
	    \pdv{}{r} h^\text{\acrshort{tt}}_{mn}(t,r) = -\partial_0 h^\text{\acrshort{tt}}_{mn} + \order{r^{-2}} = \partial^0 h^\text{\acrshort{tt}}_{mn} + \order{r^{-2}}.
	\end{align}
	Thus in the \acrshort{tt} gauge, and sufficiently far from the source, $t\ind{^{r0}}=t\ind{^{00}} + \order{r^{-3}}$, and
	\begin{align}
	    -\F = \dv{E}{t} = \ifthenelse{\boolean{NaturalUnits}}{}{c^2}\pdv{P\ind{^{0}}}{\ifthenelse{\boolean{NaturalUnits}}{t}{ct}} = - \ifthenelse{\boolean{NaturalUnits}}{}{c} \int \left(t\ind{^{00}}+\order{r^{-3}}\right) r^2 \dd{\Omega} = \ifthenelse{\boolean{NaturalUnits}}{-r^2}{\frac{-r^2}{c}} \int \expval{ \dot{h}^\text{\acrshort{tt}}_{ij} \dot{h}^{ij}_\text{\acrshort{tt}} } \dd{\Omega}, \nonumber
	\end{align}
	\begin{empheq}[box=\widefbox]{align} \label{eq:Flux:LinearTheory: hTT}
	    \F = \ifthenelse{\boolean{NaturalUnits}}{r^2}{\frac{r^2}{c^2}} \int \expval{ \dot{h}^\text{\acrshort{tt}}_{ij} \dot{h}^{ij}_\text{\acrshort{tt}} } \dd{\Omega} \stackrel{\eqref{eq:linearized:GW:QuadPole}}{=} \frac{\lambda^2}{\ifthenelse{\boolean{NaturalUnits}}{2^8\pi^2}{2^8\pi^2c}} \int \Lambda\ind{^{ij,kl}}(\tvec{n}) \expval{ \dddot{Q}\ind{_{ij}}\dddot{Q}\ind{_{kl}} } \dd{\Omega}.
	\end{empheq}
	\ifthenelse{\boolean{NaturalUnits}}{}{Note that $\dot{h}=\dv{h}{t}$ does not include any additional factors of $c$.} It is apparent that \acrshort{gw}s carry energy \emph{out} of a volume, hence the negative sign convention for $\F$. The term of order $r^{-3}$ can be neglected for large values of $r$, while in the remaining term the $r$ dependence cancels, making it non-vanishing.
\subsubsection{The Lambda tensor}
	Before computing the angular integral it is important to note that the radiation is not isotropic, but rather a function of direction $\tvec{n}$. Following the outline of \textcite{Maggiore:VolumeI} this can be accounted for by introducing the so-called \emph{Lambda tensor}
    \begin{align}
    \begin{split}\label{eq:Lambdatensor:def}
        \Lambda\ind{_{ij:kl}}(\tvec{n}) = \hspace{3pt} & \delta\ind{_{ik}} \delta\ind{_{jl}} - \frac{1}{2}\delta\ind{_{ij}} \delta\ind{_{kl}} - n\ind{_j}n\ind{_l}\delta\ind{_{ik}} - n\ind{_i} n\ind{_k} \delta\ind{_{jl}} \\
        & + \frac{1}{2}n\ind{_k} n\ind{_l} \delta\ind{_{ij}} + \frac{1}{2}n\ind{_i} n\ind{_j} \delta\ind{_{kl}} + \frac{1}{2} n\ind{_i} n\ind{_j} n\ind{_k} n\ind{_l}.
    \end{split}
	\end{align}
	
	The Lambda tensor is a projection operator which is defined in such a way that it projects a wave already in the Lorenz gauge \eqref{eq:def:DeDonder gauge} into the \acrshort{tt} gauge \eqref{eq:TT gauge condition}.
	\begin{align}
	    h^\text{\acrshort{tt}}_{ij}(\tvec{n}) = \Lambda\ind{_{ij}^{kl}}(\tvec{n}) h\ind{_{kl}},
	\end{align}
	which makes sure the wave is transverse with respect to the direction of propagation $\tvec{n}$.
	
	The Lambda tensor has the property
	\begin{align}
	    \Lambda\ind{_{ij}^{kl}} \Lambda\ind{_{kl}^{mn}} = \Lambda\ind{_{ij}^{mn}},
	\end{align}
	making it a projection operator. Thus, the contraction of two waves in the \acrshort{tt} gauge becomes
	\begin{align}
	     h^\text{\acrshort{tt}}_{ij}  h_\text{\acrshort{tt}}^{ij} = \Lambda\ind{_{ij}^{kl}} h\ind{_{kl}} \Lambda\ind{^{ij}_{mn}} h\ind{^{mn}} = \Lambda\ind{_{mn}^{ij}} \Lambda\ind{_{ij}^{kl}} h\ind{^{mn}} h\ind{_{kl}} = \Lambda\ind{_{mn}^{kl}} h\ind{^{mn}} h\ind{_{kl}},
	\end{align}
	i.e. the Lambda tensors can be used to contract tensors not in the \acrshort{tt} to get the contraction of the \acrshort{tt} gauged tensors, which is exactly what is needed for equation \eqref{eq:Flux:LinearTheory: hTT}. The last equality utilized the property $\Lambda\ind{_{ij:kl}}=\Lambda\ind{_{kl:ij}}$ of the Lambda tensor.
	
	When integrating over the Lambda tensor the following integral appears
	\begin{align}
		\label{eq:AngularIntegral:PointingVectors:generel}\int \frac{\dd{\Omega}}{4\pi} n_{i_1} n_{i_2} \cdots n_{i_{2\ell-1}} n_{i_{2\ell}} & = \frac{1}{(2\ell+1)!!} \left( \delta_{i_1 i_2} \cdots \delta_{i_{2\ell-1} i_{2\ell}} + \dots \right).
	\end{align}
	Here $\tvec{n}$ is taken to be a unit vector $\abs{\tvec{n}}=1$. To illustrate the solution take $\ell=1 \rightarrow \int \dd{\Omega}/(4\pi) n_i n_j = S_{ij}$. The trace of $S\ind{_{i}^i}=1$, because $n\ind{_i} n\ind{^i} = \abs{\tvec{n}} = 1$, reducing the integral to the scalar integral $\int \dd{\Omega}=4\pi$. The tensor structure of $S_{ij}$ should be that of the Kronecker delta, as it should be symmetric in its indices, and have a non-vanishing trace. Thus, $S_{ij}=\delta_{ij}/3$.
	
	Generalizing this argument it follows that the integral \eqref{eq:AngularIntegral:PointingVectors:generel} should be the sum of all possible Kronecker delta combinations, to maintain index symmetry. This is indicated by the dots on the \acrshort{rhs} of \eqref{eq:AngularIntegral:PointingVectors:generel}. The symbol $!!$ is here meant to imply the product of every second number: $(2\ell+1)!! = 1\cdot 3\cdots (2\ell-1) \cdot (2\ell+1)$, and $ 2\ell!! = 2\cdot 4 \cdots (2\ell-2) \cdot 2\ell $. This fixes the correct trace value, just like the factor of $1/3$ for the $\ell=1$ case. Note that all integrals over an odd number of unit vector components are 0 because the integral is also odd then.
	
	Thus, the integral of the Lambda tensor becomes
	\begin{align}
	    \nonumber \int \frac{\dd{\Omega}}{4\pi} \Lambda\ind{_{ij:kl}} = \hspace{3pt} & \delta\ind{_{ik}} \delta\ind{_{jl}} - \frac{1}{2}\delta\ind{_{ij}} \delta\ind{_{kl}} - \delta\ind{_{ik}} \int \frac{\dd{\Omega}}{4\pi} n\ind{_j}n\ind{_l} - \delta\ind{_{jl}} \int \frac{\dd{\Omega}}{4\pi} n\ind{_i} n\ind{_k}\\ 
	    \nonumber & + \frac{1}{2} \delta\ind{_{ij}} \int\frac{\dd{\Omega}}{4\pi} n\ind{_k} n\ind{_l} + \frac{1}{2} \delta\ind{_{kl}} \int\frac{\dd{\Omega}}{4\pi} n\ind{_i} n\ind{_j} + \frac{1}{2} \int\frac{\dd{\Omega}}{4\pi} n\ind{_i} n\ind{_j} n\ind{_k} n\ind{_l} \\
	    \nonumber = \hspace{3pt} & \delta\ind{_{ik}} \delta\ind{_{jl}} - \frac{1}{2}\delta\ind{_{ij}} \delta\ind{_{kl}} - \delta\ind{_{ik}} \frac{1}{3} \delta\ind{_{jl}} - \delta\ind{_{jl}} \frac{1}{3} \delta\ind{_{ik}} +  \frac{1}{2}\delta\ind{_{ij}} \frac{1}{3} \delta\ind{_{kl}} + \frac{1}{2} \delta\ind{_{kl}} \frac{1}{3} \delta\ind{_{ij}}\\
	    \nonumber & + \frac{1}{2} \frac{1}{3\cdot5} \left( \delta\ind{_{ik}}\delta\ind{_{jl}} + \delta\ind{_{ij}}\delta\ind{_{kl}} + \delta\ind{_{il}}\delta\ind{_{kj}} \right) \\
	    = \hspace{3pt} & \frac{11}{30} \delta\ind{_{ik}}\delta\ind{_{jl}} -\frac{2}{15} \delta\ind{_{ij}}\delta\ind{_{kl}} + \frac{1}{30} \delta\ind{_{il}}\delta\ind{_{kj}}. \label{eq:AngularIntLambda}
	\end{align}
	
	Utilizing this result for \eqref{eq:Flux:LinearTheory: hTT}, and the fact that $Q\ind{_{ij}}=Q\ind{_{ji}}$, yields the flux in terms of the mass quadrupole
	\begin{empheq}[box=\widefbox]{align} \label{eq:energyflux:from Quad}
	    \F = \frac{\lambda^2}{\ifthenelse{\boolean{NaturalUnits}}{2^5\cdot 5\pi}{2^5 \cdot 5 \pi c}} \expval{ \dddot{Q}\ind{_{ij}}\dddot{Q}\ind{^{ij}} - \frac{1}{3} \dddot{Q}^2 } = \ifthenelse{\boolean{NaturalUnits}}{\frac{1}{5}}{\frac{G}{5c^5}} \expval{ \dddot{Q}\ind{_{ij}}\dddot{Q}\ind{^{ij}} - \frac{1}{3} \dddot{Q}^2 },
	\end{empheq}
	with $Q\equiv Q\ind{_i^i}$. 
	
	This is an important dynamical property of \acrshort{gw}s, that they can carry energy, momentum, and angular momentum out of a system. For a compact binary this influences the orbital energy, extracting energy from it over time, make them spiral in towards each other. The fact that there is no analytical exact solution to the relativistic binary problem is often attributed to this effect. The orbit is affected by \acrshort{gw} radiation, which is determined by the orbit, making the problem complicated.
\section{Illustrative example: Binary system with circular orbits} \label{sec:Example:Binary}
    As an illustrative example this section will calculate the exact form of \acrshort{gw} produced by binary systems in circular orbits, according to linearized theory \eqref{eq:h2Lagrangian:full}.
    
    Those already familiar with \acrshort{gw}s may skip this section, as this system will be covered in greater detail in the following chapters, and is here mearly used as an example to better illustrate the result \eqref{eq:linearized:GW:QuadPole} and \eqref{eq:energyflux:from Quad}.
    
    \begin{figure}
        \centering
        \includegraphics[width=0.75\textwidth]{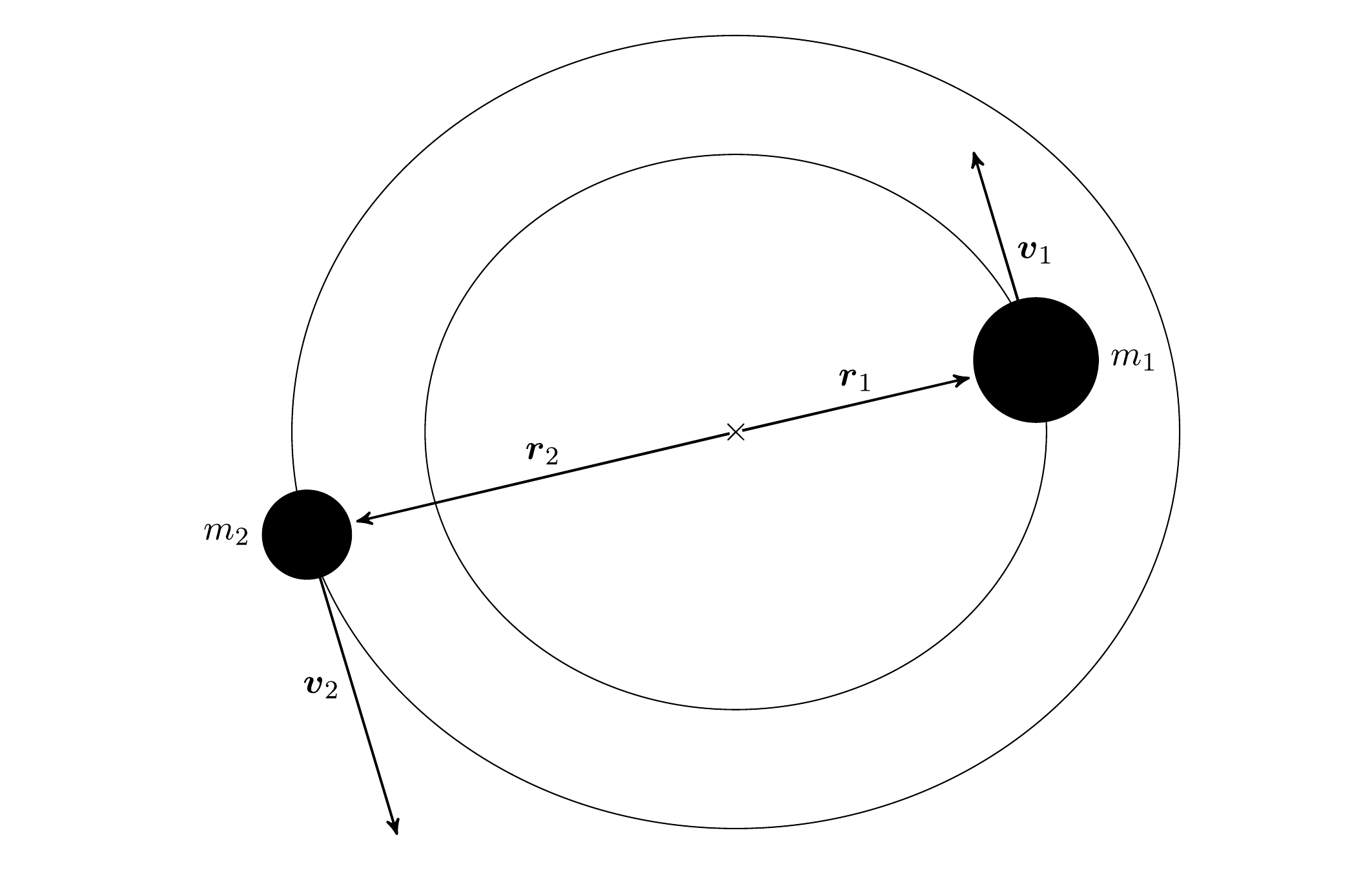}
        \caption{Diagram of a binary system.}
        \label{fig:BinarySystem}
    \end{figure}
	
	Taking the centre of mass as the origin, the positions of the stars are
	\begin{subequations}
	\begin{align}
		&\tvec{r}_1 = 
		\begin{pmatrix}
			r_1 \cos(\omega t), & r_1 \sin(\omega t), & 0
		\end{pmatrix},\\
		&\tvec{r}_2 = -
		\begin{pmatrix}
			r_2 \cos(\omega t), & r_2 \sin(\omega t), & 0
		\end{pmatrix}.
	\end{align}
	\end{subequations}
	
	This determines the mass density as
	\begin{align}
		\rho(t,\tvec{r}) = \sum_{a} m_a \dirac{3}{\tvec{r}-\tvec{r}_a(t)}
	\end{align}
	
	The quadrupole moment of the binary system is thus 
	\begin{align}
		\begin{split}
			Q\ind{_{ij}}(t) & = \int \rho x\ind{_i}x\ind{_j} \dd[3]{x} = \sum_{a} m_a (x_a)\ind{_i}(x_a)\ind{_j} \\
			& = (m_1r_1^2 + m_2r_2^2) \begin{pmatrix}
				\cos[2](\omega t) & \cos(\omega t) \sin(\omega t) & 0 \\
				\cos(\omega t)\sin(\omega t) & \sin[2](\omega t) & 0 \\
				0 & 0 & 0
			\end{pmatrix}\\
			& = \frac{\mu r^2}{2} \begin{pmatrix}
				\cos(2\omega t) & \sin(2\omega t) & 0 \\
				\sin(2\omega t) & -\cos(2\omega t) & 0 \\
				0 & 0 & 0
			\end{pmatrix}
		\end{split}\\
		\Rightarrow \quad \ddot{Q}\ind{_{ij}}(t) & = 2 \mu (\omega r)^2 \begin{pmatrix}
			-\cos(2\omega t) & -\sin(2\omega t) & 0 \\
			-\sin(2\omega t) & \cos(2\omega t) & 0 \\
			0 & 0 & 0
		\end{pmatrix}. \label{eq:Quad:double time deriv}
	\end{align}
	Details of mass and trigonometric term manipulations can be found in \cref{app:equivOneBodyAndMassTerms} and \ref{app:trig} respectively.
	
	Thus, the \acrshort{gw}s produced by a circular binary system is determined by \eqref{eq:linearized:GW:QuadPole} and \eqref{eq:Quad:double time deriv} to be
	\begin{empheq}[box=\widefbox]{align}\label{eq:Linearized:GWfromNewtonBinary:FinalExpression}
		h\ind{_{ij}}(t,\tvec{R}) = -\frac{\lambda}{8\pi}\frac{\mu v^2}{R} \begin{pmatrix}
			\cos(2\omega t_\text{ret}) & \sin(2\omega t_\text{ret}) & 0 \\
			\sin(2\omega t_\text{ret}) & -\cos(2\omega t_\text{ret}) & 0 \\
			0 & 0 & 0
		\end{pmatrix}
	\end{empheq}
	For convenience $r\omega=(r_1+r_2)\omega$ has been replaced by $v$, which is the sum of the velocities of the stars. This is equivalent to the frequency parameter $v^3=GM\omega$ introduced in \cref{chap:waveform}.
	
	It should be noted that the wave frequency $\omega_\text{\acrshort{gw}} = 2\omega = 2\omega_\text{s}$, with $\omega_\text{s}$ as the source frequency, is the dominant frequency for circular orbits, and that the amplitude is proportional to the frequency $v^2=(GM\omega)^{2/3}$. This makes the \acrshort{gw} highly dependent on the frequency of the source binary, both for the amplitude, and frequency spectrum.
	
	The energy flux produced by such a system is now easily computable by using equation \eqref{eq:energyflux:from Quad} together with \eqref{eq:Quad:double time deriv}.
	\begin{empheq}[box=\widefbox]{align}
    	\nonumber \F & = \frac{\lambda^2}{\ifthenelse{\boolean{NaturalUnits}}{2^5\cdot 5\pi}{2^5 \cdot 5 \pi c}} \expval{ \dddot{Q}\ind{_{ij}}\dddot{Q}\ind{^{ij}} - \frac{1}{3} \dddot{Q}^2 } \\
    	\nonumber & = \frac{\lambda^2}{\ifthenelse{\boolean{NaturalUnits}}{2^5\cdot 5\pi}{2^5 \cdot 5 \pi c}} 2^4 \mu^2 v^4 \omega^2 \expval{ 2\sin^2(2\omega t_\text{ret}) + 2\cos^2(2\omega t_\text{ret}) -0 } \\
    	& = \frac{\lambda^2 \mu^2 v^{10}}{5\pi G^2M^2 c} = \frac{ 32 }{ 5 } \frac{\eta^2}{G} \frac{v^{10}}{c^5}. \label{eq:NewtonianFlux pp}
	\end{empheq}
	This is exactly the first order term of the flux expansion \eqref{eq:Flux:Expansion:Gen}.
	
	Furthermore, circular orbits \emph{is} a solution of the Newtonian, \acrlong{pp}, action \eqref{eq:NewtonianAction}. If the source interact with the graviton field, and only the graviton field, this implies that the orbital energy of the system is $E_\text{Newt}=-\frac{1}{2}\mu v^2$ \eqref{eq:NewtonianEnergy pp}.
	
	With \eqref{eq:NewtonianFlux pp} and \eqref{eq:NewtonianEnergy pp}, the Newtonian order (0\acrshort{pn}) version of the phase \eqref{eq:PsiSPA:f} can be derived, assuming compact binaries can effectively be treated as point masses.
	
	The flux is obviously not zero, which means that energy is dissipated out of the system. However, according to the Lagrangian \eqref{eq:h2Lagrangian:full} the energy of the source $T\ind{^{\mu\nu}}$ \emph{is} conserved. Therefore, this energy radiation is an inconsistency of the theory.
	
	This hits deeply into the problem of gravity. As a field theory it couples to energy, while as a field it itself stores energy, and should therefore couple to itself. Amending this feature is the topic of the next section.
\section{Graviton action beyond quadratic order} \label{sec:GravitonBeyondQuadratic}
    In \cref{sec:Energy-Momentum tensor of GW} it was found that gravitons / \acrshort{gw}s can carry energy, and can even be given an effective energy-momentum (pseudo-)tensor of its own. But since the graviton field couples to energy-momentum tensors, should it not couple to itself?
    
    In \cref{sec:Example:Binary} it was even found that a binary system of \acrlong{pp}s radiated energy in the form of \acrshort{gw}s, but where does this energy come from? One natural assumption would be that the energy is extracted from the orbital energy associated with the system, leading to an inspiral. But this would mean that the energy-momentum tensor of the binary $T\ind{_{\mu\nu}^{,\nu}} \neq 0$, as it changes over time. Furthermore, $T\ind{^{\mu\nu}_{,\nu}} = 0$ was used to determine the graviton action in the first place, making the conservation of energy of the source, and the energy carrying capability of the graviton field a theoretical inconsistency!
    
    The solution is to \emph{change} the energy conservation criteria \eqref{eq:EoMLh2MustBeDivergenceless} to
    \begin{align}\label{eq:energy cons of T+t}
        \left( T\ind{^{\mu\nu}} + t\ind{^{\mu\nu}} \right)\ind{_{,\nu}} = 0,
    \end{align}
    which is to make the energy of the source \emph{plus} the energy stored in gravitons (i.e. in \acrshort{gw}s) conserved, together.
    
    To implement this in a new and improved Lagrangian for the graviton field the first naive idea would be to couple the graviton directly to the graviton energy-momentum tensor $\L_\text{int} \to \frac{\lambda}{2} h\ind{_{\mu\nu}}\left( T\ind{^{\mu\nu}} + t\ind{^{\mu\nu}} \right)$. This will however lead to problems for the \acrshort{eom}, as
    \begin{align}
    \begin{split}
        &\fdv{}{h\ind{_{\mu\nu}}} \frac{\lambda}{2} h\ind{_{\mu\nu}} t\ind{^{\mu\nu}} = \frac{\lambda}{2} t\ind{^{\mu\nu}} + \frac{\lambda}{2} h\ind{_{\rho\sigma}} \fdv{}{h\ind{_{\mu\nu}}} t\ind{^{\rho\sigma}} \neq \frac{\lambda}{2} t\ind{^{\mu\nu}}\\
        &\qq*{since} t\ind{^{\mu\nu}} \stackrel{\eqref{eq:GWt:nonaveraged}}{=} h\ind{_{\sigma\tau}^{,\mu}}h\ind{^{\sigma\tau,\nu}} - \frac{1}{2}h\ind{^{,\mu}}h\ind{^{,\nu}} + \L_{(2)}\eta\ind{^{\mu\nu}}.
    \end{split}
    \end{align}
    
    Since it is the \acrshort{eom} which dictates the physics of any given action, one would rather demand that
    \begin{align}
        \fdv{}{h\ind{_{\mu\nu}}} \L_\text{int}^{*} = \frac{\lambda}{2} \left( T\ind{^{\mu\nu}} + t\ind{^{\mu\nu}} \right).
    \end{align}
    
    This requires the addition of a more general cubic term to the Lagrangian $\frac{\lambda}{2}\L_{(3)}$ such that 
    \begin{align}\label{eq:condit L3}
        \fdv{}{h\ind{_{\mu\nu}}} \frac{\lambda}{2}\L_{(3)} = \frac{\lambda}{2} t\ind{^{\mu\nu}}.
    \end{align}
    
    Writing out all possible terms cubic in $h$, containing only two derivatives (necessary to obtain $t\ind{^{\mu\nu}}$), which also produces $t\ind{^{\mu\nu}}$ when varied \eqref{eq:condit L3}, together with new energy conservation condition \eqref{eq:energy cons of T+t} fixes the 18 coefficients uniquely (up to the same overall factor of $\L_{(2)}$). This is obviously a lot of work, but the result can be quoted from \textcite{Feynman:GravityLectures} (equation (6.1.13)).\footnote{Note that Feynman uses a different overall scaling factor for his graviton action ($a_1=1$, rather than $a_1=-1/2$). In \eqref{eq:L3} all barred factors of $h$ have been written out, also in contrast to (6.1.13) of \cite{Feynman:GravityLectures}.}
    \begin{align}
    \begin{split}\label{eq:L3}
        \frac{\lambda}{2}\L_{(3)} = \hspace{3pt} & \frac{\lambda}{2} \Bigl[ h\ind{^{\alpha\beta}} h\ind{^{\gamma\delta}} h\ind{_{\alpha\beta,\gamma\delta}} - \frac{1}{2} h h\ind{^{\alpha\beta}} h\ind{_{,\alpha\beta}} - \frac{1}{2} h h\ind{^{\alpha\beta}}\dalembertian h\ind{_{\alpha\beta}} - \frac{3}{8}h^2 \dalembertian h \\
        & + h\ind{_\gamma^\beta} h\ind{^{\gamma\alpha}} \dalembertian h\ind{_{\alpha\beta}} - \frac{3}{4} h\ind{_{\alpha\beta}} h\ind{^{\alpha\beta}} \dalembertian h - 2 h\ind{^{\alpha\beta}} h\ind{_{\beta\delta}} h\ind{_{\alpha\gamma}^{,\gamma\delta}} + h\ind{^{\alpha\beta}} h\ind{_{\beta\delta}} h\ind{_{,\alpha}^\delta} \\
        & +2 h\ind{_{\alpha\beta}} h\ind{^{\sigma\alpha}_{,\sigma}} h\ind{^{\tau\alpha}_{,\tau}} - 2 h\ind{_{\alpha\beta}} h\ind{^{\sigma\alpha}_{,\sigma}} h\ind{^{,\beta}} + \frac{1}{2} h\ind{_{\alpha\beta}} h\ind{^{,\alpha}} h\ind{^{,\beta}} - h h\ind{^{\sigma\alpha}_{,\sigma}} h\ind{^\tau_{\alpha,\tau}} \\
        & + h h\ind{^{\sigma\alpha}_{,\sigma}} h\ind{_{,\alpha}} - \frac{1}{4} h h\ind{_{,\alpha}} h\ind{^{,\alpha}} + \frac{1}{2} h\ind{_{\alpha\beta}} h\ind{^{\alpha\beta}} h\ind{^{\sigma\tau}_{,\sigma\tau}} + \frac{1}{4} h^2 h\ind{^{\sigma\tau}_{,\sigma\tau}} \Bigr].
    \end{split}
    \end{align}
    
    Using again Noether's theorem on this new Lagrangian $\L_{(2)} + \L_{(3)}$, the new effective energy-momentum tensor picks up a term cubic in $h$:
    \begin{align}
        t\ind{^{\mu\nu}} = t^{\mu\nu}_{(2)} + \frac{\lambda}{2} t^{\mu\nu}_{(3)}.
    \end{align}
    
    Again it can be argued that the cubic term here is not conserved, and the process of this section can be started all over again. In fact, it turns out that
    \begin{empheq}[box=\widefbox]{align} \label{eq:Graviton Action All Orders}
        \L_\text{grav} = \L_\text{int} +  \L_{(2)} + \frac{\lambda}{2} \L_{(3)} + \frac{\lambda^2}{2^2} \L_{(4)} + \dots,
    \end{empheq}
    with the \acrshort{eom}
    \begin{empheq}[box=\widefbox]{align} \label{eq:EoM Graviton All Orders}
        \dalembertian \bar{h}\ind{_{\mu\nu}} = - \frac{\lambda}{2} \left( T\ind{_{\mu\nu}} + t_{\mu\nu}^{(2)} + \frac{\lambda}{2}t_{\mu\nu}^{(3)} + \dots \right).
    \end{empheq}
    
    This makes the \acrlong{eom} nonlinear, and in general both the action and the \acrshort{eom} contain infinitely high powers of $h$. 
    
    By now the Venusians would probably be disappointed, but for terrestrial physicists this should be expected, as \acrshort{gr} is also a nonlinear theory. When expanded in powers of metric perturbations, also the \gls{Einstein-Hilbert action} contain infinitely high powers of $h$, which coincide with the action found thus far in this thesis.\footnote{There are many who argue that the procedure outlined here can be carried out ad infinitum, and then reproduces the \gls{Einstein-Hilbert action}, but this is disputed. The \gls{Einstein-Hilbert action} can definitely be expanded in this manner, but it is disputed wether or not the \glspl{field theorist}' method produces \emph{uniquely} the \gls{Einstein-Hilbert action}. \cite{Padmanabhan_2008}}
    
    An important observation is that all these corrections scale with increasing powers of $\lambda$, which is small. Thereby the more powers of $h$ it contains, the less it contributes to the final result. Thus, the theory can be perturbatively expanded in powers of $\lambda$.
\chapter{Calculating the orbital energy} \label{chap:energy}
    In this chapter the 1\acrshort{pn} energy of compact binaries in circular motion is computed. 
    
    The derivation follows closely those presented in \textcite{Porto2016}, and Goldberger \cite{Goldberger:LesHouches,Goldberger:EFT}.
    
\section{Effective field theory} \label{sec:EFT}
    In the last chapter examples focused on \acrlong{pp}s. Compact objects like \acrshort{bh}s and \acrshort{ns}s are however not point particles. But for sufficiently far separated binaries, the separation distance $r$ will be much greater than the `size' of the compact object, which can be approximated as the \gls{R_S} $\sim R_S \ll r$. Then the system can be described by an \emph{effective action}, treating the compact object as a point mass at the scale $\sim r$.
    
    On the other end of the scale the system is producing \acrshort{gw}s, carrying energy out of the system. But when calculating the energy flux in \cref{sec:Energy-Momentum tensor of GW}, the source was assumed to effectively be a point endowed with quadrupole structure (which will be expanded to a general multipole structure in \cref{chap:flux}), since the \acrshort{gw} was measured far away. That is, the flux is measured at a scale $L \gg \lambdabar \gg r$.\footnote{\label{fn:relating wavelength and orbital scale}The wavelength can be shown to be greater than the size of the binary by noticing two things. First: $\omega_\text{\acrshort{gw}} \sim 2\omega_\text{s}$, as approximated in \eqref{eq:Linearized:GWfromNewtonBinary:FinalExpression}. The reduced wavelength $\lambdabar=\lambda/2\pi$ is related to the angular frequency as $\lambdabar = \ifthenelse{\boolean{NaturalUnits}}{\omega^{-1}}{c/\omega}$. Second: Kepler's third law \eqref{eq:KeplersThirdLaw} relate $\omega$ and $r$: $\lambdabar_\text{\acrshort{gw}} = \frac{\ifthenelse{\boolean{NaturalUnits}}{1}{c}}{2\sqrt{GM}}r^{3/2} = \sqrt{\frac{r}{2R_S}} r \gg r$.} Thus, in this chapter the effective action at the scale $r$, with $R_S \ll r \ll L$, will be derived, which approximates the immediate\footnote{Immediate because over time the energy loss through \acrshort{gw} emssion can not be ignored. However over short periodes of time, the orbital energy can be approximated to be conserved.} orbital dynamics.
    
\begin{figure}[ht]
	    \centering
	    \includegraphics[width=0.9\textwidth]{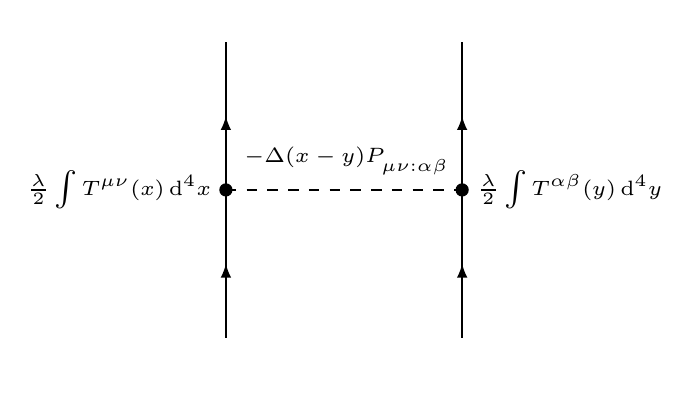}
	    \caption{Newtonian Feynman diagram / `H-diagram'.}
	    \label{fig:Feynman:H-diagram:detailed}
	\end{figure}

    To compute the 1\acrshort{pn} orbital energy the procedure of \cref{sec:GravityFromGravitons} will be expanded. In \cref{sec:GravityFromGravitons} it was found that the graviton potential could be expressed graphically like \cref{fig:Feynman:H-diagram:detailed}.
	In \cref{fig:Feynman:H-diagram:detailed} the two sources are depicted as solid lines, like in a space-time diagram, while the Green's function is depicted as a squiggly line, connecting the two point particles. Graphical representations of this kind are called \emph{Feynman diagrams}, as they were introduced by \textcite{FirstFeynmanDiagram} to illustrate expansion terms of \acrshort{qed}. Note that the Green's function has been given the subscript `inst', to remind that the Newtonian action was recovered by approximating $\dirac{4}{x^\mu-y^\mu} \approx \dirac{}{t-t'}$, making the interaction \emph{instantaneous}.
	
	For those familiar with Feynman diagrams, note that the solid lines are \emph{not propagating}, even though they are depicted the same way as propagating fermions in \acrshort{qft}. This can be confusing, but it is the convention introduced by \textcite{Goldberger:EFT}, and has by now become the standard.
\subsection{Expand the action in powers of \texorpdfstring{$h$}{h}}
	Based on the findings of \cref{sec:GravitonBeyondQuadratic}, the \gls{Einstein-Hilbert action}\footnote{See the glossary for the definition of this action.} $S_\text{\acrshort{EH}}$ can be used to find the expression of the \emph{propagator} for the $h\ind{_{\mu\nu}}$ field, by solving equation \eqref{eq:EoM Graviton All Orders}. Expanding the \gls{Einstein-Hilbert action} in powers of $h$ yields the contribution for the different levels of self interaction, which all scale with additional powers of the coupling constant $\lambda$.
	\begin{align}\label{eq:SEH:Expansion}
		S_\text{\acrshort{EH}} \sim \frac{1}{2}\int \dd[4]{x} \left[(\partial h)^2 + \lambda h(\partial h)^2 + \lambda^2 h^2(\partial h)^2 + \dots \right]
	\end{align}
	\begin{center}
	    \includegraphics[width=0.75\textwidth]{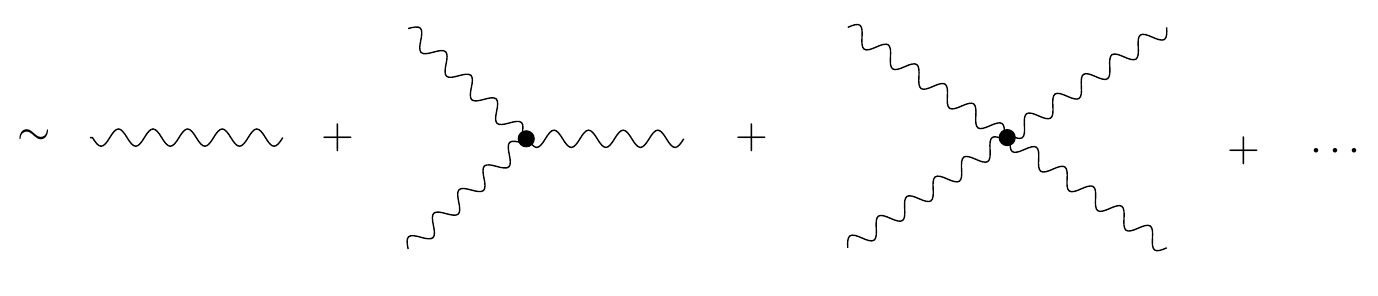}
	\end{center}
	Note that the terms in \eqref{eq:SEH:Expansion} are symbolic representations for the powers of $h\ind{_{\mu\nu}}$, and that all terms contain only two derivatives.
	
	Why does the first order correction lead to a three graviton vertex? Looking at the leading order correction in the \acrshort{eom} \eqref{eq:EoM Graviton All Orders}
	\begin{align}
	\begin{split} \label{eq:ThreePointProp:first}
	    \dalembertian h\ind{_{\mu\nu}} & = - \frac{\lambda}{2} P\ind{_{\mu\nu:\alpha\beta}} \left( t^{\alpha\beta}_{(2)} \right) \sim \frac{\lambda}{2} P\ind{_{\mu\nu:\alpha\beta}} \left( \partial h \partial h \right), \\
	    \Rightarrow \quad h\ind{_{\mu\nu}}(x) & \sim - \int \dd[4]{y_1} \dd[4]{y_2} \partial T\ind{^{\alpha\beta}}(y_1) \partial T\ind{^{\alpha\beta}}(y_2) \Delta_\text{inst}(x-y_1) \Delta_\text{inst}(x-y_2),
    \end{split}
	\end{align}
	where again $\sim$ implies symbolic relation of powers of $h\ind{_{\mu\nu}}$, rather than the exact relation. The leading order term of this three-point propagator will be derived in \cref{subsec:diagram e}. 
	
	It can however be seen from this simple relation, where the $h$'s of the \acrshort{rhs} was substituted for the linear order solution, that at this order $h$ is produced by two sources, which in principle can be at different space-time points. Therefore, an interaction $h\ind{_{\mu\nu}}T\ind{^{\mu\nu}}(x)$ in the action will to next order connect three energy-momentum tensors, using two Green's functions. This can neatly be visualized by a three graviton vertex.
	
	In \cref{sec:GravityFromGravitons} the action of \acrlong{pp}s \eqref{eq:pp action expansion} was found to also be expandable in powers of $h$, similarly to the \gls{Einstein-Hilbert action}.
	\begin{subequations} \label{eq:Spp:Expansion}
	\begin{align}
		\label{eq:Spp:Total}S_{pp} = & -m\ifthenelse{\boolean{NaturalUnits}}{}{c^2} \int \dd{\tau} \sqrt{1-\lambda h\ind{_{\mu\nu}} \ifthenelse{\boolean{NaturalUnits}}{ \dot{x}\ind{^\mu} \dot{x}\ind{^\nu} }{ \frac{\dot{x}\ind{^\mu}}{c} \frac{\dot{x}\ind{^\nu}}{c} } }\\
		\label{eq:Spp:Expansion terms}= & -m \ifthenelse{\boolean{NaturalUnits}}{}{c^2} \int \gamma^{-1} \dd{t} + \frac{\lambda}{2} \int \dd{t} h\ind{_{\mu\nu}} \gamma^{-1} m \dot{x}\ind{^\mu}\dot{x}\ind{^\nu} + \frac{m\lambda^2}{8\ifthenelse{\boolean{NaturalUnits}}{}{c^2}} \int \dd{t} \gamma^{-1} \left(h\ind{_{\mu\nu}}\dot{x}\ind{^\mu}\dot{x}\ind{^\nu}\right)^2 + \dots
	\end{align}
	\end{subequations}
	\begin{center}
	    \includegraphics[width=0.7\textwidth]{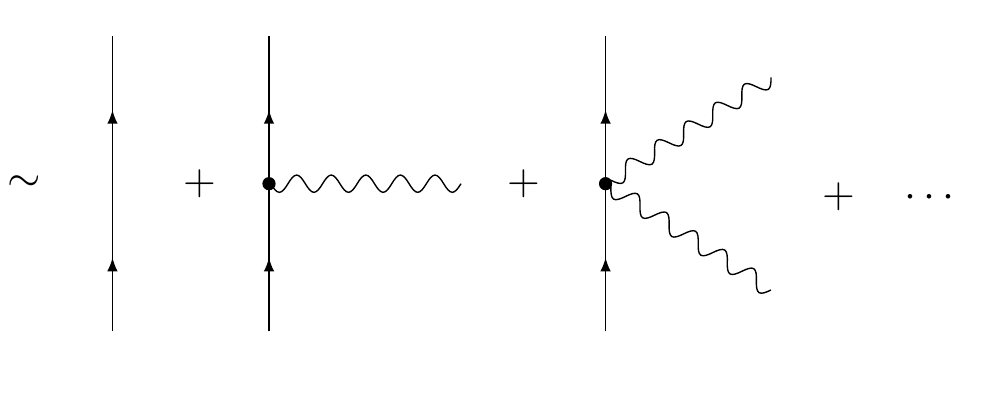}
	\end{center}
	
	Here again the choice of having two gravitons coupling to the world line for the $\sim \left(h\ind{_{\mu\nu}}\dot{x}\ind{^\mu}\dot{x}\ind{^\nu}\right)^2$ term is natural as the integral can be split in two integrals multiplied together \newline $\sim \int \dd[3]{x}\left(h\ind{_{\mu\nu}}(t,\tvec{x}) \dot{x}^\mu \dot{x}^\nu \right) \dirac{3}{\tvec{x}-\tvec{x}_a(t)} \int  \dd{y} \left(h\ind{_{\mu\nu}}(t,\tvec{y}) \dot{y}^\mu \dot{y}^\nu \dirac{3}{\tvec{y}-\tvec{x}_a(t)} \right)$, which couples this source effectively to two different gravitons.
	
	In \cref{sec:GravityFromGravitons} this was argued to be a natural choice for the \acrlong{pp} action, as it is the expansion of the action of \acrshort{gr}. It is also a natural choice for a classical field theory with a coupling to a symmetric rank 2 tensor field, and is equivalent to equation (11.40) of \textcite{SpesRel}.
\subsection{Separation of scale}
	Because of the separation of scale, $\sim r \gg R_S$ and $\sim L \gg r$, it will be useful to split the graviton field into a short-range, potential field ($H\ind{_{\mu\nu}}$) and a long range, radiation field ($\H\ind{_{\mu\nu}}$)
	\begin{empheq}[box=\widefbox]{align}
		h\ind{_{\mu\nu}} = H\ind{_{\mu\nu}} + \H\ind{_{\mu\nu}}.
	\end{empheq}
	From here on out the potential field will be drawn using dashed lines, while the radiation field will continue to be drawn using squiggly lines.
	
	The frequency of \acrshort{gw}s are proportional to the frequency of the source binary $\omega_\text{\acrshort{gw}} \simeq 2\omega = 2\omega_\text{s}$, and the relative velocity of a binary in circular motion is $v=\omega r$. Using the relation between null wave frequency and wavelength $\omega = \ifthenelse{\boolean{NaturalUnits}}{\abs{\tvec{k}}}{c\abs{\tvec{k}}} = \ifthenelse{\boolean{NaturalUnits}}{ \lambdabar^{-1} }{ c \lambdabar^{-1} }$ the wavelength of \acrshort{gw}s scale as
	\begin{align}
	    \lambdabar_\text{\acrshort{gw}} \simeq \ifthenelse{\boolean{NaturalUnits}}{ (2\omega)^{-1} }{ \frac{c}{2\omega} } = \ifthenelse{\boolean{NaturalUnits}}{ \frac{r}{2 v} }{ \frac{rc}{2v} } \gg r,
	\end{align}
	for binaries moving at non-relativistic speeds.
	
	Because $\H$ is null-like, it follows that $k\ind{_\sigma} k\ind{^\sigma} = 0$, and
	\begin{empheq}[box=\widefbox]{align} \label{eq:GradiantRelation:RadiationMode}
		\partial\ind{_\alpha}\H\ind{_{\mu\nu}} = k\ind{_\alpha} \H\ind{_{\mu\nu}} \sim \lambdabar_\text{\acrshort{gw}}^{-1} \H\ind{_{\mu\nu}} \simeq \frac{v}{r} \H\ind{_{\mu\nu}},
	\end{empheq} 
	i.e. $\H$ must be \textit{on shell}.
	
    $H$ on the other hand is a potential field, and \textit{can not} be on shell, since it shall reproduce the gravitational potential in the static limit. In \cref{sec:GravityFromGravitons} this was achieved by approximating $\dirac{4}{x-y} \approx \dirac{}{t-t'}$. This `instantaneous' propagator is \emph{not} the Green's function of the d'Alembertian operator
    \begin{align}
        \Delta_\text{inst}(x-y) = \frac{-\dirac{}{t-t'}}{4\pi \abs{\tvec{x}-\tvec{y}}} = \int \frac{ \dd[4]{k} }{(2\pi)^4} \frac{ e^{ik\ind{_{\sigma}} (x\ind{^\sigma}-y\ind{^\sigma}) } }{ -\tvec{k}^2 },
    \end{align}
    but it \emph{is} the Green's function of the \emph{Laplace operator} $\nabla^2 \equiv \partial\ind{_i} \partial{^i}$. This integral scales proportional to $\abs{\tvec{k}}$
    \begin{align}
        \int \dd[3]{k} \frac{1}{\tvec{k}^2} \sim \frac{\abs{\tvec{k}}^3}{\tvec{k^2}} \sim k\ind{_i} \sim \frac{1}{\abs{\tvec{x}-\tvec{y}}}.
    \end{align}
    
    Assuming $k\ind{_0}$ is small compared to $k\ind{_i}$, the instantaneous propagator may be obtained as a leading order term of an expansion in $k\ind{_0}/\abs{\tvec{k}}$
    \begin{align}
    \begin{split} \label{eq:Propagator:Expansion}
        \Delta_\text{inst}(k) \equiv \frac{1}{ -k\ind{_\mu}k\ind{^\mu} } & = \frac{1}{ k_0^2 - \tvec{k}^2 } = \frac{1}{-\tvec{k}^2} \frac{1}{1 - k_0^2/\tvec{k}^2 }\\
        & = \frac{1}{-\tvec{k}^2} \left( 1 + \left( \frac{k\ind{_0}}{\tvec{k}} \right)^2 + \left( \frac{k\ind{_0}}{\tvec{k}} \right)^4 +\dots \right).
    \end{split}
    \end{align}
    \begin{center}
        \includegraphics[width=0.9\textwidth]{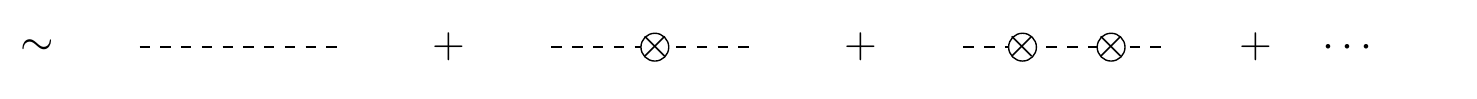}
    \end{center}
    
    Not only is this a highly constructed expansion to obtain a desired leading order term, it will be demonstrated that this expansion \emph{actually} scales as an expansion in $\ifthenelse{\boolean{NaturalUnits}}{v^2}{(v/c)^2}$, and will be drawn graphically by $\otimes$ on the propagator to show which order in the $(k\ind{_0}/\abs{\tvec{k}})^2$ expansion it represents. This also implies that $k\ind{_0}/k\ind{_i} \sim v \Rightarrow k\ind{_0} \sim v k\ind{_i} \sim v/r$, and in conclusion
	\begin{empheq}[box=\widefbox]{align} \label{eq:GradiantRelation:PotentialMode}
		\partial\ind{_0}H\ind{_{\mu\nu}} = k\ind{_0} H\ind{_{\mu\nu}} \sim \frac{v}{r} H\ind{_{\mu\nu}}, \quad \partial\ind{_i}H\ind{_{\mu\nu}} = k\ind{_i}H\ind{_{\mu\nu}} \sim \frac{1}{r} H\ind{_{\mu\nu}}.
	\end{empheq}
	
	Now Feynman diagrams may be constructed by putting together terms as presented in \eqref{eq:SEH:Expansion}, \eqref{eq:Spp:Expansion}, and \eqref{eq:Propagator:Expansion}, but the following three rules must be upheld to make sense as an expansion term in the point particle action.
	\begin{enumerate}
		\item Diagrams must remain connected if the particle lines are stripped off. \label{enu:conectedDiagrams}
		\item Diagrams may only contain internal $H\ind{_{\mu\nu}}$ lines. \label{enu:Hinternal}
		\item Diagrams may only contain external $\H\ind{_{\mu\nu}}$ lines. \label{enu:Hexternal}
	\end{enumerate}
	
	Rule \ref{enu:Hinternal}-\ref{enu:Hexternal} follow by definition of $H\ind{_{\mu\nu}}$ and $\H\ind{_{\mu\nu}}$. Rule \ref{enu:conectedDiagrams} however seems more mysterious, but it is a consequence of the solid lines \emph{not propagating}, and thus is just a requirement that the Feynman diagram is connected, as all Feynman diagrams must. Unconnected diagrams are simply separate diagrams, and represents multiple terms in the expansion at once.
	\begin{figure}[ht]
	    \centering
	    \includegraphics[width=0.5\textwidth]{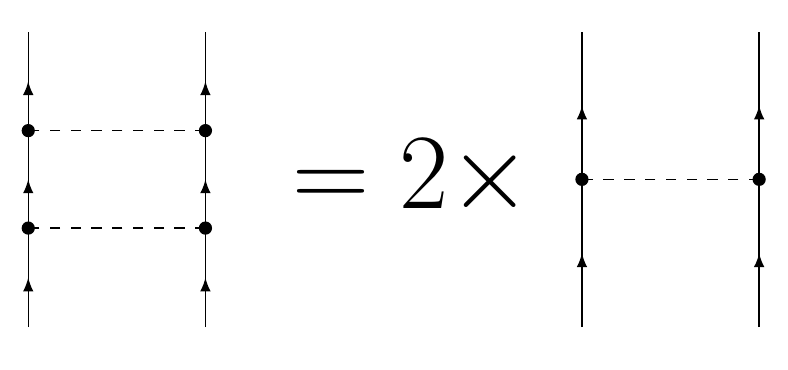}
	    \caption[`Ladder' Feynman diagram]{`Ladder' Feynman diagram. With non-propagating sources, unconnected graviton lines simply represents different diagrams.}
	    \label{fig:FeynmanDiagram:Ladder}
	\end{figure}
\section{The 1PN Lagrangian} \label{sec:1PNLagrangian}
	In order to acquire relativistic, \acrlong{pn}, corrections to the Newtonian Lagrangian \eqref{eq:NewtonianAction} it is a straightforward matter to just add additional terms of the form found in expansion \eqref{eq:SEH:Expansion}, \eqref{eq:Spp:Expansion}, and \eqref{eq:Propagator:Expansion} to the action. But \eqref{eq:SEH:Expansion} and \eqref{eq:Spp:Expansion} are expansions in $\lambda$ and $H$ rather than in $(v/c)^2$, so how can it be determined which Feynman diagrams contribute at which \acrshort{pn} order?
\subsection{Assigning PN order to Feynman diagrams} \label{sec:assigning PN to Feynman}
	Using the scaling of $k\ind{_\mu}$ for the potential field \eqref{eq:GradiantRelation:PotentialMode} and Kepler's law \eqref{eq:KeplersThirdLaw}, it turns out that it is possible to assign a power of $v$, $r$ and $m$ to each Feynman diagram, and thus select the appropriate diagrams and terms relevant to each \acrshort{pn} order, all without doing the full calculation! This is why the diagrams are introduced in the first place, as they are tools to make it easier to order expansion terms.
	
	The scaling of coordinate (integration variables) follows from the relations of $k\ind{_\mu}$.
	
	\begin{subequations}
    \begin{align}
    \begin{split}
        \int \dd{x\ind{^0}} & \int\frac{\dd{k\ind{_0}}}{2\pi} e^{ ik\ind{_0} x\ind{^0} } = \int \dd{x\ind{_0}} \dirac{}{x\ind{_0}} = 1 \sim 1, \\
        & \sim x\ind{^0} \cdot k\ind{_0} \sim x\ind{^0} \cdot \frac{v}{r}
    \end{split}\\
    \begin{split}
        \int \dd[3]{x} & \int \frac{ \dd[3]{k}}{(2\pi)^3} e^{ i \tvec{k \cdot x} } = \int \dd[3]{x} \dirac{3}{\tvec{x}} = 1 \sim 1 \\
        & \sim \left(x\ind{^i} k\ind{_i}\right)^3 \sim \left(x\ind{^i} \cdot \frac{1}{r} \right)^3
    \end{split} \\
        &\Rightarrow \quad \boxed{\quad x\ind{^0} \sim \frac{r}{v}, \quad x\ind{^i} \sim r. \quad}
	\end{align}
	\end{subequations}
	Notice that Dirac delta functions carries inverse dimension and scaling of its argument (and the exponential function is of course dimensionless)\footnote{Since $e^x = 1 + x + \frac{x^2}{2!} + \dots$, and all the terms in a sum must have the same dimension.}.
	
	For each graviton $H\ind{_{\mu\nu}}$ in a diagram, it scales as
	\begin{subequations}
	\begin{align}
	    \expval{T H\ind{_{\mu\nu}}(x) H\ind{_{\alpha\beta}}(y) } = \Delta_\text{inst}(x-y) & P\ind{_{\mu\nu:\alpha\beta}} = \frac{-\dirac{}{t-t'}}{4\pi r} P\ind{_{\mu\nu:\alpha\beta}} \sim \frac{v}{r} \cdot \frac{1}{r} \sim \frac{v}{r^2}, \\
	    \Rightarrow \quad &\boxed{\quad H\ind{_{\mu\nu}}(x) \sim \frac{\sqrt{v}}{r}.\quad}
	\end{align}
	\end{subequations}
	Note that to leading order in interaction terms $H\ind{_{\mu\nu}}$ couples only to temporal components $\mu=\nu=0$, since $\dot{x}^0 = \gamma \ifthenelse{\boolean{NaturalUnits}}{}{c} \sim \ifthenelse{\boolean{NaturalUnits}}{}{c}$, and $\dot{x}^i = \gamma v^i \sim v$, relegating spatial indices to higher \acrshort{pn} orders compared to the temporal ones.
	
	Since \eqref{eq:SEH:Expansion} and \eqref{eq:Spp:Expansion} expands in powers of $\lambda$, it would be useful to associate a scaling with the coupling constant. This can be achieved using Kepler's third law \eqref{eq:KeplersThirdLaw}, and $v=\omega r$.
	\begin{subequations}
        \begin{align}
            v^2 = \omega^2 r^2 = \frac{GM}{r^3} \hspace{3pt} & r^2 = \frac{GM}{r} = \frac{(\lambda^2 c^4/32\pi)M }{r}, \\
            \lambda^2 & \sim \frac{v^2 r}{m} \sim \frac{ (rmv) \cdot v }{m^2} \\ \Rightarrow \quad &\boxed{ \quad \lambda \sim \frac{\sqrt{Lv}}{m}.\quad}
        \end{align}
    \end{subequations}
	
	The orbital angular momentum scale $L \sim rmv$ has been introduced as a convenient scaling, as will be demonstrated shortly.
	
	It is now a straightforward exercise to assign \acrshort{pn} orders to different diagrams:
	
	\begin{minipage}{0.2\linewidth}
		\begin{center}
			\includegraphics[width=\textwidth]{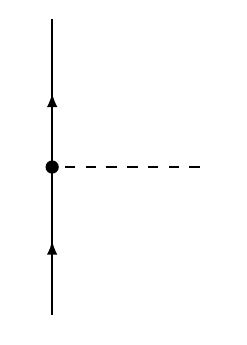}
		\end{center}
	\end{minipage}
	\begin{minipage}{0.75\linewidth}
		\begin{align} \label{eq:Interac1:Scaling}
			\hspace{-2.5cm} \sim \frac{m\lambda}{2} \int \dd{x\ind{^0}} H\ind{_{00}} \dot{x}^{0}_a \dot{x}^0_a \sim \frac{m\sqrt{Lv}}{m} \cdot \frac{r}{v} \cdot \frac{\sqrt{v}}{r} \sim \sqrt{L}.
		\end{align}
    \end{minipage}
    
	Thus the Newtonian diagram, \cref{fig:Feynman:H-diagram:detailed}, scales as \eqref{eq:Interac1:Scaling} squared, making $L$ the scaling of the 0PN order. This is why introducing $L=rmv$ as a scaling is convenient, as it easily makes the leading order scaling apparent. Now, all $1$\acrshort{pn} diagrams should scale as $Lv^2$
	
	The next interaction graph is similarly found to scale
	
	\begin{minipage}{0.2\linewidth}
		\begin{center}
			\includegraphics[width=\textwidth]{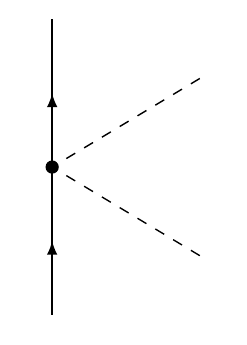}
		\end{center}
	\end{minipage}
	\begin{minipage}{0.75\linewidth}
		\begin{center}
			\begin{align} \label{eq:SecondOrderConection:Powercounting}
			\begin{split}
				\hspace{-2.5cm} \sim \frac{m\lambda^2}{8} \int \dd{x\ind{^0}} \left(H\ind{_{00}}(x) \dot{x}^{0}_a \dot{x}^0_a \right)^2 \sim \frac{(rmv)v}{m} \cdot \frac{r}{v} \cdot \frac{v}{r^2} \sim v^2.
			\end{split}
			\end{align}
		\end{center}
	\end{minipage}
	
	And so it goes on...
	
	\begin{minipage}{0.2\linewidth}
		\begin{center}
			\includegraphics[width=\textwidth]{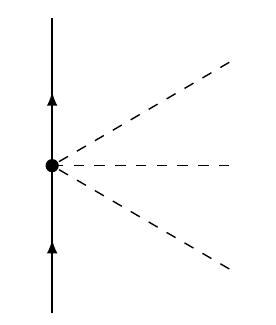}
		\end{center}
	\end{minipage}
	\begin{minipage}{0.75\linewidth}
		\begin{center}
			\begin{align} \label{eq:ThirdOrderConection:Powercounting}
			\begin{split}
				\hspace{-3cm} & \sim \frac{m\lambda^3}{16} \int \dd{x\ind{^0}} \left(H\ind{_{00}}(x) \dot{x}^{0}_a \dot{x}^0_a \right)^3 \sim m \frac{(rmv^2)^{\frac{3}{2}}}{m^3} \cdot \frac{r}{v} \cdot \frac{v^\frac{3}{2}}{r^3} \\
				& \sim v^4/\sqrt{L}.
			\end{split}
			\end{align}
		\end{center}
	\end{minipage}\vspace{0.5cm}
	
	The last type of diagram that needs to be assigned a scaling is the multi graviton vertex propagators. Note the inclusion of a three-dimensional Dirac delta, to make sure momentum ($\tvec{k}_i$ for graviton $i$) is conserved. This is because the gravitons only transfers momentum from one source to another (Newton's third law), and therefore the sum of momentum in and out of this vertex should be zero. Otherwise, the graviton field would spontaneously generate momentum and energy. The two spatial derivatives comes from the definition of $t^{\mu\nu}_{(2)}$, which consist of two derivatives $\partial_j \sim x_j^{-1} \sim k_j$, and two factors of $H$, which is the reason the propagator exists in the first place. Notice that $\partial_j \sim r^{-1}$, while $\partial_0 \sim \frac{v}{r}$, relegating temporal derivatives to higher \acrshort{pn} orders.\vspace{0.5cm}
	
	\begin{minipage}{0.25\linewidth}
		\begin{center}
			\includegraphics[width=\textwidth]{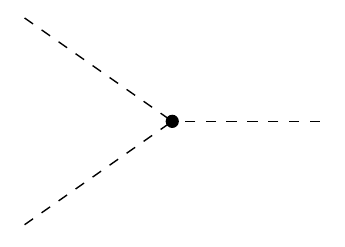}
		\end{center}
	\end{minipage}
	\begin{minipage}{0.7\linewidth}
		\begin{center}
			\begin{align}\label{eq:SecondOrderPropagator:Powercounting}
				\begin{split}
                    & \sim \lambda \int \dd{x^0} \dirac{3}{ \sum_{i=1}^3 \tvec{k}_i} \partial_j^2 \left(H\ind{_{00}}\right)^3 \\
					& \sim \frac{\sqrt{Lv}}{m} \cdot \frac{r}{v} \cdot r^3 \cdot r^{-2} \cdot \left(\frac{\sqrt{v}}{r}\right)^3 = v^2/\sqrt{L}
				\end{split}
			\end{align}
		\end{center}
	\end{minipage}\vspace{0.5cm}
	
	At the next order, the only change is an additional factor of $\lambda$, and $H$. This follows from the expansion of the graviton energy-momentum tensor \eqref{eq:EoM Graviton All Orders}. \vspace{0.5cm}
	
	\begin{minipage}{0.25\linewidth}
		\begin{center}
			\includegraphics[width=\textwidth]{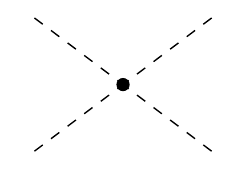}
		\end{center}
	\end{minipage}
	\begin{minipage}{0.7\linewidth}
		\begin{center}
			\begin{align}\label{eq:ThirdOrderPropagator:Powercounting}
				\begin{split}
                    & \sim \lambda^2 \int \dd{x^0} \dirac{3}{ \sum_{i=1}^3 \tvec{k}_i} \partial_j^2 \left(H\ind{_{00}}\right)^4 \\
					& \sim \frac{(rmv)v}{m^2} \cdot \frac{r}{v} \cdot r^3 \cdot r^{-2} \cdot \frac{v^2}{r^4} = v^4/L
				\end{split}
			\end{align}
		\end{center}
	\end{minipage}\vspace{0.5cm}
	
	As already mentioned, spatial indices in the interaction term between the source and graviton leads to additional powers of velocity ($H\ind{_{0i}} \dot{x}^0 \dot{x}^i = H\ind{_{0i}}\gamma c \gamma v^i$), and thus belong to higher \acrshort{pn} orders. These will graphically be represented by a $v^n$ next to the interaction vertex, with $n$ describing the power of $v$ correction the diagram represents.
	
\begin{figure}[h!]
    \begin{center}
        \begin{subfigure}[b]{0.2\textwidth}
            \centering
            \includegraphics[width=\textwidth]{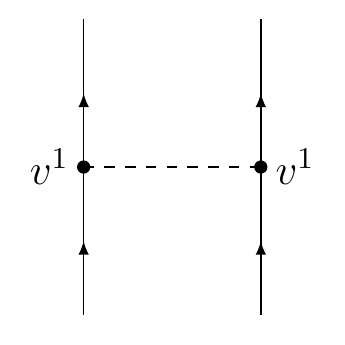}
            \caption{}
            \label{fig:Feynman:H-diagram:v1v1}
        \end{subfigure}
        \begin{subfigure}[b]{0.2\textwidth}
            \centering
            \includegraphics[width=\textwidth]{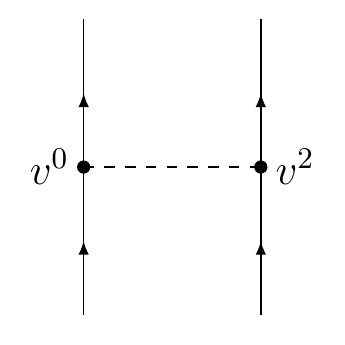}
            \caption{}
            \label{fig:Feynman:H-diagram:v0v2}
        \end{subfigure}
        \begin{subfigure}[b]{0.2\textwidth}
            \centering
            \includegraphics[width=\textwidth]{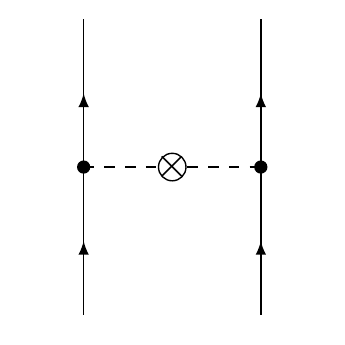}
            \caption{}
            \label{fig:Feynman:H-diagram:ox}
        \end{subfigure}
    \end{center}
    \begin{center}
        \begin{subfigure}[b]{0.2\textwidth}
            \centering
            \includegraphics[width=\textwidth]{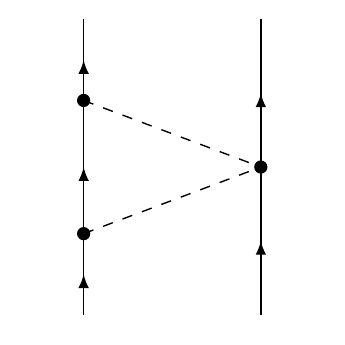}
            \caption{}
            \label{fig:Feynman:V-diagram}
        \end{subfigure}
        \begin{subfigure}[b]{0.2\textwidth}
            \centering
            \includegraphics[width=\textwidth]{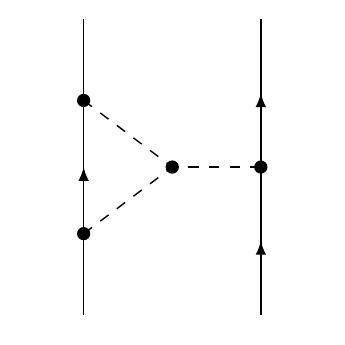}
            \caption{}
            \label{fig:Feynman:Y-diagram}
        \end{subfigure}
    \end{center}
        \caption[The Feynman diagrams contributing to 1PN order orbital energy.]{The Feynman diagrams contributing to 1\acrshort{pn} order orbital energy. The three first diagrams (\subref{fig:Feynman:H-diagram:v1v1})-(\subref{fig:Feynman:H-diagram:ox}) are Newtonian, `H'-type, diagrams with relativistic corrections to the interaction terms and propagator. Diagram (\subref{fig:Feynman:V-diagram}) and (\subref{fig:Feynman:Y-diagram}) represent new types of diagrams, and are of the `V'-type, and `Y'-type respectively.}
        \label{fig:1PN diagrams}
    \end{figure}

	All thinkable diagrams belonging to the 1\acrshort{pn} correction scale as $Lv^2$, and are depicted in \cref{fig:1PN diagrams}. Summing them all up will result with the 1\acrshort{pn} Lagrangian \eqref{eq:EIHLagrangianFull}.
\subsection{Computing Feynman diagram \texorpdfstring{(\subref{fig:Feynman:H-diagram:v1v1})}{(a)}} \label{subsec:diagram a}

    In this first diagram nothing has changed from the Newtonian diagram (see \cref{fig:Feynman:H-diagram:detailed}), except a coupling between the velocity of the two sources\ifthenelse{\boolean{NaturalUnits}}{}{\footnote{The additional factor of $c$ with the effective action is to allow $S= \int \L \dd[3]{x} \dd{t} \hspace{2pt} \to \hspace{2pt} cS=\int\L\dd[4]{x}$.}}
    
	\begin{minipage}{0.2\linewidth}
		\begin{center}
			\includegraphics[width=\textwidth]{figures/FeynmanDiagrams/Hdiagramv1v1.pdf}
		\end{center}
	\end{minipage}
	\begin{minipage}{0.8\linewidth}
		\begin{align}
		\begin{split}\label{eq:diagram:a}
			\ifthenelse{\boolean{NaturalUnits}}{}{c} S^\text{eff}_\text{(\subref{fig:Feynman:H-diagram:v1v1})} = \hspace{3pt} & \frac{ m_1\lambda}{2} \int \dd[4]{x} \dot{x}_1^i \dot{x}_1^0 \dirac{3}{\tvec{x}-x_1(x\ind{^0})} \\ 
			& \cdot \frac{m_2\lambda}{2}\int \dd[4]{y}  \frac{\dirac{}{x\ind{^{0}} - y\ind{^{0}}} P\ind{_{0i:0j}} }{4\pi \abs{\tvec{x}-\tvec{y}}} \dot{x}_2^j \dot{x}_2^0 \dirac{3}{\tvec{y}-x_2(y\ind{^0})} \quad
		\end{split}
		\end{align}
	\end{minipage}
	
	Integrals over Dirac deltas should be straightforward, leaving $x_1^i P\ind{_{0i:0j}} x_2^j$ as the only new and interesting part. Note that $P\ind{_{\mu\nu:\alpha\beta}}$ is symmetric in $\mu \leftrightarrow \nu$, $\alpha \leftrightarrow \beta$, and $\mu\nu \leftrightarrow \alpha\beta$, and therefore all permutations of the indices belong to this same diagram. The last symmetry, $\mu\nu \leftrightarrow \alpha\beta$, is just a relabelling of $m_1 m_2 \leftrightarrow m_2 m_1$, and is superfluous. The other symmetries also result in the same potential, but should be summed over, producing a factor of $2\cdot 2=4$. 
	
	Recalling the definition of the projector $P\ind{_{\mu\nu:\alpha\beta}}$, \eqref{eq:Pmunu:ab:def}\footnote{For the readers convinience: $P\ind{_{\mu\nu:\alpha\beta}} = \frac{1}{2}\left( \eta\ind{_{\mu\alpha}}\eta\ind{_{\nu\beta}} + \eta\ind{_{\mu\beta}}\eta\ind{_{\nu\alpha}} - \eta\ind{_{\mu\nu}} \eta\ind{_{\alpha\beta}} \right)$.}, this is
	\begin{align}
	    v_1^i P\ind{_{0i:0j}} v_2^j = v_1^i \left( \frac{1}{2} \cdot (-1)\cdot \eta\ind{_{ij}} \right) v_2^j = \frac{-1}{2} \tvec{v}_1 \tvec{\cdot v}_2
	\end{align}
    
    Thus the potential of diagram \ref{fig:Feynman:H-diagram:v1v1} is
	\begin{align}
	\begin{split}\label{eq:Potential:a}
		V_{(\subref{fig:Feynman:H-diagram:v1v1})} \hspace{3pt} & = \frac{m_1 m_2 \lambda^2}{4} \left( \frac{1}{4\pi r} \cdot \frac{ 4 \ifthenelse{\boolean{NaturalUnits}}{}{c} \tvec{v}_1 \tvec{\cdot} c\tvec{v}_2}{2}  \right) = \frac{m_1 m_2 c^2 \lambda^2}{32\pi r} \left( 4 \tvec{v}_1 \tvec{\cdot v}_2 \right) \\
		& = 4\frac{Gm_1m_2}{r} \ifthenelse{\boolean{NaturalUnits}}{ \left( \tvec{v}_1 \tvec{\cdot v}_2 \right) }{\frac{\tvec{v}_1 \tvec{\cdot v}_2}{c^2}} = -4V_\text{Newt} \ifthenelse{\boolean{NaturalUnits}}{ \left( \tvec{v}_1 \tvec{\cdot v}_2 \right) }{\frac{\tvec{v}_1 \tvec{\cdot v}_2}{c^2}}.
	\end{split}
	\end{align}
    
    According to the sign, this is seemingly a \emph{repulsive} force, but it is dependent on the dot product of the two velocity vectors, which for any Keplerian orbit will be negative (see \cref{fig:BinarySystem}). Of course, it also scales by an additional factor of $v^2/c^2$ compared to the Newtonian term, as a 1\acrshort{pn} term should.
    
    But for particles moving in the same direction it \emph{is} a repulsive force, making it comparable to a magnetic type force. For oppositely charged particles, which should attract in the static approximation, will also repel each other magnetically when moving in the same direction, quite analogously. The analogy goes even further, as magnetic forces can also be interpreted as relativistic corrections to the electric force \cite{relElectrodynamics}.
\subsection{Computing Feynman diagram \texorpdfstring{(\subref{fig:Feynman:H-diagram:v0v2})}{(b)}} \label{subsec:diagram b}
    This diagram mostly follows suit of \cref{subsec:diagram a}, but with a few additional details. These are
    \begin{itemize}
        \item The relativistic expansion of the free \acrlong{pp} kinetic energy is added here, as it also scales $\sim \tvec{v}_a^2$ compared to 0\acrshort{pn}.
        \item The relativistic expansion of the Lorentz factor of the energy-momentum tensor is expanded, as it also scales $\sim \tvec{v}_a^2$ compared to 0\acrshort{pn}.
        \item And of course, the $P\ind{_{00:ij}}$ and $P\ind{_{ij:00}}$ couplings belong to this diagram.
    \end{itemize}
    
	The first term is just the velocity expansion of the free point particle action 
	\begin{equation}\label{eq:ProperTimeExpansion}
		-m_a \ifthenelse{\boolean{NaturalUnits}}{}{c^2} \int \dd{\tau}_a = -m_a \ifthenelse{\boolean{NaturalUnits}}{}{c^2} \int \dd{x^0}\gamma_a^{-1} = -m_a \ifthenelse{\boolean{NaturalUnits}}{}{c^2} \int \dd{x^0} \left(1-\frac{1}{2} \ifthenelse{\boolean{NaturalUnits}}{\tvec{v}_a^2}{\frac{\tvec{v}_a^2}{c^2}} - \frac{1}{8} \ifthenelse{\boolean{NaturalUnits}}{\tvec{v}_a^4}{\frac{\tvec{v}_a^4}{c^4}} + \dots\right),
	\end{equation}
	which for the 1\acrshort{pn} expansion is $\frac{1}{8}m_a \ifthenelse{\boolean{NaturalUnits}}{\tvec{v}_a^4}{\frac{\tvec{v}_a^4}{c^2}}$, as it has an extra factor of $\ifthenelse{\boolean{NaturalUnits}}{\tvec{v}_a^2}{\frac{\tvec{v}_a^2}{c^2}}$ compared to the 0\acrshort{pn} kinetic term $\frac{1}{2}m_av_a^2$.
	
	Next notice that there is also a Lorentz factor in the energy-momentum tensor, that until now has been approximated to $1$.
	\begin{align}
	\begin{split}\label{eq:LorentzExp}
        \frac{m_a \lambda}{2\ifthenelse{\boolean{NaturalUnits}}{}{c}} \int & \dd[4]{x} \gamma_a^{-1} H\ind{_{00}} \dot{x}_a^0 \dot{x}_a^0 \dirac{3}{\tvec{x}-\tvec{x}_a(t)} = \frac{m_a \lambda}{2\ifthenelse{\boolean{NaturalUnits}}{}{c}} \int \dd[4]{x} \gamma_a H\ind{_{00}} \ifthenelse{\boolean{NaturalUnits}}{}{c^2} \dirac{3}{\tvec{x}-\tvec{x}_a(t)} \\
        & = \frac{m_a \lambda}{2\ifthenelse{\boolean{NaturalUnits}}{}{c}} \int \dd[4]{x} \left( 1 + \frac{1}{2}\ifthenelse{\boolean{NaturalUnits}}{\tvec{v}_a^2}{\frac{\tvec{v}_a^2}{c^2}} + \frac{3}{8}\ifthenelse{\boolean{NaturalUnits}}{\tvec{v}_a^4}{\frac{\tvec{v}_a^4}{c^4}} +\dots \right) H\ind{_{00}}\ifthenelse{\boolean{NaturalUnits}}{}{c^2} \dirac{3}{\tvec{x}-\tvec{x}_a(t)}.
    \end{split}
	\end{align}
	
	Other than the factor of $\frac{1}{2}\ifthenelse{\boolean{NaturalUnits}}{\tvec{v}_a^2}{\frac{\tvec{v}_a^2}{c^2}}$ this is exactly the same as the 0\acrshort{pn} potential.
	
	Lastly, the velocity dependent coupling. Using again \eqref{eq:Pmunu:ab:def}
	\begin{align}\label{eq:ijcoupling}
	    P\ind{_{ij:00}} \dot{x}_a^i \dot{x}_a^j = \frac{1}{2} \tvec{v}_a \tvec{ \cdot v}_a = \frac{1}{2}\tvec{v}_a^2.
	\end{align}
	
	To the first \acrshort{pn} order, only one particle at the time may be expanded this way, therefore the $v^0$-$v^2$ vertices in the diagram. But really both the $v^0$-$v^2$ diagram and the $v^2$-$v^0$ diagram belong to the 1\acrshort{pn} potential, therefore these will both be summed up here, under the same diagram.
	
	Summing up all these contributions to the \ref{fig:Feynman:H-diagram:v0v2} diagram yields 
	
	\begin{minipage}{0.2\linewidth}
		\begin{center}
			\includegraphics[width=\textwidth]{figures/FeynmanDiagrams/Hdiagramv0v2.pdf}
		\end{center}
	\end{minipage}
	\begin{minipage}{0.7\linewidth}
		\begin{center}
		    \begin{subequations} \label{eq:diagram:b}
			\begin{align}
                \label{eq:diagram:b:kinexp} \ifthenelse{\boolean{NaturalUnits}}{}{c} S^\text{eff}_{(\subref{fig:Feynman:H-diagram:v0v2})} = \sum_{a} \Biggl[ \int \dd{x\ind{^0}} \frac{1}{8}m_a \ifthenelse{\boolean{NaturalUnits}}{\tvec{v}_a^4}{\frac{\tvec{v}_a^4}{c^2}}& \\
                \label{eq:diagram:b:vvcoup}+\frac{m_a\lambda}{2} \int \dd{x^0} \Biggl\{ & H\ind{_{ij}}(\tvec{x}_a) \dot{x}_a^i \dot{x}_a^j\\
				\label{eq:diagram:b:v2coup}& + \frac{1}{2} \ifthenelse{\boolean{NaturalUnits}}{\tvec{v}_a^2}{\frac{\tvec{v}_a^2}{c^2}} H\ind{_{00}}(\tvec{x}_a) \dot{x}_a^0 \dot{x}_a^0 \Biggr\} \Biggr].
			\end{align}
			\end{subequations}
		\end{center}
	\end{minipage}
	
	Here \eqref{eq:diagram:b:kinexp}\footnote{It could be argued that this contributon of the action should be an expansion of the free particle diagram (with no propagators), but for streamlining it is included here.} is the result of \eqref{eq:ProperTimeExpansion}, \eqref{eq:diagram:b:v2coup} of \eqref{eq:LorentzExp}, and \eqref{eq:diagram:b:vvcoup} of \eqref{eq:ijcoupling}.
	
	Substituting $H\ind{_{00}}(x\ind{^0}, \tvec{x}_a)$ for the usual term \eqref{eq:EoM of h by Tbar} yields the potential
	\begin{empheq}{align}
		\nonumber V_{(\subref{fig:Feynman:H-diagram:v0v2})} =& -\frac{m_1 m_2\lambda^2}{2\cdot2\cdot4\pi r} \sum_{a} \frac{1}{2}\left[ \ifthenelse{\boolean{NaturalUnits}}{}{c^2\cdot} \tvec{v}_a^2 + \frac{1}{2} \ifthenelse{\boolean{NaturalUnits}}{}{c^4 \cdot} \ifthenelse{\boolean{NaturalUnits}}{\tvec{v}_a^2}{\frac{\tvec{v}_a^2}{c^2}} \right] \\ 
		\nonumber = & -\frac{m_1 m_2 \lambda^2 \ifthenelse{\boolean{NaturalUnits}}{}{c^2}}{32\pi} \frac{3}{2} \left(\tvec{v}_1^2+\tvec{v}_2^2\right)\\
		= & -\frac{G m_1 m_2}{r} \frac{3}{2} \ifthenelse{\boolean{NaturalUnits}}{ \left(\tvec{v}_1^2+\tvec{v}_2^2\right)}{ \frac{\tvec{v}_1^2+\tvec{v}_2^2}{c^2} } = V_\text{Newt} \frac{3}{2} \ifthenelse{\boolean{NaturalUnits}}{ \left(\tvec{v}_1^2+\tvec{v}_2^2\right)}{ \frac{\tvec{v}_1^2+\tvec{v}_2^2}{c^2} }. \label{eq:Potential:b}
	\end{empheq}
	
	This potential is attractive, and proportional to the Newtonian kinetic energy. Thus, this diagram can be thought of as the gravitational attraction from the kinetic energy of one of the particles on the other, showing that in \acrshort{gr} gravity is an attraction of \emph{energies}, and not just masses. Using Einstein's mass-energy equivalence, the `Newtonian kinetic mass' is $E=\frac{1}{2}mv^2=m_\text{kin}c^2$, and inserted into Newton's law of gravity produces $V_{(\subref{fig:Feynman:H-diagram:v0v2})}/3$. This analogy can not explain the missing factor of $3$, because the analogy only account for the \eqref{eq:LorentzExp} part of the potential.
	
	Notice that terms of the type \eqref{eq:LorentzExp} can in principle be expanded to infinite orders, thus there is no reason to stop expanding vertices past the $v^2$ order.
\subsection{But wait, what about 0.5PN diagrams?}
    After having computed diagram \subref{fig:Feynman:H-diagram:v1v1} and \subref{fig:Feynman:H-diagram:v0v2}, a natural idea of a $v^0$-$v^1$ coupling emerges as a 0.5\acrshort{pn} contribution.
    
	\begin{minipage}{0.2\linewidth}
		\begin{center}
			\includegraphics[width=\textwidth]{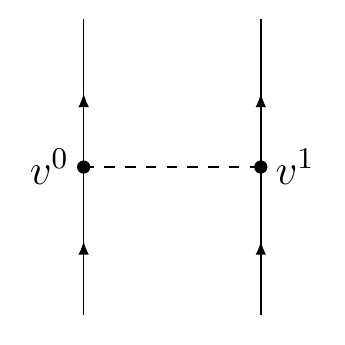}
		\end{center}
	\end{minipage}
	\begin{minipage}{0.8\linewidth}
	    \vspace{-7pt}
	    \begin{align}
		\begin{split}\label{eq:diagram:v0v1}
			\ifthenelse{\boolean{NaturalUnits}}{}{c} S^\text{eff}_\text{0.5\acrshort{pn}} = \hspace{3pt} & \sum_{a \neq b} \frac{ m_a\lambda}{2} \int \dd[4]{x} \dot{x}_a^0 \dot{x}_a^0 \dirac{3}{\tvec{x}-x_a(x\ind{^0})} \\ 
			& \cdot \frac{m_b\lambda}{2}\int \dd[4]{y}  \frac{\dirac{}{x\ind{^{0}} - y\ind{^{0}}} P\ind{_{00:0i}} }{4\pi \abs{\tvec{x}-\tvec{y}}} \dot{x}_b^i \dot{x}_b^0 \dirac{3}{\tvec{y}-x_b(y\ind{^0})} \quad
		\end{split}
		\end{align}
	\end{minipage}
	
	The situation is analogous to \cref{subsec:diagram a}, only here the connection is not symmetric.
	Thus, the potential is also here the Newtonian potential multiplied by some velocity factor.
	\begin{align}
		V_{0.5\text{\acrshort{pn}}} = \frac{m_1 m_2\lambda^2}{16\pi r} \sum_{a\neq b} \left[ \dot{x}_a^0 \dot{x}_a^0 \left( P\ind{_{00:i0}} + P\ind{_{00:0i}} \right) \dot{x}_b^i \dot{x}_b^0 \right] = 0,
	\end{align}
	which follows from $P\ind{_{00:i0}}=\frac{1}{2}\left( 2\eta\ind{_{00}}\eta\ind{_{0i}} - \eta\ind{_{00}}\eta\ind{_{i0}} \right)=P\ind{_{00:0i}} = 0$.
	
	A lesson to be taken from this is that $P\ind{_{\mu\nu:\alpha\beta}}$ only couples sources which both have an even or both an odd number of temporal indices. E.g. 00:00, 00:$ij$, and 0$i$:0$j$. Terms like 0$i$:00, or 0$i$:$ij$ will turn out to be zero. Therefore, this kind of interaction expansion can only produce even powers of $\ifthenelse{\boolean{NaturalUnits}}{v}{v/c}$.
\subsection{Computing Feynman diagram \texorpdfstring{(\subref{fig:Feynman:H-diagram:ox})}{(c)}} \label{subsec:diagram c}
	The last `H'-shaped diagram is not expanded in its vortexes, but rather the propagator of $H$ is expanded to second order in $v$, according to expansion \eqref{eq:Propagator:Expansion}. In that expansion, the propagator was only expanded in Fourier space, so it remains to determine the value of the first order correction of the propagator in real space.
	\begin{align} \label{eq:Prop:2:Fourier}
		\Delta_\text{inst}^{(2)}(x-x') \equiv \int \frac{\dd[4]{k}}{(2\pi)^4} \frac{ - k_0^2}{\tvec{k}^4} e^{ik_\nu(x^\nu-x'^\nu)}.
	\end{align}
	From the gradient relations \eqref{eq:GradiantRelation:PotentialMode} it is evident that $(k^0/\abs{\tvec{k}})^2 \sim \left(\frac{v}{r} / \frac{1}{r}\right)^2 = v^2$ and thus belong to the 1\acrshort{pn} correction, but the scaling of $k\ind{_0} \sim \frac{v}{r}$ was not truly argued for. So this section will demonstrate this scaling by computing the propagator in real space.
	\begin{minipage}{0.2\linewidth}
		\begin{center}
			\includegraphics[width=\textwidth]{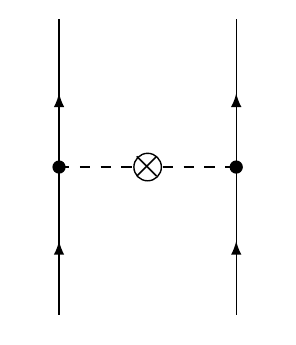}
		\end{center}
	\end{minipage}
	\begin{minipage}{0.8\linewidth}
	    \vspace{-7pt}
	    \begin{subequations}
		\begin{align}
			\begin{split}\label{eq:PropagatorActionExp}
			    \ifthenelse{\boolean{NaturalUnits}}{}{c}S^\text{eff}_{(\subref{fig:Feynman:H-diagram:ox})} = \frac{m_1 m_2\lambda^2 \ifthenelse{\boolean{NaturalUnits}}{}{c^4}}{8} & \int \dd[4]{x} \dd[4]{y} \Big\{ \dirac{3}{\tvec{x}-\tvec{x}_1(t)} \\
				& \cdot \hspace{1pt} \dirac{3}{\tvec{y}-\tvec{x}_2(t')} \Delta_\text{inst}^{(2)}(x-y) \Bigr\}
			\end{split} \\
				= \frac{m_1m_2\lambda^2 \ifthenelse{\boolean{NaturalUnits}}{}{c^4}}{8} & \int \dd{\ifthenelse{\boolean{NaturalUnits}}{}{c}t} \dd{\ifthenelse{\boolean{NaturalUnits}}{}{c}t'} \Delta_\text{inst}^{(2)}\left(t-t', \tvec{x}_1(t) - \tvec{x}_2(t')\right)
		\end{align}
		\end{subequations}
	\end{minipage}
	
	To evaluate the integral in \eqref{eq:Prop:2:Fourier} some tricks are in order. Writing out the exponent as $e^{ik_\mu(x^\mu-x'^\mu)} = e^{-ik_0(\ifthenelse{\boolean{NaturalUnits}}{}{c}t-\ifthenelse{\boolean{NaturalUnits}}{}{c}t')} e^{i\tvec{k \cdot}(\tvec{x}-\tvec{y})}$, a useful way to proceed is to rewrite the integral as $k_0^2e^{-ik_0(\ifthenelse{\boolean{NaturalUnits}}{}{c}t-\ifthenelse{\boolean{NaturalUnits}}{}{c}t')} = - \pdv{  }{\ifthenelse{\boolean{NaturalUnits}}{}{c}t}{\ifthenelse{\boolean{NaturalUnits}}{}{c}t'} e^{-ik_0(\ifthenelse{\boolean{NaturalUnits}}{}{c}t-\ifthenelse{\boolean{NaturalUnits}}{}{c}t')} $. In the following note that the time derivative is \emph{only} operating on the $k_0$ integral.
	\begin{subequations}\label{eq:PropagatorActionExp:PartialStepBefore}
		\begin{align}
			\int \frac{\dd[4]{k}}{(2\pi)^4} \frac{(k_0)^2}{\tvec{k}^4} e^{ik_\mu(x^\mu-x'^\mu)} & = - \pdv{  }{\ifthenelse{\boolean{NaturalUnits}}{}{c}t}{\ifthenelse{\boolean{NaturalUnits}}{}{c}t'} \left[ \int \frac{\dd{k_0}}{2\pi} e^{-ik_0(\ifthenelse{\boolean{NaturalUnits}}{}{c}t-\ifthenelse{\boolean{NaturalUnits}}{}{c}t')} \right] \int \frac{\dd[3]{k}}{(2\pi)^3} \frac{1}{\tvec{k}^4} e^{i\tvec{k \cdot}(\tvec{x}-\tvec{x'})} \\
			&= -\pdv{  }{\ifthenelse{\boolean{NaturalUnits}}{}{c}t}{\ifthenelse{\boolean{NaturalUnits}}{}{c}t'} \left[ \dirac{}{\ifthenelse{\boolean{NaturalUnits}}{}{c}t-\ifthenelse{\boolean{NaturalUnits}}{}{c}t'} \right] \int \frac{\dd[3]{k}}{(2\pi)^3} \frac{1}{\tvec{k}^4} e^{i\tvec{k \cdot}(\tvec{x}-\tvec{x'})} \label{eq:PropagatorActionExp:PartialStepBeforeDerivativeOfDelta}
		\end{align}
	\end{subequations}
	
	To proceed the derivative of Dirac's delta function must be determined.
	To that end, partial integration is the key.
	\begin{subequations}
		\begin{align}
			\int_{-\infty}^{\infty}& f(x) \dv{\delta(x-x_0)}{x} \dd{x} = \int_{-\infty}^{\infty} \dv{x} \left[ f(x) \delta(x-x_0) \right] \dd{x} - \int_{-\infty}^{\infty} \delta(x-x_0) \dv{f(x)}{x} \dd{x} \\
			& = \eval{ f(x) \delta(x-x_0)}_{-\infty}^\infty - \int \delta(x-x_0) f'(x)\dd{x} = - \int \delta(x-x_0) f'(x)\dd{x},
		\end{align}
	\end{subequations}
	the last equality only holding for $x_0 \notin \pm\infty$. The result generalizes for any number of derivatives as
	\begin{align} \label{eq:DiracDelta:derivative:n}
		\int_{-\infty}^{\infty} f(x) \dv[n]{\delta(x-x_0)}{x} \dd{x} = \int_{-\infty}^\infty (-1)^n \delta(x-x_0) \dv[n]{x} f(x) \dd{x}.
	\end{align}
	
	Back to equation \eqref{eq:PropagatorActionExp:PartialStepBeforeDerivativeOfDelta}, the situation is analogous, thus partial integrations can also be utilized here to rewrite the derivative of the Dirac delta function in the same manner.
	
	\begin{subequations}
		\begin{align}
			\ifthenelse{\boolean{NaturalUnits}}{}{c}S^\text{eff}_{(\subref{fig:Feynman:H-diagram:ox})} & = - \frac{m_1m_2\lambda^2 \ifthenelse{\boolean{NaturalUnits}}{}{c^4}}{8} \int \dd{\ifthenelse{\boolean{NaturalUnits}}{}{c}t} \dd{\ifthenelse{\boolean{NaturalUnits}}{}{c}t'} \dirac{}{\ifthenelse{\boolean{NaturalUnits}}{}{c}t-\ifthenelse{\boolean{NaturalUnits}}{}{c}t'} \pdv{  }{\ifthenelse{\boolean{NaturalUnits}}{}{c}t}{\ifthenelse{\boolean{NaturalUnits}}{}{c}t'} \int \frac{\dd[3]{k}}{(2\pi)^3} \frac{1}{\tvec{k}^4} e^{i\tvec{k \cdot}\left(\tvec{x}_1(t)-\tvec{x}_2(t')\right)} \\
			& = 4\pi G m_1m_2 \int \dd{\ifthenelse{\boolean{NaturalUnits}}{}{c}t} \dd{\ifthenelse{\boolean{NaturalUnits}}{}{c}t'} \dirac{}{\ifthenelse{\boolean{NaturalUnits}}{}{c}t-\ifthenelse{\boolean{NaturalUnits}}{}{c}t'} \ifthenelse{\boolean{NaturalUnits}}{v_1^i v_2^j}{\frac{v_1^i}{c} \frac{v_2^j}{c}} \int \frac{\dd[3]{k}}{(2\pi)^3} \frac{k\ind{_i}k\ind{_j}}{\tvec{k}^4} e^{i\tvec{k \cdot}\left(\tvec{x}_1(t)-\tvec{x}_2(t')\right)} \\
			& = 4\pi G m_1m_2 \int \dd{\ifthenelse{\boolean{NaturalUnits}}{}{c}t} \ifthenelse{\boolean{NaturalUnits}}{v_1^i v_2^j}{\frac{v_1^i}{c} \frac{v_2^j}{c}} \int \frac{\dd[3]{k}}{(2\pi)^3} \frac{k\ind{_i}k\ind{_j}}{\tvec{k}^4} e^{i\tvec{k \cdot r}(t)} \\
			& = -4\pi G m_1m_2 \int \dd{\ifthenelse{\boolean{NaturalUnits}}{}{c}t} \ifthenelse{\boolean{NaturalUnits}}{v_1^i v_2^j}{\frac{v_1^i}{c} \frac{v_2^j}{c}} \pdv{}{r\ind{^i}}{r\ind{^{j}}} \int \frac{\dd[3]{k}}{(2\pi)^3} \frac{1}{\tvec{k}^4} e^{i\tvec{k \cdot r}(t)}. \label{eq:PropagatorExp:AfterDelta}
		\end{align}
	\end{subequations}
	
	This final integral might not look like that much of an improvement, but these are the kinds of integrals often encountered in \acrshort{qft}, and there exists a known solution \cite{Porto2016}.
	\begin{align}
		\int \frac{\dd[d]{k}}{(2\pi)^d} \frac{1}{\tvec{k}^{2\alpha}} e^{i\tvec{k \cdot r}} = \frac{1}{(4\pi)^{d/2}} \frac{\Gamma\left(\frac{d}{2}-\alpha\right)}{\Gamma(\alpha)} \left(\frac{\tvec{r}^2}{4}\right)^{\alpha-d/2}.
	\end{align}
	
	Here $\Gamma(z)$ is the gamma-function, and $d$ is the spatial dimension. In the integral of equation \eqref{eq:PropagatorExp:AfterDelta}, $d=3$ and $\alpha=2$. Recalling that $\Gamma(1)=1$, $\Gamma\left(\frac{1}{2}\right)=\sqrt{\pi}$, and $z\Gamma(z)=\Gamma(z+1)$, it follows that $\Gamma(2)=1\cdot \Gamma(1)=1=\Gamma(\alpha)$. In the same fashion $-\frac{1}{2}\Gamma\left(-\frac{1}{2}\right) = \Gamma\left(\frac{1}{2}\right)=\sqrt{\pi}$ $\rightarrow \Gamma\left(\frac{d}{2}-\alpha\right) = \Gamma\left(-\frac{1}{2}\right) = -2\sqrt{\pi}$.
	
	Then, the final expression of the potential follows as
	\begin{subequations}
		\begin{align}
			V_{(\subref{fig:Feynman:H-diagram:ox})} & = 4\pi G m_1m_2 \ifthenelse{\boolean{NaturalUnits}}{v_1^i v_2^j}{\frac{v_1^i}{c} \frac{v_2^j}{c}} \pdv{}{r\ind{^i}}{r\ind{^{j}}} \left[ \frac{1}{8 \pi^{3/2}}\frac{-2\pi^{1/2}}{1}\frac{\abs{\tvec{r}}}{2} \right] \\
			& = -\frac{G m_1m_2}{2} \ifthenelse{\boolean{NaturalUnits}}{v_1^i v_2^j}{\frac{v_1^i}{c} \frac{v_2^j}{c}} \frac{1}{r^3} \left( r^2 \delta_{ij} - r_i r_j \right) \label{eq:potential:projector} \\
			& = -\frac{G m_1m_2}{2r} \left( \ifthenelse{\boolean{NaturalUnits}}{ \tvec{v}_1\tvec{ \cdot v}_2 - \frac{\left(\tvec{v}_1 \tvec{\cdot r}\right)\left(\tvec{v}_2 \tvec{\cdot r}\right)}{r^2} }{ \frac{\tvec{v}_1\tvec{ \cdot v}_2}{c^2} - \frac{\left(\tvec{v}_1 \tvec{\cdot r}\right)\left(\tvec{v}_2 \tvec{\cdot r}\right)}{c^2 r^2} } \right) \\
			& = V_\text{Newt} \frac{1}{2\ifthenelse{\boolean{NaturalUnits}}{}{c^2}}\left( \tvec{v}_1\tvec{ \cdot v}_2 - \frac{\left(\tvec{v}_1 \tvec{\cdot r}\right)\left(\tvec{v}_2 \tvec{\cdot r}\right)}{r^2}\right). \label{eq:Potential:c}
		\end{align}
	\end{subequations}
	
	Again, this is obviously of first \acrlong{pn} order, as it has an additional factor of \ifthenelse{\boolean{NaturalUnits}}{$v^2$}{$\frac{v^2}{c^2}$} compared to the Newtonian potential. From the procedure it is hopefully clear how additional powers of $(k\ind{_0}/\tvec{k})^2$ leads to additional time-derivatives\ifthenelse{\boolean{NaturalUnits}}{}{, additional powers of $c^{-2}$}, and additional powers of $v^2$. Thereby the scaling $k\ind{_0}\sim \frac{v}{r}$ should be justified.
	
	Like diagram \ref{fig:Feynman:H-diagram:v1v1}, this potential is also dependent on $\tvec{v}_1\tvec{ \cdot v}_2$, but has the opposite sign. Thus, it is attractive for particles moving in the same direction, and repulsive for particles moving in opposite directions.
	
	Some insight might be gained from the \emph{projector} in line \eqref{eq:potential:projector}
	\begin{align}
	    P\ind{_{ij}}(\tvec{n}) \equiv \delta\ind{_{ij}} - n\ind{_i}n\ind{_j},
    \end{align}
    where $\tvec{n}=\tvec{r}/r$. Contracting $P\ind{_{ij}}(\tvec{n})$ with a vector $\tvec{x}$ has the effect of projecting $\tvec{x}$ onto the orthogonal plane of $\tvec{n}$
    \begin{align*}
        P\ind{_{ij}}(\tvec{n}) x\ind{^j} & = x\ind{_i} - \left( \tvec{ n \cdot x} \right)n\ind{_i}, \\
        \Rightarrow \quad n\ind{^i} P\ind{_{ij}}(\tvec{n}) x\ind{^j} & = \left( \tvec{ n \cdot x} \right) - \left( \tvec{ n \cdot x} \right) = 0.
    \end{align*}
    Ergo, $P\ind{_{ij}}(\tvec{n}) x\ind{^j}$ is orthogonal to $\tvec{n}$.
    
    With this in mind, the potential only couples velocities orthogonal to $\tvec{r}$, and is equivalent to
    \begin{align}
    \begin{split}
        V_{(\subref{fig:Feynman:H-diagram:ox})} & = V_\text{Newt} \frac{(\tvec{r \cp v}_1) \tvec{\cdot} (\tvec{r \cp v}_2) }{2r^2 \ifthenelse{\boolean{NaturalUnits}}{}{c^2}} = \frac{G}{2r^3 \ifthenelse{\boolean{NaturalUnits}}{}{c^2}} (\tvec{r \cp p}) \tvec{\cdot} (\tvec{r \cp p}) \\
        & = \frac{G\mu}{r \ifthenelse{\boolean{NaturalUnits}}{}{c^2}} \left( \frac{1}{2} \mu r^2 \omega^2 \right),
    \end{split}
    \end{align}
    where $\tvec{v}_i$ was exchanged for $\tvec{v}$ according to \eqref{eq:relVelocity:exchange}, and $\tvec{p} \equiv \mu \tvec{v}$. In the last line, $\tvec{\omega} = \tvec{r \cp v}/r^2$ was used, and the term inside the parenthesis is the kinetic energy associated with rotational motion of a particle with effective mass $\mu$.
    
    A last observation is that the Lambda tensor, introduced in \cref{sec:Energy-Momentum tensor of GW} \eqref{eq:Lambdatensor:def}, can be defined using this projection operator (\textcite{Maggiore:VolumeI})
    \begin{align}
        \Lambda\ind{_{ij:kl}}(\tvec{n}) \equiv P\ind{_{ik}}P\ind{_{jl}} - \frac{1}{2}P\ind{_{ij}}P\ind{_{kl}}.
    \end{align}
\subsection{Computing Feynman diagram \texorpdfstring{(\subref{fig:Feynman:V-diagram})}{(d)}} \label{subsec:diagram d}
	The last two diagrams have a higher order of $H$'s and are thus \emph{non-linear} in $\lambda$. Therefore, these diagrams should be proportional to $G^2$. Diagram \ref{fig:Feynman:V-diagram} represent a second order in $H$ coupling between the graviton field and the \acrlong{pp}, while the last diagram represents the higher order propagator, where $H$ couples with itself. It was argued in equation \eqref{eq:SecondOrderConection:Powercounting} and \eqref{eq:SecondOrderPropagator:Powercounting} that these diagrams belong to the 1\acrshort{pn} correction, but then only coupling to the $00$ component of $T\ind{^{\mu\nu}}$. Thus, the same expansions preformed for diagram (\subref{fig:Feynman:H-diagram:v1v1})-(\subref{fig:Feynman:H-diagram:ox}) will need to be implemented to these non-linear diagrams, when computing higher order \acrshort{pn} corrections.
	
	\begin{minipage}{0.2\linewidth}
		\begin{center}
			\includegraphics[width=\textwidth]{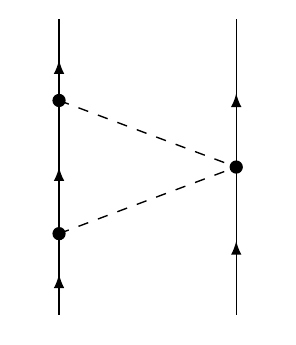}
		\end{center}
	\end{minipage}
	\begin{minipage}{0.8\linewidth}
		\begin{align}
			\begin{split}
				\ifthenelse{\boolean{NaturalUnits}}{}{c}S^\text{eff}_{(\subref{fig:Feynman:V-diagram})} = & \sum_{a\neq b} \frac{m_a^2\lambda^2}{4} \frac{m_b \lambda^2}{8} \int \dd{x^0} \dd{\tilde{x}^0} \dd{y^0} \ifthenelse{\boolean{NaturalUnits}}{}{c^6} P^2_{00:00} \\
				& \cdot \Delta_\text{inst}\left(\tvec{x}_a(x^0) - \tvec{x}_b(y^0) \right) \cdot \Delta_\text{inst}\left(\tvec{x}_a(\tilde{x}^0) - \tvec{x}_b(y^0) \right)
			\end{split}
		\end{align}
	\end{minipage}
	
	Notice that particle $a$ interacts with the graviton field \emph{twice}, and possibly at different times, while particle $b$ only interacts once, but to a higher order, thus connecting it to the other particle through two graviton propagators, as depicted in the diagram. The sum adds the mirrored diagram as well.
	
	There is nothing surprising in this diagram, and the integrals can be carried out without any fuzz, eliminating two of the three time integrals, making both graviton exchanges instantaneous, and simultaneous.
	\begin{subequations}\label{eq:Potential:d}
    \begin{align}
        V_{(\subref{fig:Feynman:V-diagram})} & = -\sum_{a\neq b} \frac{m_a^2 m_b \lambda^4 \ifthenelse{\boolean{NaturalUnits}}{}{c^6}}{4\cdot 8 \cdot 4} \frac{1}{4\pi r}\frac{1}{4\pi r} = -\sum_{a\neq b} \frac{G^2 m_a^2 m_b}{2r^2\ifthenelse{\boolean{NaturalUnits}}{}{c^2}} \\
        & = -\frac{G^2m_1m_2(m_1+m_2)}{2r^2\ifthenelse{\boolean{NaturalUnits}}{}{c^2}} = V_\text{Newt} \frac{GM}{2r\ifthenelse{\boolean{NaturalUnits}}{}{c^2}}
	\end{align}
	\end{subequations}
	
	One way to interpret this result is as a gravitational coupling between the total mass of the system, and half the Newtonian potential energy itself
	\begin{align}
	    V_{(\subref{fig:Feynman:V-diagram})} = -\frac{GM (\abs{V_\text{Newt}}/2\ifthenelse{\boolean{NaturalUnits}}{}{c^2})}{r}.
	\end{align}
	This does not however make much sense as being separated the distance $r$, but it is an interesting analogy.
	
\subsection{Computing Feynman diagram \texorpdfstring{(\subref{fig:Feynman:Y-diagram})}{(e)} } \label{subsec:diagram e}
    Note that in this section spacial indices are suppressed, as subscripts are used to enumerate vector variables.
    
	The last diagram makes use of the three graviton propagator, and is the first one not to use the linear propagator.
	
	\begin{minipage}{0.2\linewidth}
		\begin{center}
			\includegraphics[width=\textwidth]{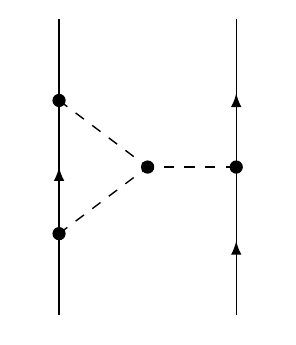}
		\end{center}
	\end{minipage}
	\begin{minipage}{0.8\linewidth}
	    \vspace{-10pt}
		\begin{align}
			\begin{split}
				\ifthenelse{\boolean{NaturalUnits}}{}{c} S_{(\subref{fig:Feynman:Y-diagram})} = \frac{m_1^2 m_2 \lambda^3}{8} \int \dd[4]{x} \dd[4]{\tilde{x}} \dd[4]{y} \left( \dirac{3}{ \tvec{x} - \tvec{x}_1(t_1) } +\dots \right) \ifthenelse{\boolean{NaturalUnits}}{}{c^6} \\
				\int  \expval{T\left\{ H_{00}(t_1,\tvec{k}_1) H_{00}(t_2,\tvec{k}_2) H_{00}(t_3,\tvec{k}_3) \right\}} \prod_{j=1}^{3} \frac{\dd[3]{k_j}}{(2\pi)^3} \\
				+ \qq*{ the mirrored diagram.}
			\end{split}
		\end{align}
	\end{minipage}
	Here the propagator in the second line needs a lot of work. It can be read off from \textcite{Goldberger:EFT}, equation (37)-(39). But it will also be derived in the following.
	
	Much of the structure is already given in \eqref{eq:ThreePointProp:first}, which was found inserting leading order results\footnote{Leading order here referes to the \acrshort{pn} expansion, not the $\lambda$ expansion of the \acrshort{eom} \eqref{eq:EoM Graviton All Orders}.} for $h$ in the expanded \acrshort{eom}.
	\begin{align}
	    H\ind{_{\mu\nu}}(y) = -\frac{\lambda}{2} P\ind{_{\mu\nu:\alpha\beta}} \int \frac{\dd[4]{k_1}\dd[4]{k_2}\dd[4]{k_3}}{(2\pi)^{12}} \frac{-e^{ik_1 y}}{\tvec{k}_1^2} P\ind{_{\rho\sigma:\tau\delta}} k_1^\alpha H\ind{_{\tvec{k}_2 \hspace{1pt} \sigma\rho}} k_1^\beta H\ind{_{\tvec{k}_3 \hspace{1pt}\tau\delta}} e^{ik_2 x} e^{ik_3 \tilde{x}}
	\end{align}
	
	To make sure momentum and energy is not spontaneously generated by the propagator, one should demand the momentum vectors $\tvec{k}_i$ sum to zero. This is achieved by multiplying the integral by $(2\pi)^4\dirac{4}{\sum_i \tvec{k}_i}$.
	
	\begin{align}
	\begin{split}
        H\ind{_{00}}(y) & = -\frac{\lambda}{4} P\ind{_{00:\alpha\beta}} \int \left[ \prod_{j=1}^3 \frac{\dd[4]{k_j}}{(2\pi)^{4}} \right] \frac{-e^{ik_1 y}}{\tvec{k}_1^2}  \frac{-e^{ik_2 x}}{\tvec{k}_2^2}  \frac{-e^{ik_3 \tilde{x}}}{\tvec{k}_3^2} k_1^\alpha  k_1^\beta (2\pi)^4\dirac{4}{\sum_i k_i} \\
        & = -\frac{\lambda}{4} \int \left[ \prod_{j=1}^3 \frac{\dd[4]{k_j}}{(2\pi)^{4}} \right] \frac{-e^{ik_1 y}}{\tvec{k}_1^2}  \frac{-e^{ik_2 x}}{\tvec{k}_2^2}  \frac{-e^{ik_3 \tilde{x}}}{\tvec{k}_3^2} \frac{(k_1^0)^2+\tvec{k}_1^2}{2} (2\pi)^4\dirac{4}{\sum_i k_i}.
    \end{split}
	\end{align}
	
	The factor of $(k_1^0)^2+\tvec{k}_1^2 = \tvec{k}^2_1 \left( 1 + (k^0_1)^2/\tvec{k}_1^2 \right)$, which was found to induce factors of $v^2$ past leading order in \cref{subsec:diagram c}. Therefore, this factor will be approximated as $\tvec{k}_1^2$.
	
	Furthermore, when one of the $k_i$ integrals are preformed; the Dirac delta function will eliminate that vector by $k_1 + k_2 + k_3 = 0$. For example preforming the $k_1$ integral yields
	\begin{subequations}
    \begin{align}
        H\ind{_{00}}(y) = \hspace{3pt}& \frac{\lambda}{8} \int \left[ \prod_{j=2}^3 \frac{\dd[4]{k_j}}{(2\pi)^{4}} \right] e^{-i(k_2+k_3) y}  \frac{e^{ik_2 x}}{\tvec{k}_2^2}  \frac{e^{ik_3 \tilde{x}}}{\tvec{k}_3^2} \\
        = \frac{\lambda}{8} \int \frac{ \dd{k_2^0} \dd{k_3^0} }{(2\pi)^2} e^{-ik_2^0 (x^0-y^0)} &e^{-ik_3^0 (\tilde{x}^0-y^0) } \int \frac{ \dd[3]{k_2} }{(2\pi)^3} \frac{e^{i\tvec{k}_2 \tvec{ \cdot }(\tvec{x}_2 - \tvec{y})}}{\tvec{k}_2^2} \int \frac{ \dd[3]{k_3} }{(2\pi)^3} \frac{e^{i\tvec{k}_3 \tvec{ \cdot }(\tvec{\tilde{x}}_2 - \tvec{y})}}{\tvec{k}_3^2} \\
        & = \frac{\lambda}{8} \frac{\dirac{}{x^0-y^0}}{4\pi \abs{\tvec{x}_2-\tvec{y}}} \frac{\dirac{}{\tilde{x}^0-y^0}}{4\pi \abs{\tvec{x}_3-\tvec{y}}} \\
        = \frac{\lambda}{2} P\ind{_{00:00}} & \Delta_\text{inst}(x_2-y) \cdot P\ind{_{00:00}}\Delta_\text{inst}(x_3-y)
    \end{align}
	\end{subequations}
	
	It is the product of two propagators, connecting two different points to the same third point, just like in the diagram. This result could also have been argued to result from \emph{Wick's theorem}, like  \textcite{Porto2016} does, but then it would be all pairwise combinations of points, including $\Delta_\text{inst}(x_2-x_3)$. It was discarder here because the $k_1$ propagator was eliminated by the derivative, which again followed from $k_1$ being the main transform, i.e. the lone graviton going in the final interaction term of the action. In \cite{Porto2016} this `missing' contribution was handled using \emph{dimensional regularization}, and turns out to be zero, as demanded by our result.
	
	The potential should now follow straightforwardly as
    \begin{align}
        \begin{split} \label{eq:Potential:e}
		V_{(\subref{fig:Feynman:Y-diagram})} & = \sum_{a\neq b} \frac{m_a^2 m_b \lambda^4 \ifthenelse{\boolean{NaturalUnits}}{}{c^6}}{2^{10} \pi^2 r^2} \\
		& = \frac{G^2 m_1m_2(m_1+m_2)}{r^2\ifthenelse{\boolean{NaturalUnits}}{}{c^2}} = -V_\text{Newt} \frac{GM}{r\ifthenelse{\boolean{NaturalUnits}}{}{c^2}} \\
		& = -2V_{(\subref{fig:Feynman:V-diagram})}.
	\end{split}
	\end{align}
	
	Surprisingly, this diagram has the same potential as diagram \ref{fig:Feynman:V-diagram}, times negative $2$. Thus, the joined effect of these two last diagrams is a positive, and thus repulsive, potential proportional to the total mass and the Newtonian potential.
	
	It is surprising that the effect of non-linear terms is to weaken the static force, but this is the case.
	
\subsection{The total 1PN Lagrangian}
    Summing up all the potentials $V_{(\subref{fig:Feynman:H-diagram:v1v1})}$-$V_{(\subref{fig:Feynman:Y-diagram})}$ (\eqref{eq:Potential:a}, \eqref{eq:Potential:b}, \eqref{eq:Potential:c}, \eqref{eq:Potential:d}, and \eqref{eq:Potential:e}), and remembering to add the kinetic energy expansion \eqref{eq:diagram:b:kinexp}, the final 1\acrshort{pn} Lagrangian will be the result
	\begin{empheq}[box=\boxed]{align}
		\begin{split}\label{eq:EIHLagrangianFull}
			\hspace{3pt} L_{1\text{\acrshort{pn}}} = L_\text{\acrshort{EIH}} = &\frac{1}{8} \sum_{a} m_a \ifthenelse{\boolean{NaturalUnits}}{v_a^4}{\frac{v_a^4}{c^2}} + \frac{G m_1 m_2}{2r\ifthenelse{\boolean{NaturalUnits}}{}{c^2}} \left[ 3(\tvec{v}_1^2 + \tvec{v}_2^2) - 7 \tvec{v}_1 \tvec{\cdot v}_2 - \frac{(\tvec{v}_1 \tvec{\cdot r})(\tvec{v}_2 \tvec{\cdot r})}{r^2\ifthenelse{\boolean{NaturalUnits}}{}{c^2}} \right] \hspace{3pt} \\
			&- \frac{G^2 m_1 m_2 (m_1 + m_2)}{2r^2\ifthenelse{\boolean{NaturalUnits}}{}{c^2}},
		\end{split}
	\end{empheq}
	\begin{align}
		\qq*{and with total effective action} S^\text{eff} = \int \dd{t} \left\{ L_{\text{0\acrshort{pn}}} + L_{\text{1\acrshort{pn}}} + \dots \right\}
	\end{align}
	
	This is the \textit{\acrlong{EIH}} Lagrangian from 1938 \cite{EinsteinInfeldHoffmann}, derived in an entirely different manner, warranting some confidence in the result.
	
	In order to determine the orbital energy of the binary system it is useful to reduce this Lagrangian to its equivalent one body problem.
	
	As it is already \emph{just} spatially dependent on the relative displacement of the two bodies $r$, the last thing needed is just to express their velocities by the relative velocity.
	
	Using the centre of mass frame the position of each body can be expressed through the relative displacement $\tvec{r}$ as\footnote{For details on how to derive these relations, and how to do the soon to come mass-term manipulation, see \cref{app:equivOneBodyAndMassTerms}.}
		\begin{align}
			\tvec{r} = \tvec{r}_1 - \tvec{r}_2, \quad \tvec{r}_1 = \frac{m_2}{M} \tvec{r}, \quad \tvec{r}_2 = -\frac{m_1}{M}\tvec{r}.
		\end{align}
	
	The velocity of particle number $i$ is defined as the time derivative of its position $\tvec{v}_i \equiv \dot{\tvec{r}}_i$. Defining the time derivative of the relative displacement as the relative velocity one finds
	\begin{align} \label{eq:relationv1v2v}
		\tvec{v} \equiv \dot{\tvec{r}}, \quad  \tvec{v}_1 = \frac{m_2}{M}\tvec{v}, \quad \tvec{v}_2 = -\frac{m_1}{M} \tvec{v}.
	\end{align}
	
	Substituting \eqref{eq:relationv1v2v} into the 1\acrshort{pn} \eqref{eq:EIHLagrangianFull}, and the 0\acrshort{pn} \eqref{eq:NewtonianAction} Lagrangian, the equivalent, total, one-body version is obtained
	
	\begin{subequations}
		\begin{align}
			\begin{split}
				L^\text{eff} = \hspace{3pt} & \frac{1}{2}\left( m_1\frac{m_2^2}{M^2} + m_2\frac{m_1^2}{M^2} \right)\tvec{v}^2 + \frac{Gm_1 m_2}{r} + \frac{1}{8\ifthenelse{\boolean{NaturalUnits}}{}{c^2}}\left(m_1\frac{m_2^4}{M^4} + m_2\frac{m_1^4}{M^4}\right)\tvec{v}^4 \\
				& + \frac{Gm_1m_2}{2r\ifthenelse{\boolean{NaturalUnits}}{}{c^2}}\left[ 3\left(\frac{m_2^2+m_1^2}{M^2}\right)\tvec{v}^2 + 7\frac{m_2m_1}{M^2}\tvec{v}^2 + \frac{m_2m_1}{M^2}\tvec{v}^2\left(\tvec{\hat{v}\cdot\hat{r}}\right)^2 \right] \\
				&- \frac{G^2m_1m_2(m_1+m_2)}{2r^2\ifthenelse{\boolean{NaturalUnits}}{}{c^2}}
			\end{split}\\
			\begin{split}\label{eq:Lagrangian:1PN:ReducedOneBodyProblem}
			    = \hspace{3pt} & \frac{\mu}{2} \tvec{v}^2 + \frac{GM\mu}{r} + \frac{\mu}{8\ifthenelse{\boolean{NaturalUnits}}{}{c^2}}\left(1-3\eta\right) \tvec{v}^4 \\
		        & + \frac{GM\mu}{2r}\left[ 3 + \eta \left(1+\left(\tvec{\hat{v}\cdot\hat{r}}\right)^2\right) \right] \ifthenelse{\boolean{NaturalUnits}}{\tvec{v}^2}{\frac{\tvec{v}^2}{c^2}} - \frac{G^2M^2\mu}{2r^2\ifthenelse{\boolean{NaturalUnits}}{}{c^2}}.
		    \end{split}
		\end{align}
	\end{subequations}
	
\section{Computing the 1PN equations of motion and energy} \label{subsec:1PNEnergy}
	Equipped with the effective Lagrangian up to first \acrlong{pn} order \eqref{eq:Lagrangian:1PN:ReducedOneBodyProblem}, all that remains is to determine the \acrlong{eom} and associated energy.
\subsection{Finding the associated equations of motion}
	The corresponding \acrlong{eom} can be obtained by finding the extremum of the action. Using polar coordinates it is obvious that $\theta$ is a cyclic coordinate, as it does not appear in the Lagrangian.
	
	\begin{subequations}
	\begin{align}
		&\dv{t} \pdv{L}{\omega} = \pdv{L}{\theta} = 0,\\
		\begin{split}
			\ell \equiv \pdv{L}{\omega} = \hspace{4pt} & \mu r^2 \omega + \frac{\mu}{2\ifthenelse{\boolean{NaturalUnits}}{}{c^2}}(1-3\eta)(r^4\omega^3+\dot{r}^2r^2\omega) + \frac{GM\mu}{r\ifthenelse{\boolean{NaturalUnits}}{}{c^2}}(3+\eta)r^2\omega\\
			= \hspace{4pt} & \mu r^2\omega \left[ 1+ \ifthenelse{\boolean{NaturalUnits}}{}{\frac{1}{c^2}}\left\{ \frac{1-3\eta}{2} (r^2\omega^2 + \dot{r}^2) + \frac{GM}{r}(3+\eta) \right\} \right].\label{eq:Lagrangian:1PN:CanonicalAngularMomentum}
		\end{split}\\
		\begin{split}
			\dv{t} \pdv{L}{\omega} = \hspace{4pt} & \mu(r^2\dot{\omega}+ 2r\omega\dot{r})\left[ 1+ \ifthenelse{\boolean{NaturalUnits}}{}{\frac{1}{c^2}} \left\{ \frac{1-3\eta}{2} (r^2\omega^2 + \dot{r}^2) + \frac{GM}{r}(3+\eta) \right\} \right] \\
			& + \ifthenelse{\boolean{NaturalUnits}}{}{\frac{1}{c^2}} \mu r^2\omega \left[ (1-3\eta)(r\omega^2\dot{r} +r^2\omega\dot{\omega} + \dot{r}\ddot{r}) - \frac{GM}{r^2}(3+\eta)\dot{r} \right].
		\end{split}
	\end{align}
	\end{subequations}
	
	Approximating $\ell \approx \mu r^2 \omega + \order{\frac{1}{c^2}}$ it is clear that $\ell$ is the angular momentum of Newtonian theory, with a 1\acrshort{pn} correction.
	
	The radial \acrlong{eom} is similarly obtained by
	\begin{subequations}
	\begin{align}
		& \pdv{L}{r} - \dv{t} \pdv{L}{\dot{r}} = 0. \\
		\begin{split}
			\pdv{L}{r} = \hspace{3pt} & \mu r\omega^2 -\frac{GM\mu}{r^2} + \frac{\mu}{2\ifthenelse{\boolean{NaturalUnits}}{}{c^2}}(1-3\eta)(r^3\omega^4 + r\omega^2 \dot{r}^2)\\
			&+ \frac{GM\mu}{2\ifthenelse{\boolean{NaturalUnits}}{}{c^2}}\left[ (3+\eta)\omega^2 - (3+2\eta) \left(\frac{\dot{r}}{r}\right)^2 + \frac{2GM}{r^3} \right],
		\end{split}\\
		\begin{split}
			\pdv{L}{\dot{r}} = \hspace{3pt} & \mu\dot{r} + \frac{\mu}{2\ifthenelse{\boolean{NaturalUnits}}{}{c^2}}(1-3\eta)(\dot{r}^3 + r^2\omega^2\dot{r}) + \ifthenelse{\boolean{NaturalUnits}}{GM\mu}{\frac{GM\mu}{c^2}}(3+2\eta)\frac{\dot{r}}{r} \\
			= \hspace{3pt} & \mu\dot{r} \left[ 1 + \ifthenelse{\boolean{NaturalUnits}}{}{\frac{1}{c^2}}\left\{ \frac{1-3\eta}{2}(\dot{r}^2 + r^2\omega^2) + \frac{GM}{r}(3+2\eta) \right\} \right], \label{eq:Lagrangian:1PN:CanonicalRadialMomentum}
		\end{split}\\
		\begin{split}
			\dv{t} \pdv{L}{\dot{r}} = \hspace{3pt} & \mu\ddot{r}\left[ 1 + \ifthenelse{\boolean{NaturalUnits}}{}{\frac{1}{c^2}}\left\{ \frac{1-3\eta}{2}(\dot{r}^2 + r^2\omega^2) + \frac{GM}{r}(3+2\eta) \right\} \right]\\
			& + \ifthenelse{\boolean{NaturalUnits}}{}{\frac{1}{c^2}} \mu \dot{r} \left[ \left(1-3\eta\right)\left( \dot{r}\ddot{r} + r\omega^2\dot{r} + r^2\omega \dot{\omega} \right) - \frac{GM}{r^2}\dot{r}\left( 3+2\eta \right) \right].
		\end{split}
	\end{align}
	\end{subequations}
	
	Imposing circular motion entails $\dot{r}=\ddot{r}=0$, thus the \acrshort{eom} simplifies to
	\begin{align}
		0 = \left\{ \frac{G^2M^2}{r^3\ifthenelse{\boolean{NaturalUnits}}{}{c^2}} - \frac{GM}{r^2} \right\} + \left\{ r+\frac{GM(3+\eta)}{2\ifthenelse{\boolean{NaturalUnits}}{}{c^2}} \right\} \omega^2 + \left\{\frac{r^3(1-3\eta)}{2\ifthenelse{\boolean{NaturalUnits}}{}{c^2}}\right\} (\omega^2)^2
	\end{align}
	
	The solution to this equation follows as
	\begin{subequations}\label{eq:RelativisticCorrection:KeplersLawsToR}
		\begin{align}
			&\omega^2 = \frac{GM}{r^3}\left\{ 1 - (3-\eta) \frac{GM}{r\ifthenelse{\boolean{NaturalUnits}}{}{c^2}} + \order{\frac{1}{c^4}} \right\},\label{eq:KeplersLaw:Relativistic} \\
			\Rightarrow \quad &v^2 = r^2 \omega^2 = \frac{GM}{r} \left\{ 1 - (3-\eta)\frac{GM}{r\ifthenelse{\boolean{NaturalUnits}}{}{c^2}} + \order{\frac{1}{c^4}} \right\}. \label{eq:relativistic:v2 of GMr}
		\end{align}
	\end{subequations}
	Notice that to the 0\acrshort{pn} order equation \eqref{eq:KeplersLaw:Relativistic} reduces to Kepler's third law for circular orbits \eqref{eq:KeplersThirdLaw}.
	
	Using \eqref{eq:RelativisticCorrection:KeplersLawsToR}, $r$, $\omega$, and $v$ can be related with 1\acrshort{pn} corrections. To compute the other relations, organize the equation into a quadratic equation, and solve for the quadratic parameter (e.g. solve for $GM/r$ in \eqref{eq:relativistic:v2 of GMr} to obtain $GM/r$ as a function of $v^2$).
	
	The 1\acrshort{pn} correct relations of $\omega$, $r$, and $v$ turns out to be
    \begin{subequations} \label{eq:Relativistic: omega}
        \begin{align}
            \omega^2 = \hspace{3pt} & \frac{GM}{r^3}\left\{ 1 - (3-\eta) \frac{GM}{r\ifthenelse{\boolean{NaturalUnits}}{}{c^2}} + \order{\frac{1}{c^4}} \right\}, \label{eq:Relativistic: omega2 from GMr3} \\
            GM\omega = \hspace{3pt}& v^3\left\{ 1 + (3-\eta)\ifthenelse{\boolean{NaturalUnits}}{v^2}{\frac{v^2}{c^2}} + \order{\frac{1}{c^4}}\right\}. \label{eq:Relativistic: GMomega from v3}
        \end{align}
    \end{subequations}
    \begin{subequations} \label{eq:Relativistic: GMr}
        \begin{align}
            \frac{GM}{r} = \hspace{3pt} & (GM\omega)^{2/3}\left\{ 1 + \left( 1-\frac{\eta}{3} \right) \ifthenelse{\boolean{NaturalUnits}}{(GM\omega)^{2/3}}{\frac{(GM\omega)^{2/3}}{c^2}} + \order{\frac{1}{c^4}} \right\}, \label{eq:Relativistic: GMr from GMomega23}\\
            \frac{GM}{r} = \hspace{3pt} & v^2\left\{ 1 + (3-\eta)\ifthenelse{\boolean{NaturalUnits}}{v^2}{\frac{v^2}{c^2}} + \order{\frac{1}{c^4}}\right\}. \label{eq:Relativistic: GMr from v2}
        \end{align}
    \end{subequations}
    \begin{subequations} \label{eq:Relativistic: v2}
        \begin{align}
            v^2 = \hspace{3pt} & (GM\omega)^{2/3} \left\{ 1 - \left(2-\frac{2}{3}\eta\right) \ifthenelse{\boolean{NaturalUnits}}{(GM\omega)^{2/3}}{\frac{(GM\omega)^{2/3}}{c^2}} + \order{\frac{1}{c^4}} \right\}, \label{eq:Relativistic: v2 from GMomega 23}\\
            v^2 = \hspace{3pt} & \frac{GM}{r} \left\{ 1 - (3-\eta)\frac{GM}{r\ifthenelse{\boolean{NaturalUnits}}{}{c^2}} + \order{\frac{1}{c^4}} \right\}. \label{eq:Relativistic: v2 from GMr}
        \end{align}
    \end{subequations}
	To get the relations of other order of the \acrshort{lhs}, is simply a matter of raising the equation to the desired power, and then Taylor expanding away terms that are not of first \acrlong{pn} order. But the relations as written here are those that usually come up, many having already been used in this thesis at the 0\acrshort{pn} approximation.
	
	One illustrative example to get a sense of the scale of the 1\acrshort{pn} correction is to use \eqref{eq:Relativistic: GMr from GMomega23} to compute the correction of the lunar orbital distance. The Moon does not follow a circular orbit, but rather has an eccentricity of $0.02 < e_{\Moon} < 0.08$, so it will not be an exact approximation. It can however give an idea of the scale of the effect.
	
	The Moon has a (sidereal) period of $T_{\Moon} = 27.32$ days, and mass of $m_{\Moon} = 1.23 \cdot 10^{-2} M_{\Earth}$, where the Earth mass is $M_{\Earth} = 5.97 \cdot 10^{24}$ kg \cite{physicsFormula}.
	\begin{subequations}
    \begin{align}
        r_{\Moon} & = \frac{GM}{(GM\omega_{\Moon})^{2/3}} \left\{ 1 - \left(1-\frac{\eta}{3} \right) \frac{(GM\omega_{\Moon})^{2/3}}{c^2} \right\} \\
    \begin{split}
        & \simeq 3.85 \cdot 10^8 \left\{ 1 - 1.17\cdot10^{-11} \right\} \text{m} \\
        & \simeq 3.85 \cdot 10^8 \text{ m } - 4.47\cdot10^{-3} \text{ m}
    \end{split}
    \end{align}
    \end{subequations}
    
    Thus, even though the approximation is crude, this shows that the correction is in order of \emph{millimetres} for the Earth-Moon system. It is perhaps not surprising considering that the Moon is not exactly moving at relativistic speeds.
    
    However, due to reflective mirrors left by the Apollo missions, the Earth-Moon distance \emph{is} measured at a millimetre precision \cite{LunarLaserRanging}.
    
    The best models for the Earth-Moon system operates at this precision, and thus needs to account for 1\acrshort{pn} corrections like this one \cite{JPL_model}.

\subsection{Computing the Hamiltonian}
	To obtain the orbital energy it will suffice to derive the corresponding Hamiltonian of the 1\acrshort{pn} Lagrangian \eqref{eq:Lagrangian:1PN:ReducedOneBodyProblem} by Legendre transformation \eqref{eq:Hamiltonian:LegendreTransform:def}. Utilizing the results of equations \eqref{eq:Lagrangian:1PN:CanonicalAngularMomentum} and \eqref{eq:Lagrangian:1PN:CanonicalRadialMomentum} the Hamiltonian is found to be
	\begin{subequations}
		\begin{align}
			H(r, v) = \hspace{3pt} & \dot{r} \pdv{L}{\dot{r}} + \omega \pdv{L}{\omega} - L\\
			\begin{split}
				= \hspace{3pt} & \mu\dot{r}^2 \left[ 1 + \ifthenelse{\boolean{NaturalUnits}}{}{\frac{1}{c^2}} \left\{ \frac{1-3\eta}{2}(\dot{r}^2 + r^2\omega^2) + \frac{GM}{r}(3+2\eta) \right\} \right]\\
				& + \mu r^2\omega^2 \left[ 1+ \ifthenelse{\boolean{NaturalUnits}}{}{\frac{1}{c^2}} \left\{ \frac{1-3\eta}{2} (r^2\omega^2 + \dot{r}^2) + \frac{GM}{r}(3+\eta) \right\} \right] -L
			\end{split}\\
			= \hspace{3pt} & \mu v^2 + \frac{4\mu}{8}(1-3\eta) \ifthenelse{\boolean{NaturalUnits}}{v^4}{\frac{v^4}{c^2}} + 2\frac{GM}{2r}\left[ 3+\eta \left(1+\frac{\dot{r}^2}{v^2}\right) \right] \ifthenelse{\boolean{NaturalUnits}}{v^2}{\frac{v^2}{c^2}} - L\\
			\begin{split}
				= \hspace{3pt} & \frac{\mu}{2}v^2 - \frac{GM\mu}{r} + \frac{3\mu}{8}(1-3\eta) \ifthenelse{\boolean{NaturalUnits}}{v^4}{\frac{v^4}{c^2}} \\
				& + \frac{GM\mu}{2r}\left[ 3+\eta\left(1 + \frac{\dot{r}^2}{v^2}\right) \right] \ifthenelse{\boolean{NaturalUnits}}{v^2}{\frac{v^2}{c^2}} + \frac{G^2M^2\mu}{2r^2\ifthenelse{\boolean{NaturalUnits}}{}{c^2}}.
			\end{split}
		\end{align}
	\end{subequations}
	
	The Hamiltonian is expressed in terms of the relative velocity $v^2 = \dot{r}^2 + r^2\omega^2$ instead of the canonical momentum because the end goal is simply to obtain the 1\acrshort{pn} energy in terms of the frequency. Note that this expression is valid for all type of motion, not just circular.
	
	Imposing circular motion again, the relations between $v, r, \omega$ from \eqref{eq:Relativistic: omega}-\eqref{eq:Relativistic: v2} may be used to express the energy in terms of one of these variables. The most commonly used variable is the frequency $\omega$, as it is most directly related to the observable: the \acrshort{gw} frequency. However here the velocity will be used for more convenient calculations.
	\begin{subequations}
		\begin{align}
			E = \hspace{3pt} & \frac{\mu}{2}v^2 - \left(v^2 + (3-\eta)\frac{v^4}{c^2} \right)\mu + \frac{3\mu}{8}(1-3\eta)\frac{v^4}{c^2} + \frac{\mu}{2}\frac{v^4}{c^2}(3+\eta) + \frac{\mu}{2}\frac{v^4}{c^2} \\
			= \hspace{3pt} & -\frac{\mu}{2}v^2 + \frac{\mu}{2}\left[ \left(-6+\frac{3}{4}+3+1 \right) + \left( 2 - \frac{9}{4} + 1 \right)\eta  \right]\frac{v^4}{c^2} \\
			= \hspace{3pt} & -\frac{\mu}{2}v^2 \left[ 1 + \left\{ \frac{5}{4} - \frac{3}{4}\eta \right\}\frac{v^2}{c^2} \right]. \label{eq:1PN:Energy:v}
		\end{align}
	\end{subequations}
	
	This is \emph{not} the energy expansion \eqref{eq:Energy:Expansion:Gen} presented in \cref{chap:waveform}, so what is going on?
	
	Recalling that in \cref{chap:waveform} $v^*$ was only used as a proxy variable for the orbital frequency, and was defined as $v^*\equiv(GM\omega)^{1/3}$. In the Newtonian theory $v=v^*$, but at 1\acrshort{pn} the relative velocity and orbital frequency are related according to \eqref{eq:Relativistic: v2 from GMomega 23}, hence $v\neq v^*$. Therefore, \eqref{eq:1PN:Energy:v} is the orbital energy in terms of the \emph{actual} relative velocity.
	
	Using equation \eqref{eq:Relativistic: v2 from GMomega 23} to transform $v \to \omega$ the energy in terms of frequency is obtained to be
	\begin{empheq}[box=\widefbox]{align}\label{eq:1PN:Energy}
		E = -\frac{\mu}{2} (GM\omega)^{2/3} \left\{ 1 + \left\{ -\frac{3}{4}-\frac{1}{12}\eta \right\} \ifthenelse{\boolean{NaturalUnits}}{(GM\omega)^{2/3}}{\frac{(GM\omega)^{2/3}}{c^2}} + \order{\frac{1}{c^3}} \right\}.
	\end{empheq}
	
	\emph{This} is the energy presented in \eqref{eq:Energy:Expansion:Gen}, where $(GM\omega)^{1/3}$ was named $v$, somewhat confusingly from the point of view of this chapter.
	
	Beware that in the literature, energy and flux can, and are, presented in terms of $\frac{GM}{r}$, $(GM\omega)^{1/3}$, or $v$. But they are all the frequency energy/flux, relabelled using the 0\acrshort{pn} approximation of the relations \eqref{eq:Relativistic: omega} - \eqref{eq:Relativistic: v2}. This is of course since they are ultimately used to compute waveforms, which are computed from differential equations of the frequency. And in the end, the frequency is the directly observable parameter.
	
	To get a sense of the scale of this energy correction, lets use this on the Earth-Moon system.
    \begin{align}
    \begin{split}
        E_{\Moon} & \simeq -3.81 \cdot 10^{28} \left(1 - 8.76 \cdot 10^{-12} \right) \text{J} \\
        & \simeq -3.81 \cdot 10^{28} \text{J} + 3.34 \cdot 10^{17} \text{ J}.
    \end{split}
    \end{align}
    
    Which is of course comparably tiny. Using the mass-energy equivalence, the correction is comparable to $\sim 3\text{kg}$, of an $\sim 4\cdot 10^8$ metric tonnes 0\acrshort{pn} energy.
\chapter{Calculating the energy flux} \label{chap:flux}
	In order to fully describe the 1\acrshort{pn} dynamics of the compact binary the energy dissipation by generated \acrshort{gw}s need to be accounted for. In this section this total radiated power is to be calculated.
	
	Derivations presented here closely follows those presented in \textcite{Maggiore:VolumeI}, \textcite{Porto2016}, and \textcite{Ross2012Multipole}.
\section{The graviton field evaluated at large scales}
\subsection{Separation of scales}
    In \cref{sec:EFT} it was argued that the binary system could be separated into three different length scales $\sim L$, $\sim r$, and $\sim R_S$, related by $L \gg r \gg R_S$. In \cref{chap:energy} the $\sim R_S$ scale was `integrated out', leaving \acrshort{bh}s and \acrshort{ns}s only with a point mass structure at the scale of the orbit $\sim r$. Similarly, in \cref{sec:Energy-Momentum tensor of GW} the total energy flux of a system was found evaluating the graviton field at a scale $L \gg \lambdabar_\text{\acrshort{gw}}$, leaving the source effectively as a point source, endowed with a quadrupole structure.
    
    By the requirement of evaluating at a scale $L\gg \lambdabar$, it is also automatically realized to be evaluated at a scale much larger than the binary system that created it, $r \gg \lambdabar$.\footnote{To see why this relation holds for the inspiral, see footnote \ref{fn:relating wavelength and orbital scale} from \cref{chap:energy}.}
    
\subsection{Modifying the source of gravitational waves}
    In \cref{sec:Energy-Momentum tensor of GW} the solution of the graviton field was found to be \eqref{eq:Linearized:GW:SourceTerm}, which reads
    \begin{align}
        \Bar{\H}_{ij}^\text{\acrshort{tt}}(t,\tvec{R}) = \frac{\lambda}{8\pi R} \Lambda\ind{_{ij}^{kl}} \int_\V  T\ind{_{kl}}(t_\text{ret},\tvec{x}) \dd[3]{x}, \qq{where} t_\text{ret} = t - \ifthenelse{\boolean{NaturalUnits}}{\abs{\tvec{R}-\tvec{x}}}{\frac{\abs{\tvec{R}-\tvec{x}}}{c}}.
    \end{align}
    With $\V\sim L^3$, such that $T\ind{_{ij}}$ evaluated at $\partial \V$ is zero.
    
    On the other hand, in the far region $\dalembertian\H=0$, which admits solution of the general form
    \begin{align}
    \begin{split}
        \bar{\H}\ind{_{ij}}(t,\tvec{R}) & = \frac{F_{kl}(t-R/c)}{R} - \partial\ind{_{i_1}} \left[ \frac{F\ind{_{kl}^{i_1}}(t-R/c)}{R} \right] + \frac{1}{2} \partial\ind{_{i_1}} \partial\ind{_{i_2}} \left[ \frac{F\ind{_{kl}^{{i_1}i_2}}(t-R/c)}{R} \right] + \dots \\
        & \equiv \sum_{\ell=0}^\infty \frac{(-1)^\ell}{\ell!} \partial\ind{_L} \left[ \frac{F^L_{kl}(t-R/c)}{R} \right],\\ &\qq{with}
        \dalembertian \left[ \frac{F^L_{ij}(t-R/c)}{R} \right] = 0.
    \end{split}
    \end{align}
    Here the \emph{multi index notation} has been introduced, which is to say a capital letter index $L$ represent a number of $\ell$ indices.
    
    By comparing these two expressions, which should be equivalent for $R \gg r$, $F^L$ can be demonstrated to be \cite{DamourMultipole}
    \begin{align}
        F^L_{ij}(t_\text{ret}) = \int \dd[3]{x} x^L_\text{\acrshort{stf}} \sum_{p=0}^\infty \frac{(2\ell+1)!!}{2^p p! (2\ell+2p+1)!!} \left(\frac{\abs{\tvec{x}}}{c}\pdv{}{t} \right)^{2p} T\ind{_{ij}}(t_\text{ret},\tvec{x}).
    \end{align}
    
    Here $t_\text{ret}=t-R/c$. The subscript \acrshort{stf} stands for \acrlong{stf}. This is the only part that is not eliminated by the Lambda tensor \eqref{eq:Lambdatensor:def}. Recall that in the flux, the Lambda tensor eliminated the trace of the quadrupole moment. 
    
    As a remainder $\ell!! = \ell \cdot (\ell-2) \cdot (\ell-4) \cdots 2 \text{ or } 1$, depending on whether $\ell$ is even or odd respectively.
    
    Before proceeding, it will be useful to investigate \acrshort{stf} tensors.
\subsection{STF tensor decomposition}
    The \acrshort{stf} part of a tensor is the irreducible representation of the tensor under rotations. Therefore, in \acrshort{gw} physics it represents the physical degrees of freedom, where the other terms can be gauged away.
    
    As an example, a rank two tensor can be decomposed into three parts
    \begin{align}
        T\ind{_{ij}} = T\ind{_{[ij]}} + T\ind{_{\{ij\}}} = A\ind{_{[ij]}} + S\ind{_{\{ij\}}} = \frac{1}{3} S\ind{_k^k} \delta\ind{_{ij}} + \varepsilon\ind{_{ijk}} A\ind{^k} + \left( S\ind{_{\{ij\}}} - \frac{1}{3} S\ind{_k^k} \delta\ind{_{ij}} \right).
    \end{align}
    
    The first term in the last equality is the trace part, the second the anti-symmetric part, and finally the third term is the \acrshort{stf} part.
    
    Now, noticing $\Lambda\ind{^{ij}_{kl}}\delta^{kl}=0$, and since the Lambda tensor is symmetric in $k\leftrightarrow l$ $\Lambda\ind{^{ij}_{kl}} \varepsilon\ind{^{kl}_m}=0$. This is why only the \acrshort{stf} part of $F^L_{ij}$ contribute to the final flux.
    
    The \acrshort{stf} part of a rank $n$ tensor can be obtained by \cite{Thorn1980Multipole}
	\begin{align}\label{eq:STF:def}
		\begin{split}
			T^{i_1 \cdots i_n}_\text{STF} &= \sum_{p=0}^{\floor{n/2}} c_p^{(n)} \delta^{ \{i_1 i_2 }  \cdots \delta^{ i_{2p-1} i_{2p} } T^{i_{2p+1} \cdots i_{n}\} a_1a_1\cdots a_p a_p}, \\
			c_p^{(n)} &\equiv (-1)^p\frac{n!(2n-4p+1)!!}{(n-2p)!(2n-2p+1)!!(2p)!!}.
		\end{split}
	\end{align}
	
	The construction of this expression is not self-evident, but calculating it for the quadrupole and octupole moments will be instructive. The operator $\floor{x}$ rounds $x$ down to the closest integer, and is called the floor function. E.g. $\floor{3/2} = 1 $.
	
	First notice that the $p=0$ term always correspond to the symmetric version of the tensor in question
	\begin{align}
	    \text{for $p=0$:} \quad \quad \frac{n!(2n+1)!!}{n!(2n+1)!!} T^{\{i_1 \cdots i_n\}} = T_\text{sym}^{i_1\cdots i_n}.
	\end{align}
	Not surprisingly, the terms of $p>0$ in \eqref{eq:STF:def} is used to subtract all possible traces, thus making the expression symmetric and trace free. E.g. the quadrupole moment becomes
	\begin{align}\label{eq:STFoperation:quadrupole}
		Q^{ij}_\text{\acrshort{stf}} = Q^{\{ij\}} - \frac{2!(1)!!}{0!(3)!!(2)!!} \delta^{ij} Q\ind{^a_a} = Q^{ij}_\text{sym} - \frac{1}{3} \delta^{ij} \tr(Q),
	\end{align}
	which is equivalent to the expression used in the quadrupole radiation \eqref{eq:energyflux:from Quad}. 
	
	The octupole moment follows similarly as
	\begin{align}
		\begin{split}\label{eq:STFoperation:octupole}
			O^{ijk}_\text{\acrshort{stf}} = \hspace{3pt} & O^{\{ijk\}} - \frac{3!(3)!!}{(1)!(5)!! (2)!!} \frac{1}{3}\left( \delta^{ij} O\ind{^{ka}_a} + \delta^{ik} O\ind{^{ja}_a} + \delta^{jk} O\ind{^{ia}_a} \right) \\
			= \hspace{3pt} & O^{ijk}_\text{sym} - \frac{1}{5}\left( \delta^{ij} O\ind{^{ka}_a} + \delta^{ik}O\ind{^{ja}_a} + \delta^{jk} O\ind{^{ia}_a} \right).
		\end{split}
	\end{align}
	The factor of $1/3$ comes from the symmetrizing of $\delta\ind{^{\{ij}} O\ind{^{k\}a}_a} = \frac{1}{3!} \left( \left( \delta^{ij} + \delta^{ji} \right)O\ind{^{ka}_a} + \dots \right) = \frac{1}{3} \left( \delta^{ij}O\ind{^{ka}_a} +\dots \right)$. Hopefully these examples provide some familiarity with formula \eqref{eq:STF:def}.
\subsection{The multipole structure of GWs}
    Working the expression further, following the somewhat complicated steps of \textcite{Ross2012Multipole} the result is
    \begin{align}
    \begin{split}
        \bar{\H}_{ij}^\text{\acrshort{tt}} = -\frac{4G}{Rc^2} \Lambda\ind{_{ij:k_{\ell-1}k_\ell}} \sum_{\ell=2}^\infty \frac{1}{\ell!} \Biggl[ & n\ind{_{L-2}} \partial_0^\ell I\ind{_{ij}^{L-2}}(t_\text{ret}) \\
        &- \frac{2\ell}{\ell+1} \varepsilon\ind{^a_{b\{k_{\ell-1} }} n\ind{_{a L-2}} \partial_0^\ell J\ind{_{k_{\ell}\}}^{L-2}}(t_\text{ret}) \Biggr]. \label{eq:H:rad:fromMultipoles}
    \end{split}
    \end{align}
    
    Here $I$ is the mass multipole, while $J$ is current multipole, defined as
    \begin{align}
		\begin{split} \label{eq:EnergyMultipole:allOrders}
			&I^{L}(t) = \sum_{p=0}^{\infty} \\
			&\frac{(2\ell+1)!!}{(2p)!!(2\ell + 2p +1)!!} \left(1 + \frac{8p(\ell+p+1)}{(\ell+1)(\ell+2)}\right) \left[ \int \dd[3]{x} \partial_0^{2p} \T\ind{^{00}}(t,\tvec{x}) r^{2p} x^L \right]_\text{\acrshort{stf}} \\
			& + \frac{(2\ell+1)!!}{(2p)!!(2\ell+2p+1)!!} \left(1 + \frac{4p}{(\ell+1)(\ell+2)}\right) \left[ \int \dd[3]{x} \partial_0^{2p} \T\ind{^k_k}(t,\tvec{x}) r^{2p} x^L \right]_\text{\acrshort{stf}} \\
			& - \frac{(2\ell+1)!! 4}{(2p)!!(2\ell+2p+1)!!(\ell+1)} \left( 1+ \frac{2p}{(\ell+2)} \right) \left[ \int \dd[3]{x} \partial_0^{2p+1} \T\ind{^0_i}(t,\tvec{x}) r^{2p} x^{L} x^i \right]_\text{\acrshort{stf}} \\
			& + \frac{(2\ell+1)!!}{(2p)!!(2\ell+2p+1)!!} \left(\frac{2}{(\ell+1)(\ell+2)}\right) \left[ \int \dd[3]{x} \partial_0^{2p+2} \T\ind{_{ij}}(t,\tvec{x}) r^{2p} x^{L} x^ix^j \right]_\text{\acrshort{stf}}.
		\end{split}
		\end{align}
		
		\begin{align}
		\begin{split} \label{eq:CurrentMultipole:allOrders}
			&J^{L}(t) = \sum_{p=0}^{\infty} \\
			&\frac{(2\ell+1)!!}{(2p)!!(2\ell + 2p +1)!!} \left( 1+\frac{2p}{(\ell+2)} \right) \left[ \int \dd[3]{x} \epsilon\ind{^{k_\ell}_{mn}} \partial_0^{2p} \T\ind{^{0m}}(t,\tvec{x}) r^{2p} x^{L-1} x^{n} \right]_\text{\acrshort{stf}} \\
			& - \frac{(2\ell+1)!!}{(2p)!!(2\ell + 2p +1)!!(\ell+2)} \left[ \int \dd[3]{x} \epsilon\ind{^{k_\ell}_{ms}} \partial_0^{2p+1} \T\ind{^{mn}}(t,\tvec{x}) r^{2p} x^{L-1} x_{n}x^s \right]_\text{\acrshort{stf}}.
		\end{split}
	\end{align}
	Here $\T\ind{^{\mu\nu}}$ is the energy-momentum tensor of the source.
	
	Now, using \eqref{eq:Flux:LinearTheory: hTT} the total energy flux is determined as \cite{Thorn1980Multipole}
	\begin{subequations}
	\begin{align}
        \F = \hspace{3pt} & \frac{R^2}{c^2} \int \expval{ \dot{\H}_{ij}^\text{\acrshort{tt}} \dot{\H}^{ij}_\text{\acrshort{tt}} } \dd{\Omega} \\
    \begin{split} \label{eq:Flux:GeneralTerm:AllOrders}
        = \hspace{3pt} & \frac{G}{c^3}\sum_{\ell=2}^{\infty} \frac{(\ell+1)(\ell+2)}{ \ell(\ell-1) \ell! (2\ell+1)!!} \expval{ \left( \dv[\ell+1]{I^L(t)}{(ct)} \right)^2 }\\
        &+ \frac{4\ell(\ell+2)}{ (\ell-1)(\ell+1)!(2\ell+1)!! } \expval{ \left( \dv[\ell+1]{J^L(t)}{(ct)} \right)^2 },
    \end{split}
	\end{align}
	\end{subequations}
	where the bracket is understood as averaging over time.
\section{The 1PN flux terms}
    To assign \acrshort{pn} orders to the different terms, notice that every time derivative contributes with a factor of $c^{-1}$, $\partial_0^{2p} \sim 1/c^{2p}$. Also since $T^{\mu\nu} \sim m \dot{x}^\mu \dot{x}^\nu$, and $\dot{x}^{0} \sim c$, we can expect every spatial index of the energy-momentum tensor to contribute with a factor of $c^{-1}$ compared to the 00 term. Utilizing these observations it should be clear at which \acrshort{pn} order the various terms of \eqref{eq:EnergyMultipole:allOrders} and \eqref{eq:CurrentMultipole:allOrders} enter.
	
\subsection{Leading order term, the quadrupole moment}\label{subsec:LOTQuad}
	For the leading order term, only moments with the lowest power of $(c^{-1})^n$ can contribute. From the general flux expression \eqref{eq:Flux:GeneralTerm:AllOrders} it is clear that every derivative of the multipole moments contributes with additional factors of $c^{-1}$, thus the leading order term must be of only two indices, a quadrupole. Because the leading order term in the energy-momentum tensor is the point particle contributions, and since the point particle energy-momentum tensor is proportional to the tensor product of the particle's four velocity \eqref{eq:pp EnergyMomentumTensor} any spatial index of $\T^{\mu\nu}$ contributes with an additional factor of $c^{-1}$. This excludes the current multipole \eqref{eq:CurrentMultipole:allOrders} entirely, and all but the first line of the mass multipole \eqref{eq:EnergyMultipole:allOrders}.
	
	Since the $ct$ derivatives contribute with superfluous factors of $c^{-1}$, only the $p=0$ term of the first line of \eqref{eq:EnergyMultipole:allOrders} for $L=2$ contributes to the leading order energy flux.\\
	
	Using this leading order term of \eqref{eq:EnergyMultipole:allOrders} and \eqref{eq:pp EnergyMomentumTensor} for the $\T^{00}$ term the resulting leading order\footnote{To leading order $\gamma_a=1$. Recall that $\gamma_a = (1-v^2/c^2)^{-1/2} \approx 1 + \frac{1}{2} v^2/c^2 + \frac{3}{8} v^4/c^4 +\dots$.} expression for the mass quadrupole moment is
	\begin{align}
		\begin{split}
			I^{ij}_{(0)}(t) & = \frac{(5)!!}{(5)!!} \int \dd[3]{x} \T\ind{^{00}}(t,\tvec{x}) \left[ x^i x^j \right]_\text{\acrshort{stf}}\\
			& = \int \dd[3]{x} \sum_{a} \gamma_a m_a c^2 \delta^{(3)}(\tvec{x}-\tvec{x}_a(t)) \left[ x^i x^j - \frac{1}{3}r^2 \delta^{ij} \right] \\
			& = \mu c^2r^2 \left[n^in^j -\frac{1}{3}\delta^{ij} \right]
		\end{split}
	\end{align}
	In the last line the masses were rewritten to the reduced mass (see \cref{app:equivOneBodyAndMassTerms} for more details).
	
	To compute the resulting flux, equation \eqref{eq:Flux:GeneralTerm:AllOrders} requires the third time derivative of this term. Applying circular motion implies $n^x=\cos(\omega t)$, $n^y=\sin(\omega t)$, and $n^z=0$. Thus, after consulting \eqref{eq:App:Trig:Square} for how to rewrite squared trigonometric functions, the result is
	\begin{align}
		\nonumber \dv[3]{I^{ij}_{(0)}}{(ct)} & = \frac{\mu r^2}{2c} \dv[3]{}{t} \begin{pmatrix}
			\frac{1}{3} + \cos(2\omega t) & \sin(2\omega t) & 0 \\
			\sin(2\omega t) & \frac{1}{3} - \cos(2\omega t) & 0 \\
			0 & 0 & -\frac{2}{3}
		\end{pmatrix} \\
		& = \frac{2^2\mu r^2 \omega^3}{c} \begin{pmatrix}
			\sin(2\omega t) & -\cos(2\omega t) & 0 \\
			-\cos(2\omega t) & -\sin(2\omega t) & 0 \\
			0 & 0 & 0
		\end{pmatrix}. \label{eq:Quadrupole:firstOrder:thirdDerivative}
	\end{align}
	
	Notice that $I^{ij}$ is indeed symmetric and trace free. To calculate the energy flux the sum over squares of each component is needed.
	\begin{align}
		\nonumber \expval{\dv[3]{I^{ij}_{(0)}}{(ct)} \dv[3]{{I_{(0)}}_{ij}}{(ct)}} & = \frac{2^4\mu^2r^4\omega^6}{c^2} \left( 2\sin[2](2\omega t) + 2\cos[2](2\omega t) \right) \\
		& = 2^5\mu^2 \frac{v^4 \omega^2}{c^2} = \frac{2^5\eta^2}{G^2} \frac{v^{10}}{c^2}. \label{eq:Flux:0PN}
	\end{align}
	In the last line $v=\omega r$ was used, and finally Kepler's third law \eqref{eq:KeplersThirdLaw} to exchange $\omega$ for $v$. 
	
	Then the leading order term of the energy flux is
	\begin{align}
		\F_\text{Newt} = \frac{G}{c^3}\frac{ 3\cdot4 }{ 2\cdot2!\cdot (5)!! } \expval{ \dddot{I}^{ij}\dddot{I}_{ij} } = \frac{2^5}{5} \frac{\eta^2}{G} \frac{v^{10}}{c^5} = \frac{32}{5} \frac{\eta^2}{G} \frac{v^{10}}{c^5} \equiv F_\text{Newt} v^{10}
	\end{align}
	which is the well established result \eqref{eq:NewtonianFlux pp}. It is worth noting that $[c^5/G]$ does indeed have the dimension of energy per time, as expected from the energy flux term.\\
	
	For the next to leading order correction the octupole moment $I^{ijk}(t)$ and the current quadrupole moment $J^{ij}(t)$ must be added, and also there are relativistic corrections to the quadrupole formula used here for the 0\acrshort{pn} flux term, like the other terms in \eqref{eq:EnergyMultipole:allOrders} and relativistic corrections to $\T^{\mu\nu}$.
	
\subsection{Next to leading order term, the octupole moment}
	Using \eqref{eq:EnergyMultipole:allOrders} and \eqref{eq:pp EnergyMomentumTensor} the mass octupole moment reads
	\begin{align}
		\begin{split}
			I^{ijk}_{(2)}(t) & = \frac{(7)!!}{(7)!!} \cdot (1) \cdot \int \dd[3]{x} \T\ind{^{00}}(t,\tvec{x}) \left[x^i x^j x^k\right]_\text{\acrshort{stf}}\\
			& = \int \dd[3]{x} \sum_{a} \gamma_a m_a c^2 \delta^{(3)}(\tvec{x}-\tvec{x}_a(t)) \left[ x^i x^j x^k - \frac{r^2}{5}\left( \delta^{ij}x^k + \delta^{ik} x^j + \delta^{jk} x^i \right)\right] \\
			& = \mu c^2 r^3 \sqrt{1-4\eta} \left[n^{i}n^{j}n^{k} - \frac{1}{5}\left(\delta^{ij}n^k + \delta^{ik} n^j + \delta^{jk} n^i \right) \right].
		\end{split}
	\end{align}
	
	Inserting circular motion ($n^x = \cos(\omega t)$, $n^y = \sin(\omega t)$ and $n^z=0$) and then taking the 4$\textsuperscript{th}$ time derivative, as necessitated by equation \eqref{eq:Flux:GeneralTerm:AllOrders} produces
	
    \begin{subequations}\label{eq:Octupole:derivatives}
        \begin{align}
            & \dv[4]{I^{xxx}_{(2)}}{(ct)} = \frac{\mu r^3}{c^2} \sqrt{1-4\eta} \left[\frac{(3\omega)^4}{4}\cos(3\omega t) + \frac{3\omega^4}{20} \cos(\omega t)\right], \label{eq:Octupole:derivatives:xxx} \\
            & \dv[4]{I^{xyy}_{(2)}}{(ct)} = \frac{\mu r^3}{c^2} \sqrt{1-4\eta} \left[-\frac{(3\omega)^4}{4}\cos(3\omega t) + \frac{\omega^4}{20} \cos(\omega t)\right], \label{eq:Octupole:derivatives:xyy} \\
            & \dv[4]{I^{xzz}_{(2)}}{(ct)} = \frac{\mu r^3}{c^2} \sqrt{1-4\eta} \left[-\frac{\omega^4}{5} \cos(\omega t)\right]. \label{eq:Octupole:derivatives:xzz} \\ &\nonumber \\
            & \dv[4]{I^{yyy}_{(2)}}{(ct)} = \frac{\mu r^3}{c^2} \sqrt{1-4\eta} \left[\frac{(3\omega)^4}{4}\sin(3\omega t) + \frac{3\omega^4}{20} \sin(\omega t)\right], \label{eq:Octupole:derivatives:yyy} \\
            & \dv[4]{I^{yxx}_{(2)}}{(ct)} = \frac{\mu r^3}{c^2} \sqrt{1-4\eta} \left[-\frac{(3\omega)^4}{4}\sin(3\omega t) + \frac{\omega^4}{20} \sin(\omega t)\right], \label{eq:Octupole:derivatives:yxx} \\
            & \dv[4]{I^{yzz}_{(2)}}{(ct)} = \frac{\mu r^3}{c^2} \sqrt{1-4\eta} \left[-\frac{\omega^4}{5} \sin(\omega t)\right]. \label{eq:Octupole:derivatives:yzz}
        \end{align}
    \end{subequations}
	
	Because of the symmetric property of $I^L$, terms like $I^{xyx}$ are equivalent to $I^{yxx}$. Thus, any term not appearing in equation \eqref{eq:Octupole:derivatives} are either equivalent to one of the listed terms by symmetry, or zero (like odd numbers of $z$-indices). Notice also that the sum of \eqref{eq:Octupole:derivatives:xxx}-\eqref{eq:Octupole:derivatives:xzz} is zero. Similarly the sum of \eqref{eq:Octupole:derivatives:yyy}-\eqref{eq:Octupole:derivatives:yzz} is also zero, as they should be since $I^{L}$ is trace free.\\
	
	Note that for the square sum $\expval{\cos(n\omega t)\cos(m\omega t)}=\delta_{nm}/2$, thus all contributing terms will be of the form $\sin[2](n\omega t)$, and $\cos[2](n\omega t)$. For example $\left( \dv[4]{I^{xxx}_{(2)}}{(ct)}\right)^2 = \frac{\mu^2r^6\omega^8}{c^4}(1-4)\left( \frac{3^8}{2^4}\cos[2](3\omega t) + \frac{3^2}{2^4\cdot5^2}\cos[2](\omega t) \right)$, i.e. cross terms can be dropped.
	
	\begin{align}
		\begin{split}
			\dv[4]{I_{(2)}^{ijk}}{(ct)} \dv[4]{{I_{(2)}}_{ijk} }{(ct)} & = \frac{\mu^2}{c^4} (1-4\eta) r^6 \omega^8 \frac{(3^8 + 3^9) \cdot 5^2 + (3^2+3) + 3\cdot2^4}{2^4\cdot5^2} \\
			& = \frac{ 2 \cdot 3 \cdot 1367 }{5} \frac{\eta^2 (1-4\eta)}{G^2} \frac{v^{12}}{c^4},
		\end{split} \\
		\nonumber\Rightarrow \quad \F_{(2)}^{\text{oct.}} = \frac{G}{c^3}&\frac{1}{3^3\cdot7} \frac{ 2 \cdot 3 \cdot 1367 }{5} \frac{\eta^2 (1-4\eta)}{G^2} \frac{v^{12}}{c^4} = \frac{ 2 \cdot 1367 }{3^2\cdot5\cdot7} \frac{\eta^2 (1-4\eta)}{G} \frac{v^{12}}{c^7} \\
		= &\F_\text{Newt} \frac{ 1367 }{2^4\cdot3^2\cdot7} (1-4\eta) \frac{v^2}{c^2}. \label{eq:Flux:Oct}
	\end{align}
	
\subsection{Next to leading order term, the current quadrupole moment}
	Using \eqref{eq:CurrentMultipole:allOrders} and \eqref{eq:pp EnergyMomentumTensor} the current quadrupole moment reads
	\begin{align}
		\nonumber J^{ij}_{(2)}(t) & = \frac{(5)!!}{(5)!!} \left[\int \dd[3]{x} \epsilon\ind{^{j}_{mn}} \T\ind{^{0m}}(t,\tvec{x}) x^i x^n\right]_\text{\acrshort{stf}} = \sum_{a} m_a c \left[ x^i_a \epsilon\ind{^{j}_{mn}} v^m_a x^n_a \right]_\text{\acrshort{stf}} \\
		\nonumber & = \sum_a m_a c \left[ x^i_a \epsilon\ind{^{j}_{mn}} \left( \epsilon\ind{^m_{kl}}\omega^k x^l_a  \right) x^n_a \right]_\text{\acrshort{stf}} = \sum_a m_a c \left[ x^i_a \left( \delta\ind{_{nk}}\delta\ind{^j_{l}} - \delta\ind{_{nl}}\delta\ind{^j_k} \right) \omega^k x^l_a x^n_a \right]_\text{\acrshort{stf}} \\
		& = -\sum_a m_a c r^2_a \omega \frac{1}{2} \left( x^i_a \delta\ind{^j_z} + x^j_a \delta\ind{^i_z} \right) = -\frac{\mu c r^3 \omega}{2} \sqrt{1-4\eta} \left[ n^i \delta\ind{^j_z} + n^j \delta\ind{^i_z} \right].
	\end{align}
	
	Notice that circular motion is here already assumed as $\tvec{v}_a = \tvec{\omega} \times \tvec{x}_a$ is used to simplify the expression of the first equality of the second line. Again, using $n^x=\cos(\omega t)$, $n^y = \sin(\omega t)$, and $n^z = 0$, and taking the third time derivative as instructed by formula \eqref{eq:Flux:GeneralTerm:AllOrders} results in
	\begin{align}
		\nonumber \dv[3]{J_{(2)}^{ij}(t)}{(ct)} & = -\frac{\mu r^3\omega}{2c^2}\sqrt{1-4\eta} \dv[3]{}{t} \begin{pmatrix}
			0 & 0 & \cos(\omega t) \\
			0 & 0 & \sin(\omega t) \\
			\cos(\omega t) & \sin(\omega t) & 0
		\end{pmatrix} \\
		& = \frac{\mu r^3\omega^4}{2c^2}\sqrt{1-4\eta} \begin{pmatrix}
			0 & 0 & -\sin(\omega t) \\
			0 & 0 & \cos(\omega t) \\
			-\sin(\omega t) & \cos(\omega t) & 0
		\end{pmatrix}.
	\end{align}
	
	Notice that $J^{ij}_{(2)}$ is trace free and symmetric, as it should be. The flux is determined by the sum of squares of all the tensor components, which is
	\begin{align}
		\nonumber \dv[3]{J_{(2)}^{ij}(t)}{(ct)} \dv[3]{ {J_{(2)}}_{ij}(t)}{(ct)} & = \frac{\mu^2 r^6 \omega^8}{2^2 c^4}\left(1-4\eta\right) \left(2\sin[2](\omega t) + 2\cos[2](\omega t)\right) \\
		& = \frac{ \mu^2 (1-4\eta) }{2c^4} v^6 \omega^2 = \frac{\eta^2 (1-4\eta) }{2G^2} \frac{v^{12}}{c^4} \\
		\nonumber \Rightarrow \quad \F_{(2)}^\text{curr.quad} & = \frac{2^4 G}{3^2\cdot 5c^3} \frac{\eta^2 (1-4\eta) }{2G^2} \frac{v^{12}}{c^4} = \frac{8}{45} \frac{\eta^2(1-4\eta)}{G} \frac{v^{12}}{c^7} \\
		& = \F_\text{Newt} \frac{1}{2^2\cdot3^2} (1-4\eta) \frac{v^2}{c^2}. \label{eq:Flux:CurrQuad}
	\end{align}
	
\subsection{Next to leading order term, the quadrupole moment corrections}
	Circling back to the mass quadrupole moment, all first order assumption that went into \ref{subsec:LOTQuad} must now be expanded to next to leading order. This primarily means 3 things:
	\begin{enumerate}
		\item Relativistic corrections to $\T^{\mu\nu}$, like kinetic energy, and gravitational energy.\label{enu:HowToExpandI2:ExpandT}
		\item Including other terms from \eqref{eq:EnergyMultipole:allOrders}, like $p=1$, $\T^{0i}$ and $\T\ind{^k_k}$. \label{enu:HowToExpandI2:ExpandIL}
		\item Relativistic corrections to the inserted motion of the source. For \glspl{QCO} the motion does not change, but the relation between $v$, $\omega$, and $r$ pick up some relativistic corrections \eqref{eq:Relativistic: omega}-\eqref{eq:Relativistic: v2}.\label{enu:HowToExpandI2:ExpandOmega}
	\end{enumerate}
	
	Starting with the relativistic corrections to $\T^{00}$, and $\T\ind{^k_k}$ (it will be clear in a moment why these are lumped together) recall that \eqref{eq:pp EnergyMomentumTensor}
	\begin{align}
		\begin{split}
			T^{00}_{\acrshort{pp}}(t,\tvec{x}) & = \sum_{a} \gamma_a m_a c^2 \delta^{(3)}(\tvec{x}-\tvec{x}_a(t))\\
			& = \sum_{a} \left(1 + \frac{1}{2} \frac{v^2_a}{c^2} + \frac{3}{8}\frac{v^4_a}{c^4}+\dots\right) m_a c^2 \delta^{(3)}(\tvec{x}-\tvec{x}_a(t)) \\
			& = \sum_{a} \left(m_a c^2 + \frac{1}{2} m_a v^2_a + \frac{3}{8} m_a \frac{v^4_a}{c^2}+\dots\right) \delta^{(3)}(\tvec{x}-\tvec{x}_a(t)),
		\end{split}
	\end{align}
	simply Taylor expanding $\gamma_a = \left(1-v_a^2/c^2\right)^{-1/2}$ around $v/c=0$. The next to leading order terms in $\T^{00}$ is thus proportional to $(c^{-1})^0$. From the Virial theorem, or equivalently from \eqref{eq:Relativistic: GMr from v2}, the leading order (Newtonian) term of the gravitational potential scales also as $v^2$, and should therefore also be included. This concludes point \ref{enu:HowToExpandI2:ExpandT}., the relativistic correction of $\T^{\mu\nu}$. Finally, to leading order $\T\ind{^k_k}$ is the point particle tensor \eqref{eq:pp EnergyMomentumTensor}
	\begin{align}
		{T_{\acrshort{pp}}}\ind{^k_k} = \sum_{a} \gamma_a m_a v_a^k {v_a}_k \delta^{(3)}(\tvec{x}-\tvec{x}_a(t)) \simeq \sum_{a} m_a v_a^2 \delta^{(3)}(\tvec{x}-\tvec{x}_a(t)),
	\end{align}
	which is just twice the kinetic energy. Thus, the trace of $\T\ind{^\mu_\mu}$ part of the quadrupole reads
	\begin{align}
		\begin{split}
			\text{Tr. part of }I^{ij}_{(2)} & = \int \dd[3]{x} \left(\T^{00} + \T\ind{^k_k}\right) \left[x^i x^j\right]_\text{\acrshort{stf}} \\
			& = \left[n^in^j\right]_\text{\acrshort{stf}}\sum_{a} m_a c^2 r_a^2\left( 1 + \frac{3}{2}\frac{v^2_a}{c^2} - \sum_{b>a} \frac{Gm_b}{ c^2 \abs{\tvec{x}_a-\tvec{x}_b}} \right) \\
			& = \mu c^2 r^2 \left( 1 + \frac{3}{2}(1-3\eta)\frac{v^2}{c^2} - (1-2\eta) \frac{GM}{rc^2} \right) \left[n^in^j\right]_\text{\acrshort{stf}} \\
			& = \mu c^2 r^2 \left( 1 + \frac{1}{2}(1-5\eta)\frac{v^2}{c^2} \right) \left( n^in^j-\frac{1}{3}\delta^{ij} \right).
		\end{split}
	\end{align}
	
	Since the time dependent part ($n^i$) is equivalent to the first order term of the quadrupole formula, the third derivative of the expression correct to first order in $(v^2/c^2)^1$ can be inferred directly from \eqref{eq:Quadrupole:firstOrder:thirdDerivative}
	\begin{align}
		\dv[3]{I^{ij}_{(2)}}{(ct)} = \frac{2^2\mu r^2 \omega^3}{c} \left( 1 + \frac{1-5\eta}{2}\frac{v^2}{c^2} \right) \begin{pmatrix}
			\sin(2\omega t) & -\cos(2\omega t) & 0 \\
			-\cos(2\omega t) & -\sin(2\omega t) & 0 \\
			0 & 0 & 0
		\end{pmatrix}.
	\end{align}
	
	For the $p=0$ terms this only leaves the $\partial_0 \T^{0k}$ term in line three of \eqref{eq:EnergyMultipole:allOrders}. The last line containing $\partial_0^2\T^{ij}$ will not contribute as it scales as $(\partial_0^2\T^{ij})/\T^{00} \sim c^{-4}$. To leading order also this energy-momentum tensor component is the free point particle tensor, and thus
	\begin{align}
		\text{$0k$-part of } I^{ij}_{(2)}(t) = - \frac{4}{3} \int \dd[3]{x} \partial_0 \sum_a m_a c v_a^k x_k \delta^{(3)}(\tvec{x}-\tvec{x}_a(t)) \left[x^i x^j\right]_\text{\acrshort{stf}}.
	\end{align}
	For circular motion $\tvec{v}_a = \pm \omega r_a \begin{pmatrix} -\sin(\omega t), & \cos(\omega t), & 0 \end{pmatrix}$, and is thus orthogonal to $\tvec{x}_a$: $\tvec{v}_a \tvec{\cdot} \tvec{x}_a = 0$. Therefore the leading order contribution of the $\T^{0k}$-part of $I^L$ is $0$.\\
	
	To finish point \ref{enu:HowToExpandI2:ExpandIL}. only accounting for the $p=1$ term remains. This term follows as
	\begin{align}
		\nonumber \text{$p=1$ part of } I^{ij}_{(2)} & = \frac{5!!}{(2)!!(7)!!} \left(1+\frac{8}{3}\right) \int\dd[3]{x} \partial_0^2 \T^{00}(t,\tvec{x}) r^2\left[ x^i x^j \right]_\text{\acrshort{stf}} \\
		\nonumber & = \frac{11}{42} \sum_a m_a c^2 r_a^4 \partial_0^2 \left[ n^i n^j \right]_\text{\acrshort{stf}} \\
		\nonumber & = \frac{11}{2^2\cdot3\cdot7} \mu(1-3\eta) r^4 \dv[2]{}{t} \begin{pmatrix}
			\frac{1}{3} + \cos(2\omega t) & \sin(2\omega t) & 0 \\
			\sin(2\omega t) & \frac{1}{3} - \cos(2\omega t) & 0 \\
			0 & 0 & -\frac{2}{3}
		\end{pmatrix} \\
		& = -\frac{11}{3\cdot7} \mu (1-3\eta) r^4 \omega^2 \begin{pmatrix}
			\cos(2\omega t) & \sin(2\omega t) & 0 \\
			\sin(2\omega t) & - \cos(2\omega t) & 0 \\
			0 & 0 & 0
		\end{pmatrix},
	\end{align}
	\begin{align}
	    \dv[3]{I^{ij}_{(2)}}{(ct)} = \frac{2^2 \mu r^2 \omega^3}{c} \left( -\frac{2\cdot11}{3\cdot7}(1-3\eta) \frac{v^2}{c^2} \right) \begin{pmatrix}
			\sin(2\omega t) & -\cos(2\omega t) & 0 \\
			-\cos(2\omega t) & -\sin(2\omega t) & 0 \\
			0 & 0 & 0
		\end{pmatrix}.
	\end{align}
	
	This leaves the totally 1\acrshort{pn} correct third derivative of the quadrupole moment
	\begin{align}\label{eq:ThirdDerivativeQuad:MixedROmega}
		\dv[3]{I^{ij}_{(2)}}{(ct)} = \frac{2^2\mu r^2\omega^3}{c} \left( 1 - \frac{23-27\eta}{42}\frac{v^2}{c^2}  \right) \begin{pmatrix}
			\sin(2\omega t) & -\cos(2\omega t) & 0 \\
			-\cos(2\omega t) & -\sin(2\omega t) & 0 \\
			0 & 0 & 0
		\end{pmatrix}.
	\end{align}
	
	Here the subtlety of point \ref{enu:HowToExpandI2:ExpandOmega}., corrections to the equations of motion, enters. At this stage equations \eqref{eq:Relativistic: omega}-\eqref{eq:Relativistic: v2} must be used to convert between $r$, $\omega$, and $v$. One might expect at this point to use $\omega r = v$ and \eqref{eq:Relativistic: v2 from GMomega 23} to convert the last factor of $\omega=v^3/GM$, but this is not the case. Recalling that $v$ is just a proxy variable for the frequency, one should expand the flux in terms of $\omega$, and perhaps change $\omega$ to $v=(GM\omega)^{2/3}$.
	
	Doing that
	\begin{align*}
		\frac{2^2\mu r^2 \omega^3}{c} & = \frac{2^2\mu}{c} \omega v^2 = \frac{2^2\mu}{c} \omega (GM\omega)^{2/3} \left\{ 1- \left(2 - \frac{2}{3}\eta\right) \frac{(GM\omega)^{2/3}}{c^2} \right\} \\
		& = \frac{4\mu}{GMc} v^5 \left\{ 1 - \left(2 - \frac{2}{3}\eta\right) \frac{v^2}{c^2} \right\}.
	\end{align*}
	Inserting this into \eqref{eq:ThirdDerivativeQuad:MixedROmega}, discarding terms $\order{\frac{v^4}{c^4}}$, provides the final result
	\begin{align}
		&\dv[3]{I^{ij}_{(2)}}{(ct)} = \frac{2^2\eta}{G}\frac{v^5}{c} \left( 1 - \frac{107-5\cdot11\eta}{2\cdot3\cdot7}\frac{v^2}{c^2}  \right) \begin{pmatrix}
			\sin(2\omega t) & -\cos(2\omega t) & 0 \\
			-\cos(2\omega t) & -\sin(2\omega t) & 0 \\
			0 & 0 & 0
		\end{pmatrix}, \\
		\Rightarrow \hspace{3pt} &\dv[3]{I^{ij}_{(2)}}{(ct)} \dv[3]{{I_{(2)}}_{ij} }{(ct)} = \frac{2^5\eta^2}{G^2}\frac{v^{10}}{c^2} \left( 1 - \frac{107-5\cdot11 \eta}{3\cdot7} \frac{v^2}{c^2} \right).
	\end{align}
	And thus the energy flux from the quadrupole at next to leading order is
	\begin{align}\label{eq:Flux:Quad:1PN}
		\F_{(2)}^\text{quad} = \frac{2^5\eta^2}{5G}\frac{v^{10}}{c^5} \left( 1 - \frac{107-5\cdot11 \eta}{3\cdot7} \frac{v^2}{c^2} \right) = \F_\text{Newt} \left( 1 - \frac{107-5\cdot11 \eta}{3\cdot7} \frac{v^2}{c^2} \right).
	\end{align}
\subsection{The total 1PN energy flux}
	The total energy flux correct to 1\acrshort{pn} is then the sum of \eqref{eq:Flux:0PN}, \eqref{eq:Flux:Oct}, \eqref{eq:Flux:CurrQuad}, and \eqref{eq:Flux:Quad:1PN},
	\begin{empheq}[box=\widefbox]{align} \label{eq:1PN:Flux}
		\F = \frac{32}{5}\frac{\eta^2}{G} \frac{v^{10}}{c^5} \left\{ 1 - \left(\frac{1247}{336} + \frac{35}{12}\eta \right)\frac{v^2}{c^2} + \order{\frac{1}{c^3}} \right\}.
	\end{empheq}
	Which is exactly the flux \eqref{eq:Flux:Expansion:Gen} presented in \cref{chap:waveform}.
	
    To get some perspective, lets consider the Earth-Moon system again. According to \eqref{eq:1PN:Flux}, the energy flux due to \acrshort{gw}s is
    \begin{align}
        \F_{\Moon} = 6.03 \cdot 10^{-4} \left( 1 - 4.37 \cdot 10^{-11} \right) \text{J}/\text{s}
    \end{align}
	
	Using equation \eqref{eq:waveform:omegaOfTau} for the change in orbital frequency, and relation \eqref{eq:Relativistic: GMr from GMomega23} of $r$ and $\omega$, the time evolution of the relative separation due to \acrshort{gw} emission is found to be
	\begin{subequations}
	\begin{align}
	    r_{\Moon}(\tau) & = \frac{GM}{(GM\omega)^{2/3}} \left\{ 1- \left(1-\frac{\eta}{3}\right)\frac{(GM\omega)^{2/3}}{c^2} + \order{(GM\omega)^{3/2}} \right\} \\
	    & \approx \frac{4(GM)^{1/3}}{5^{2/3}} \left( \frac{5G\M}{c^3} \right)^{\frac{5}{12}} \tau^{1/4}, \\
	    \Rightarrow \quad\dot{r}_{\Moon}(\tau) & \approx -\frac{(GM)^{1/3}}{5^{2/3}} \left( \frac{5G\M}{c^3} \right)^{\frac{5}{12}} \tau^{-3/4} = \frac{G^3M^3 \eta}{5c^5} r^{-3} \\
	    & \approx 1.14\cdot 10^{-27} \frac{\text{m}}{{s}} \approx 3.60 \cdot 10^{-17} \frac{\text{mm}}{\text{year}}.
	\end{align}
	\end{subequations}
	This is, not surprisingly, very slowly. At this rate, it would take $\sim 2.8\cdot10^{16}$ years until the effect would be in the range of millimetres.
	
\chapter{Discussion and conclusion} \label{chap:conclusion}
    We have now seen how the 1\acrshort{pn} \acrlong{gw}form \eqref{eq:waveform:Phi-tau} can be obtained from the 1\acrshort{pn} energy \eqref{eq:1PN:Energy} and flux \eqref{eq:1PN:Flux}, assuming \glspl{QCO} and separation of scales. We have also demonstrated how the 1\acrshort{pn} energy can be determined using Feynman diagrams (\cref{chap:energy}), and how the 1\acrshort{pn} flux can be computed using multipoles (\cref{chap:flux}), all based on an \acrlong{eft} of gravity as a gauge field (\cref{chap:fieldtheory}).

    In his text \cite{Porto2016}, Porto claimed

    ``[...] that adopting an \acrshort{eft} framework, when possible, greatly simplifies the computations and provides the required intuition for `physical understanding'.''

    In my own experience, the computations do not seem all that more simplified compared to the more traditional geometrical approach (see \textcite{Maggiore:VolumeI} for an outline, or \textcite{Blanchet2014} for more details). For someone without a deep background in \acrshort{eft}, like a master's student, any simplification of the calculation is outweighed by the work of familiarizing oneself with standard results and conventions from \acrshort{qft}.

    Of course, if one \emph{does} have a deep familiarity with \acrshort{eft}s, the \gls{field theorist} approach presented in this thesis is a great way to transfer those skills to \acrlong{gw} physics. These are after all powerful tools for handling perturbative phenomena. The use of Feynman diagrams makes the terms in the perturbation series more manageable, and can give intuition for what kind of physical effects the different terms account for. In this sense, the computations can be considered to have been `simplified', and provided the intuition for `physical understanding'.

    It is also possible that the \glspl{field theorist}' approach becomes significantly simpler than the \glspl{relativist}' approach at higher \acrshort{pn} orders. In order to verify this, I would need to compute higher order corrections.

    Even so, for \glspl{relativist}, some familiarity with the field theory way of thinking of gravitational dynamics is helpful for deepening their understanding of gravity. As Feynman once said at a Cornell lecture during his gravity phase:

    ``Every theoretical physicist who is any good knows six or seven different theoretical representations for exactly the same physics. He knows that they are all equivalent, and that nobody is ever going to be able to decide which one is right at that level, but he keeps them in his head, hoping that they will give him different ideas for guessing.'' - \textcite{CharOfPhysical}

    For such reasons, it is valuable to have alternative ways of thinking about gravity and \acrlong{gw}s. Some extensions of, and alternative theories for, Einstein's theory of gravity might present themselves more naturally in the language of field theory, rather than differential geometry. E.g. quantum `loop' corrections of gravity \cite{QuntGrav}. With \acrlong{gw} data imposing some of the strongest constraints on gravity theories, having an alternative route for translating theories of gravity to \acrlong{gw}forms is a useful tool.

    In conclusion, this \acrlong{eft} approach to computing \acrlong{gw}forms will probably not replace the more traditional \gls{relativist} approach as the standard or introductory way of deriving these results any time soon. As a method it is however worth developing, as it provides an alternative perspective on the physics of \acrlong{gw}s. It might also provide a shorter path for some alternative theories of gravity to testable predictions, and can be used by physicists with a heavier \acrlong{qft} background to simplify the computations of such theories.

\printbibliography

@book{Maggiore:VolumeI,
	author = "Maggiore, Michele",
	title = "{Gravitational Waves. Vol. 1: Theory and Experiments}",
	isbn = {9780198570745},
	publisher = "Oxford University Press",
	series = "Oxford Master Series in Physics",
	year = "2007",
	doi = {10.1093/acprof:oso/9780198570745.001.0001}
}

@book{Feynman:GravityLectures,
	author = "Feynman, R.P.",
	editor = "Morinigo, F.B. and Wagner, W.G. and Hatfield, B.",
	title = "{Feynman lectures on gravitation}",
	month = "12",
	year = "1996",
	doi = "10.1201/9780429502859",
	ISBN = "9780429502859"
}

@book{Gron,
	author = "Gr\o{}n, \O{}yvind and Hervik, Sigbj\o{}r",
	title = {Einstein's General Theory of Relativity},
	publisher = {Springer-Verlag},
	year = {2007},
	isbn = {9780387691992},
	doi = {10.1007/978-0-387-69200-5},
	% language = {english},
	location = {New York},
	% pagetotal = {538},
}

@book{Kachelriess:2017cfe,
	author = "Kachelrie\ss{}, Michael",
	title = "{Quantum Fields}: {From the Hubble to the Planck Scale}",
	isbn = "9780198802877",
	publisher = "Oxford University Press",
	series = "Oxford Graduate Texts",
	month = "10",
	year = "2017",
	doi = {10.1093/oso/9780198802877.001.0001},
	%url = {http://web.phys.ntnu.no/~mika/QF.html}
}

@book{SpesRel,
	author = {Gourgoulhon, Éric},
	title = {Special Relativity in General Frames},
	publisher = {Springer-Verlag},
	year = {2013},
	isbn = {9783642372759},
	doi = {10.1007/978-3-642-37276-6},
	language = {english},
	location = {Berlin Heidelberg},
	% pagetotal = {784},
	series = {Graduate Texts in Physics},
	% origlanguage = {french}
}

@book{GoldsteinMech,
    author = {Goldstein, Herbert and Poole, Charles and Safko, John},
    year = {2014},
    month = {06},
    publisher = {Pearson Education Limited},
    % pagetotal = {638},
    title = {Classical Mechanics (3rd Edition)},
    isbn = {9781292026558}
}

@book{NoNonsenseQFT,
  title={No-Nonsense Quantum Field Theory: A Student-Friendly Introduction},
  author={Schwichtenberg, Jakob},
  url={https://nononsensebooks.com/qft/},
  year={2020},
  publisher={No-Nonsense Books},
  isbn = {9783948763015}
}

@book{CharOfPhysical,
    author = {Feynman, Richard},
    title = "{The Character of Physical Law}",
    publisher = {The MIT Press},
    year = {2017},
    month = {03},
    isbn = {9780262533416},
    doi = {https://doi.org/10.7551/mitpress/11068.001.0001},
}

@book{relElectrodynamics,
  title={Electrodynamics},
  author={Page, L. and Adams, N.I.},
  isbn={9780598854773},
  lccn={40012238},
  series={Dover books on physics and mathematical physics},
  url={https://books.google.no/books?id=o7\_VAAAAMAAJ},
  year={1940},
  publisher={D. Van Nostrand Company, Incorporated}
}

@book{physicsFormula,
    place = {Cambridge}, 
    title = {The Cambridge Handbook of Physics Formulas},
    DOI = {10.1017/CBO9780511755828}, 
    publisher = {Cambridge University Press}, 
    author = {Woan, Graham}, 
    year = {2000},
    isbn = {9780521575072}
}

@article{Porto2016,
	title={The effective field theorist’s approach to gravitational dynamics},
	volume={633},
	ISSN={0370-1573},
	%url={http://dx.doi.org/10.1016/j.physrep.2016.04.003},
	DOI={10.1016/j.physrep.2016.04.003},
	journal={Physics Reports},
	publisher={Elsevier BV},
	author={Porto, Rafael A.},
	year={2016},
	month={05},
	pages={1–104}
}

@misc{Goldberger:LesHouches,
	author = {Goldberger, Walter},
	year = {2007},
	month = {02},
%	pages = {},
	title = {Les Houches Lectures on Effective Field Theories and Gravitational Radiation},
	eprint = "hep-ph/0701129",
    archivePrefix = "arXiv",
    %primaryClass = "physics.hep-ph",
}

@article{Goldberger:EFT,
   title={Effective field theory of gravity for extended objects},
   volume={73},
   ISSN={1550-2368},
   % url={http://dx.doi.org/10.1103/PhysRevD.73.104029},
   DOI={10.1103/physrevd.73.104029},
   number={10},
   journal={Physical Review D},
   publisher={American Physical Society (APS)},
   author={Goldberger, Walter D. and Rothstein, Ira Z.},
   year={2006},
   month={05}
}

@article{Ross2012Multipole,
	author = {Ross, Andreas},
	year = {2012},
	month = {02},
	pages = {},
	title = {Multipole expansion at the level of the action},
	volume = {85},
	journal = {Physical Review D},
	doi = {10.1103/PhysRevD.85.125033}
}

@article{Thorn1980Multipole,
  title = {Multipole expansions of gravitational radiation},
  author = {Thorne, Kip S.},
  journal = {Rev. Mod. Phys.},
  volume = {52},
  issue = {2},
  pages = {299--339},
  numpages = {0},
  year = {1980},
  month = {04},
  publisher = {American Physical Society},
  doi = {10.1103/RevModPhys.52.299},
  %url = {https://link.aps.org/doi/10.1103/RevModPhys.52.299}
}

@article{DamourMultipole,
  title = {Multipole analysis for electromagnetism and linearized gravity with irreducible Cartesian tensors},
  author = {Damour, T. and Iyer, B. R.},
  journal = {Phys. Rev. D},
  volume = {43},
  issue = {10},
  pages = {3259--3272},
  numpages = {0},
  year = {1991},
  month = {05},
  publisher = {American Physical Society},
  doi = {10.1103/PhysRevD.43.3259},
  % url = {https://link.aps.org/doi/10.1103/PhysRevD.43.3259}
}

@article{QuntGrav,
    author = "Bjerrum-Bohr, N. Emil J. and Damgaard, Poul H. and Plant\'e, Ludovic and Vanhove, Pierre",
    title = "{Classical Gravity from Loop Amplitudes}",
    eprint = "2104.04510",
    archivePrefix = "arXiv",
    primaryClass = "hep-th",
    reportNumber = "IPhT-t21/015, CERN-TH-2021-052",
    month = "4",
    year = "2021"
}

@article{EinsteinInfeldHoffmann,
    ISSN = {0003486X},
	URL = {http://www.jstor.org/stable/1968714},
	author = {A. Einstein and L. Infeld and B. Hoffmann},
	journal = {Annals of Mathematics},
	number = {1},
	pages = {65--100},
	publisher = {Annals of Mathematics},
	title = {The Gravitational Equations and the Problem of Motion},
	year = {1938},
	volume = {39}
}

@article{FeynmanReview,
    author = "Di Mauro, Marco and Esposito, Salvatore and Naddeo, Adele",
    title = "{A roadmap for Feynman's adventures in the land of gravitation}",
    eprint = "2102.11220",
    archivePrefix = "arXiv",
    primaryClass = "physics.hist-ph",
    month = "02",
    year = "2021"
}

@article{Szabados2009,
    author = {Szabados, László B.},
    title = {Quasi-Local Energy-Momentum and Angular Momentum in General Relativity},
    journal = {Living Reviews in Relativity},
    year = {2009},
    month = {06},
    day = {19},
    doi = {https://doi.org/10.12942/lrr-2009-4},
    volume = {12},
    number = {4}
}

@misc{deharo2021noethers,
      title={Noether's Theorems and Energy in General Relativity}, 
      author={Sebastian De Haro},
      year={2021},
      eprint={2103.17160},
      archivePrefix={arXiv},
      primaryClass={physics.hist-ph}
}

@article{Fierz:1939ix,
    author = "Fierz, M. and Pauli, W.",
    title = "{On relativistic wave equations for particles of arbitrary spin in an electromagnetic field}",
    doi = "10.1098/rspa.1939.0140",
    journal = "Proc. Roy. Soc. Lond. A",
    volume = "173",
    pages = "211--232",
    year = "1939"
}

@article{Padmanabhan_2008,
   title={FROM GRAVITONS TO GRAVITY: MYTHS AND REALITY},
   volume={17},
   ISSN={1793-6594},
  % url={http://dx.doi.org/10.1142/S0218271808012085},
   DOI={10.1142/s0218271808012085},
   number={03n04},
   journal={International Journal of Modern Physics D},
   publisher={World Scientific Pub Co Pte Lt},
   author={Padmanabhan, T.},
   year={2008},
   month={03},
 %  pages={367–398}
}

@article{FirstFeynmanDiagram,
  title = {Space-Time Approach to Quantum Electrodynamics},
  author = {Feynman, R. P.},
  journal = {Phys. Rev.},
  volume = {76},
  issue = {6},
  pages = {769--789},
%  numpages = {0},
  year = {1949},
  month = {09},
  publisher = {American Physical Society},
  doi = {10.1103/PhysRev.76.769},
%  url = {https://link.aps.org/doi/10.1103/PhysRev.76.769}
}

@article{LunarLaserRanging,
	doi = {10.1086/596748},
	url = {https://doi.org/10.1086/596748},
	year = 2009,
	month = {01},
	publisher = {{IOP} Publishing},
	volume = {121},
	number = {875},
	pages = {29--40},
	author = {J. B. R. Battat and T. W. Murphy and E. G. Adelberger and B. Gillespie and C. D. Hoyle and R. J. McMillan and E. L. Michelsen and K. Nordtvedt and A. E. Orin and C. W. Stubbs and H. E. Swanson},
	title = {{T}he {A}pache {P}oint {O}bservatory {L}unar {L}aser-ranging {O}peration ({APOLLO}): {T}wo {Y}ears of {M}illimeter-{P}recision {M}easurements of the {E}arth-{M}oon {R}ange1},
	journal = {Publications of the Astronomical Society of the Pacific}
}

@article{JPL_model,
	doi = {10.3847/1538-3881/abd414},
	url = {https://doi.org/10.3847/1538-3881/abd414},
	year = 2021,
	month = {02},
	publisher = {American Astronomical Society},
	volume = {161},
	number = {3},
	pages = {105},
	author = {Ryan S. Park and William M. Folkner and James G. Williams and Dale H. Boggs},
	title = {The {JPL} Planetary and Lunar Ephemerides {DE}440 and {DE}441},
	journal = {The Astronomical Journal}
}

@article{Arun2008higherorder,
    author = "Arun, K. G. and Buonanno, Alessandra and Faye, Guillaume and Ochsner, Evan",
    title = "{Higher-order spin effects in the amplitude and phase of gravitational waveforms emitted by inspiraling compact binaries: Ready-to-use gravitational waveforms}",
    eprint = "0810.5336",
    archivePrefix = "arXiv",
    primaryClass = "gr-qc",
    doi = "10.1103/PhysRevD.79.104023",
    journal = "Phys. Rev. D",
    volume = "79",
%    pages = "104023",
    year = "2009",
    note = "[Erratum: Phys.Rev.D 84, 049901 (2011)]"
}

@article{Blanchet2014,
    author = {Blanchet, Luc},
    year = {2014},
    month = {12},
    title = {{G}ravitational {R}adiation from {P}ost-{N}ewtonian {S}ources and {I}nspiralling {C}ompact {B}inaries},
    journal = {Living Reviews in Relativity},
    volume = {17},
    issue = {2},
    doi = {10.12942/lrr-2014-2}
}

@article{SPAarticle,
    author = "Cutler, Curt and Flanagan, Eanna E.",
    title = "{Gravitational waves from merging compact binaries: How accurately can one extract the binary's parameters from the inspiral wave form?}",
    eprint = "gr-qc/9402014",
    archivePrefix = "arXiv",
    reportNumber = "GRP-369",
    doi = "10.1103/PhysRevD.49.2658",
    journal = "Phys. Rev. D",
    volume = "49",
    pages = "2658--2697",
    year = "1994"
}

@article{CompOfPNandNR,
    author = "Borhanian, Ssohrab and Arun, K. G. and Pfeiffer, Harald P. and Sathyaprakash, B. S.",
    title = "{Comparison of post-Newtonian mode amplitudes with numerical relativity simulations of binary black holes}",
    eprint = "1901.08516",
    archivePrefix = "arXiv",
    primaryClass = "gr-qc",
    doi = "10.1088/1361-6382/ab6a21",
    journal = "Class. Quant. Grav.",
    volume = "37",
    number = "6",
    pages = "065006",
    year = "2020"
}

@article{5PN,
author = {Foffa, Stefano},
year = {2013},
month = {09},
pages = {},
title = {{G}ravitating binaries at 5{PN} in the post-{M}inkowskian approximation},
volume = {89},
journal = {Physical Review D},
doi = {10.1103/PhysRevD.89.024019}
}

@article{FirstGW,
  title = {Observation of Gravitational Waves from a Binary Black Hole Merger},
  author = {Abbott, B. P. and Abbott, R. and Abbott, T. D. and Abernathy, M. R. and Acernese, F. and Ackley, K. and Adams, C. and Adams, T. and Addesso, P. and Adhikari, R. X. and Adya, V. B. and Affeldt, C. and Agathos, M. and Agatsuma, K. and Aggarwal, N. and Aguiar, O. D. and Aiello, L. and Ain, A. and Ajith, P. and Allen, B. and Allocca, A. and Altin, P. A. and Anderson, S. B. and Anderson, W. G. and Arai, K. and Arain, M. A. and Araya, M. C. and Arceneaux, C. C. and Areeda, J. S. and Arnaud, N. and Arun, K. G. and Ascenzi, S. and Ashton, G. and Ast, M. and Aston, S. M. and Astone, P. and Aufmuth, P. and Aulbert, C. and Babak, S. and Bacon, P. and Bader, M. K. M. and Baker, P. T. and Baldaccini, F. and Ballardin, G. and Ballmer, S. W. and Barayoga, J. C. and Barclay, S. E. and Barish, B. C. and Barker, D. and Barone, F. and Barr, B. and Barsotti, L. and Barsuglia, M. and Barta, D. and Bartlett, J. and Barton, M. A. and Bartos, I. and Bassiri, R. and Basti, A. and Batch, J. C. and Baune, C. and Bavigadda, V. and Bazzan, M. and Behnke, B. and Bejger, M. and Belczynski, C. and Bell, A. S. and Bell, C. J. and Berger, B. K. and Bergman, J. and Bergmann, G. and Berry, C. P. L. and Bersanetti, D. and Bertolini, A. and Betzwieser, J. and Bhagwat, S. and Bhandare, R. and Bilenko, I. A. and Billingsley, G. and Birch, J. and Birney, R. and Birnholtz, O. and Biscans, S. and Bisht, A. and Bitossi, M. and Biwer, C. and Bizouard, M. A. and Blackburn, J. K. and Blair, C. D. and Blair, D. G. and Blair, R. M. and Bloemen, S. and Bock, O. and Bodiya, T. P. and Boer, M. and Bogaert, G. and Bogan, C. and Bohe, A. and Bojtos, P. and Bond, C. and Bondu, F. and Bonnand, R. and Boom, B. A. and Bork, R. and Boschi, V. and Bose, S. and Bouffanais, Y. and Bozzi, A. and Bradaschia, C. and Brady, P. R. and Braginsky, V. B. and Branchesi, M. and Brau, J. E. and Briant, T. and Brillet, A. and Brinkmann, M. and Brisson, V. and Brockill, P. and Brooks, A. F. and Brown, D. A. and Brown, D. D. and Brown, N. M. and Buchanan, C. C. and Buikema, A. and Bulik, T. and Bulten, H. J. and Buonanno, A. and Buskulic, D. and Buy, C. and Byer, R. L. and Cabero, M. and Cadonati, L. and Cagnoli, G. and Cahillane, C. and Bustillo, J. Calder\'on and Callister, T. and Calloni, E. and Camp, J. B. and Cannon, K. C. and Cao, J. and Capano, C. D. and Capocasa, E. and Carbognani, F. and Caride, S. and Diaz, J. Casanueva and Casentini, C. and Caudill, S. and Cavagli\`a, M. and Cavalier, F. and Cavalieri, R. and Cella, G. and Cepeda, C. B. and Baiardi, L. Cerboni and Cerretani, G. and Cesarini, E. and Chakraborty, R. and Chalermsongsak, T. and Chamberlin, S. J. and Chan, M. and Chao, S. and Charlton, P. and Chassande-Mottin, E. and Chen, H. Y. and Chen, Y. and Cheng, C. and Chincarini, A. and Chiummo, A. and Cho, H. S. and Cho, M. and Chow, J. H. and Christensen, N. and Chu, Q. and Chua, S. and Chung, S. and Ciani, G. and Clara, F. and Clark, J. A. and Cleva, F. and Coccia, E. and Cohadon, P.-F. and Colla, A. and Collette, C. G. and Cominsky, L. and Constancio, M. and Conte, A. and Conti, L. and Cook, D. and Corbitt, T. R. and Cornish, N. and Corsi, A. and Cortese, S. and Costa, C. A. and Coughlin, M. W. and Coughlin, S. B. and Coulon, J.-P. and Countryman, S. T. and Couvares, P. and Cowan, E. E. and Coward, D. M. and Cowart, M. J. and Coyne, D. C. and Coyne, R. and Craig, K. and Creighton, J. D. E. and Creighton, T. D. and Cripe, J. and Crowder, S. G. and Cruise, A. M. and Cumming, A. and Cunningham, L. and Cuoco, E. and Canton, T. Dal and Danilishin, S. L. and D'Antonio, S. and Danzmann, K. and Darman, N. S. and Da Silva Costa, C. F. and Dattilo, V. and Dave, I. and Daveloza, H. P. and Davier, M. and Davies, G. S. and Daw, E. J. and Day, R. and De, S. and DeBra, D. and Debreczeni, G. and Degallaix, J. and De Laurentis, M. and Del\'eglise, S. and Del Pozzo, W. and Denker, T. and Dent, T. and Dereli, H. and Dergachev, V. and DeRosa, R. T. and De Rosa, R. and DeSalvo, R. and Dhurandhar, S. and D\'{\i}az, M. C. and Di Fiore, L. and Di Giovanni, M. and Di Lieto, A. and Di Pace, S. and Di Palma, I. and Di Virgilio, A. and Dojcinoski, G. and Dolique, V. and Donovan, F. and Dooley, K. L. and Doravari, S. and Douglas, R. and Downes, T. P. and Drago, M. and Drever, R. W. P. and Driggers, J. C. and Du, Z. and Ducrot, M. and Dwyer, S. E. and Edo, T. B. and Edwards, M. C. and Effler, A. and Eggenstein, H.-B. and Ehrens, P. and Eichholz, J. and Eikenberry, S. S. and Engels, W. and Essick, R. C. and Etzel, T. and Evans, M. and Evans, T. M. and Everett, R. and Factourovich, M. and Fafone, V. and Fair, H. and Fairhurst, S. and Fan, X. and Fang, Q. and Farinon, S. and Farr, B. and Farr, W. M. and Favata, M. and Fays, M. and Fehrmann, H. and Fejer, M. M. and Feldbaum, D. and Ferrante, I. and Ferreira, E. C. and Ferrini, F. and Fidecaro, F. and Finn, L. S. and Fiori, I. and Fiorucci, D. and Fisher, R. P. and Flaminio, R. and Fletcher, M. and Fong, H. and Fournier, J.-D. and Franco, S. and Frasca, S. and Frasconi, F. and Frede, M. and Frei, Z. and Freise, A. and Frey, R. and Frey, V. and Fricke, T. T. and Fritschel, P. and Frolov, V. V. and Fulda, P. and Fyffe, M. and Gabbard, H. A. G. and Gair, J. R. and Gammaitoni, L. and Gaonkar, S. G. and Garufi, F. and Gatto, A. and Gaur, G. and Gehrels, N. and Gemme, G. and Gendre, B. and Genin, E. and Gennai, A. and George, J. and Gergely, L. and Germain, V. and Ghosh, Abhirup and Ghosh, Archisman and Ghosh, S. and Giaime, J. A. and Giardina, K. D. and Giazotto, A. and Gill, K. and Glaefke, A. and Gleason, J. R. and Goetz, E. and Goetz, R. and Gondan, L. and Gonz\'alez, G. and Castro, J. M. Gonzalez and Gopakumar, A. and Gordon, N. A. and Gorodetsky, M. L. and Gossan, S. E. and Gosselin, M. and Gouaty, R. and Graef, C. and Graff, P. B. and Granata, M. and Grant, A. and Gras, S. and Gray, C. and Greco, G. and Green, A. C. and Greenhalgh, R. J. S. and Groot, P. and Grote, H. and Grunewald, S. and Guidi, G. M. and Guo, X. and Gupta, A. and Gupta, M. K. and Gushwa, K. E. and Gustafson, E. K. and Gustafson, R. and Hacker, J. J. and Hall, B. R. and Hall, E. D. and Hammond, G. and Haney, M. and Hanke, M. M. and Hanks, J. and Hanna, C. and Hannam, M. D. and Hanson, J. and Hardwick, T. and Harms, J. and Harry, G. M. and Harry, I. W. and Hart, M. J. and Hartman, M. T. and Haster, C.-J. and Haughian, K. and Healy, J. and Heefner, J. and Heidmann, A. and Heintze, M. C. and Heinzel, G. and Heitmann, H. and Hello, P. and Hemming, G. and Hendry, M. and Heng, I. S. and Hennig, J. and Heptonstall, A. W. and Heurs, M. and Hild, S. and Hoak, D. and Hodge, K. A. and Hofman, D. and Hollitt, S. E. and Holt, K. and Holz, D. E. and Hopkins, P. and Hosken, D. J. and Hough, J. and Houston, E. A. and Howell, E. J. and Hu, Y. M. and Huang, S. and Huerta, E. A. and Huet, D. and Hughey, B. and Husa, S. and Huttner, S. H. and Huynh-Dinh, T. and Idrisy, A. and Indik, N. and Ingram, D. R. and Inta, R. and Isa, H. N. and Isac, J.-M. and Isi, M. and Islas, G. and Isogai, T. and Iyer, B. R. and Izumi, K. and Jacobson, M. B. and Jacqmin, T. and Jang, H. and Jani, K. and Jaranowski, P. and Jawahar, S. and Jim\'enez-Forteza, F. and Johnson, W. W. and Johnson-McDaniel, N. K. and Jones, D. I. and Jones, R. and Jonker, R. J. G. and Ju, L. and Haris, K. and Kalaghatgi, C. V. and Kalogera, V. and Kandhasamy, S. and Kang, G. and Kanner, J. B. and Karki, S. and Kasprzack, M. and Katsavounidis, E. and Katzman, W. and Kaufer, S. and Kaur, T. and Kawabe, K. and Kawazoe, F. and K\'ef\'elian, F. and Kehl, M. S. and Keitel, D. and Kelley, D. B. and Kells, W. and Kennedy, R. and Keppel, D. G. and Key, J. S. and Khalaidovski, A. and Khalili, F. Y. and Khan, I. and Khan, S. and Khan, Z. and Khazanov, E. A. and Kijbunchoo, N. and Kim, C. and Kim, J. and Kim, K. and Kim, Nam-Gyu and Kim, Namjun and Kim, Y.-M. and King, E. J. and King, P. J. and Kinzel, D. L. and Kissel, J. S. and Kleybolte, L. and Klimenko, S. and Koehlenbeck, S. M. and Kokeyama, K. and Koley, S. and Kondrashov, V. and Kontos, A. and Koranda, S. and Korobko, M. and Korth, W. Z. and Kowalska, I. and Kozak, D. B. and Kringel, V. and Krishnan, B. and Kr\'olak, A. and Krueger, C. and Kuehn, G. and Kumar, P. and Kumar, R. and Kuo, L. and Kutynia, A. and Kwee, P. and Lackey, B. D. and Landry, M. and Lange, J. and Lantz, B. and Lasky, P. D. and Lazzarini, A. and Lazzaro, C. and Leaci, P. and Leavey, S. and Lebigot, E. O. and Lee, C. H. and Lee, H. K. and Lee, H. M. and Lee, K. and Lenon, A. and Leonardi, M. and Leong, J. R. and Leroy, N. and Letendre, N. and Levin, Y. and Levine, B. M. and Li, T. G. F. and Libson, A. and Littenberg, T. B. and Lockerbie, N. A. and Logue, J. and Lombardi, A. L. and London, L. T. and Lord, J. E. and Lorenzini, M. and Loriette, V. and Lormand, M. and Losurdo, G. and Lough, J. D. and Lousto, C. O. and Lovelace, G. and L\"uck, H. and Lundgren, A. P. and Luo, J. and Lynch, R. and Ma, Y. and MacDonald, T. and Machenschalk, B. and MacInnis, M. and Macleod, D. M. and Maga\~na-Sandoval, F. and Magee, R. M. and Mageswaran, M. and Majorana, E. and Maksimovic, I. and Malvezzi, V. and Man, N. and Mandel, I. and Mandic, V. and Mangano, V. and Mansell, G. L. and Manske, M. and Mantovani, M. and Marchesoni, F. and Marion, F. and M\'arka, S. and M\'arka, Z. and Markosyan, A. S. and Maros, E. and Martelli, F. and Martellini, L. and Martin, I. W. and Martin, R. M. and Martynov, D. V. and Marx, J. N. and Mason, K. and Masserot, A. and Massinger, T. J. and Masso-Reid, M. and Matichard, F. and Matone, L. and Mavalvala, N. and Mazumder, N. and Mazzolo, G. and McCarthy, R. and McClelland, D. E. and McCormick, S. and McGuire, S. C. and McIntyre, G. and McIver, J. and McManus, D. J. and McWilliams, S. T. and Meacher, D. and Meadors, G. D. and Meidam, J. and Melatos, A. and Mendell, G. and Mendoza-Gandara, D. and Mercer, R. A. and Merilh, E. and Merzougui, M. and Meshkov, S. and Messenger, C. and Messick, C. and Meyers, P. M. and Mezzani, F. and Miao, H. and Michel, C. and Middleton, H. and Mikhailov, E. E. and Milano, L. and Miller, J. and Millhouse, M. and Minenkov, Y. and Ming, J. and Mirshekari, S. and Mishra, C. and Mitra, S. and Mitrofanov, V. P. and Mitselmakher, G. and Mittleman, R. and Moggi, A. and Mohan, M. and Mohapatra, S. R. P. and Montani, M. and Moore, B. C. and Moore, C. J. and Moraru, D. and Moreno, G. and Morriss, S. R. and Mossavi, K. and Mours, B. and Mow-Lowry, C. M. and Mueller, C. L. and Mueller, G. and Muir, A. W. and Mukherjee, Arunava and Mukherjee, D. and Mukherjee, S. and Mukund, N. and Mullavey, A. and Munch, J. and Murphy, D. J. and Murray, P. G. and Mytidis, A. and Nardecchia, I. and Naticchioni, L. and Nayak, R. K. and Necula, V. and Nedkova, K. and Nelemans, G. and Neri, M. and Neunzert, A. and Newton, G. and Nguyen, T. T. and Nielsen, A. B. and Nissanke, S. and Nitz, A. and Nocera, F. and Nolting, D. and Normandin, M. E. N. and Nuttall, L. K. and Oberling, J. and Ochsner, E. and O'Dell, J. and Oelker, E. and Ogin, G. H. and Oh, J. J. and Oh, S. H. and Ohme, F. and Oliver, M. and Oppermann, P. and Oram, Richard J. and O'Reilly, B. and O'Shaughnessy, R. and Ott, C. D. and Ottaway, D. J. and Ottens, R. S. and Overmier, H. and Owen, B. J. and Pai, A. and Pai, S. A. and Palamos, J. R. and Palashov, O. and Palomba, C. and Pal-Singh, A. and Pan, H. and Pan, Y. and Pankow, C. and Pannarale, F. and Pant, B. C. and Paoletti, F. and Paoli, A. and Papa, M. A. and Paris, H. R. and Parker, W. and Pascucci, D. and Pasqualetti, A. and Passaquieti, R. and Passuello, D. and Patricelli, B. and Patrick, Z. and Pearlstone, B. L. and Pedraza, M. and Pedurand, R. and Pekowsky, L. and Pele, A. and Penn, S. and Perreca, A. and Pfeiffer, H. P. and Phelps, M. and Piccinni, O. and Pichot, M. and Pickenpack, M. and Piergiovanni, F. and Pierro, V. and Pillant, G. and Pinard, L. and Pinto, I. M. and Pitkin, M. and Poeld, J. H. and Poggiani, R. and Popolizio, P. and Post, A. and Powell, J. and Prasad, J. and Predoi, V. and Premachandra, S. S. and Prestegard, T. and Price, L. R. and Prijatelj, M. and Principe, M. and Privitera, S. and Prix, R. and Prodi, G. A. and Prokhorov, L. and Puncken, O. and Punturo, M. and Puppo, P. and P\"urrer, M. and Qi, H. and Qin, J. and Quetschke, V. and Quintero, E. A. and Quitzow-James, R. and Raab, F. J. and Rabeling, D. S. and Radkins, H. and Raffai, P. and Raja, S. and Rakhmanov, M. and Ramet, C. R. and Rapagnani, P. and Raymond, V. and Razzano, M. and Re, V. and Read, J. and Reed, C. M. and Regimbau, T. and Rei, L. and Reid, S. and Reitze, D. H. and Rew, H. and Reyes, S. D. and Ricci, F. and Riles, K. and Robertson, N. A. and Robie, R. and Robinet, F. and Rocchi, A. and Rolland, L. and Rollins, J. G. and Roma, V. J. and Romano, J. D. and Romano, R. and Romanov, G. and Romie, J. H. and Rosi\ifmmode \acute{n}\else \'{n}\fi{}ska, D. and Rowan, S. and R\"udiger, A. and Ruggi, P. and Ryan, K. and Sachdev, S. and Sadecki, T. and Sadeghian, L. and Salconi, L. and Saleem, M. and Salemi, F. and Samajdar, A. and Sammut, L. and Sampson, L. M. and Sanchez, E. J. and Sandberg, V. and Sandeen, B. and Sanders, G. H. and Sanders, J. R. and Sassolas, B. and Sathyaprakash, B. S. and Saulson, P. R. and Sauter, O. and Savage, R. L. and Sawadsky, A. and Schale, P. and Schilling, R. and Schmidt, J. and Schmidt, P. and Schnabel, R. and Schofield, R. M. S. and Sch\"onbeck, A. and Schreiber, E. and Schuette, D. and Schutz, B. F. and Scott, J. and Scott, S. M. and Sellers, D. and Sengupta, A. S. and Sentenac, D. and Sequino, V. and Sergeev, A. and Serna, G. and Setyawati, Y. and Sevigny, A. and Shaddock, D. A. and Shaffer, T. and Shah, S. and Shahriar, M. S. and Shaltev, M. and Shao, Z. and Shapiro, B. and Shawhan, P. and Sheperd, A. and Shoemaker, D. H. and Shoemaker, D. M. and Siellez, K. and Siemens, X. and Sigg, D. and Silva, A. D. and Simakov, D. and Singer, A. and Singer, L. P. and Singh, A. and Singh, R. and Singhal, A. and Sintes, A. M. and Slagmolen, B. J. J. and Smith, J. R. and Smith, M. R. and Smith, N. D. and Smith, R. J. E. and Son, E. J. and Sorazu, B. and Sorrentino, F. and Souradeep, T. and Srivastava, A. K. and Staley, A. and Steinke, M. and Steinlechner, J. and Steinlechner, S. and Steinmeyer, D. and Stephens, B. C. and Stevenson, S. P. and Stone, R. and Strain, K. A. and Straniero, N. and Stratta, G. and Strauss, N. A. and Strigin, S. and Sturani, R. and Stuver, A. L. and Summerscales, T. Z. and Sun, L. and Sutton, P. J. and Swinkels, B. L. and Szczepa\ifmmode \acute{n}\else \'{n}\fi{}czyk, M. J. and Tacca, M. and Talukder, D. and Tanner, D. B. and T\'apai, M. and Tarabrin, S. P. and Taracchini, A. and Taylor, R. and Theeg, T. and Thirugnanasambandam, M. P. and Thomas, E. G. and Thomas, M. and Thomas, P. and Thorne, K. A. and Thorne, K. S. and Thrane, E. and Tiwari, S. and Tiwari, V. and Tokmakov, K. V. and Tomlinson, C. and Tonelli, M. and Torres, C. V. and Torrie, C. I. and T\"oyr\"a, D. and Travasso, F. and Traylor, G. and Trifir\`o, D. and Tringali, M. C. and Trozzo, L. and Tse, M. and Turconi, M. and Tuyenbayev, D. and Ugolini, D. and Unnikrishnan, C. S. and Urban, A. L. and Usman, S. A. and Vahlbruch, H. and Vajente, G. and Valdes, G. and Vallisneri, M. and van Bakel, N. and van Beuzekom, M. and van den Brand, J. F. J. and Van Den Broeck, C. and Vander-Hyde, D. C. and van der Schaaf, L. and van Heijningen, J. V. and van Veggel, A. A. and Vardaro, M. and Vass, S. and Vas\'uth, M. and Vaulin, R. and Vecchio, A. and Vedovato, G. and Veitch, J. and Veitch, P. J. and Venkateswara, K. and Verkindt, D. and Vetrano, F. and Vicer\'e, A. and Vinciguerra, S. and Vine, D. J. and Vinet, J.-Y. and Vitale, S. and Vo, T. and Vocca, H. and Vorvick, C. and Voss, D. and Vousden, W. D. and Vyatchanin, S. P. and Wade, A. R. and Wade, L. E. and Wade, M. and Waldman, S. J. and Walker, M. and Wallace, L. and Walsh, S. and Wang, G. and Wang, H. and Wang, M. and Wang, X. and Wang, Y. and Ward, H. and Ward, R. L. and Warner, J. and Was, M. and Weaver, B. and Wei, L.-W. and Weinert, M. and Weinstein, A. J. and Weiss, R. and Welborn, T. and Wen, L. and We\ss{}els, P. and Westphal, T. and Wette, K. and Whelan, J. T. and Whitcomb, S. E. and White, D. J. and Whiting, B. F. and Wiesner, K. and Wilkinson, C. and Willems, P. A. and Williams, L. and Williams, R. D. and Williamson, A. R. and Willis, J. L. and Willke, B. and Wimmer, M. H. and Winkelmann, L. and Winkler, W. and Wipf, C. C. and Wiseman, A. G. and Wittel, H. and Woan, G. and Worden, J. and Wright, J. L. and Wu, G. and Yablon, J. and Yakushin, I. and Yam, W. and Yamamoto, H. and Yancey, C. C. and Yap, M. J. and Yu, H. and Yvert, M. and Zadro\ifmmode \dot{z}\else \.{z}\fi{}ny, A. and Zangrando, L. and Zanolin, M. and Zendri, J.-P. and Zevin, M. and Zhang, F. and Zhang, L. and Zhang, M. and Zhang, Y. and Zhao, C. and Zhou, M. and Zhou, Z. and Zhu, X. J. and Zucker, M. E. and Zuraw, S. E. and Zweizig, J.},
  collaboration = {LIGO Scientific Collaboration and Virgo Collaboration},
  journal = {Phys. Rev. Lett.},
  volume = {116},
%  issue = {6},
%  pages = {061102},
%  numpages = {16},
  year = {2016},
  month = {02},
  day = {11},
  publisher = {American Physical Society},
  doi = {10.1103/PhysRevLett.116.061102},
  %url = {https://link.aps.org/doi/10.1103/PhysRevLett.116.061102}
}

@article{Blumlein:2021txe,
	author = {Bl\"umlein, J. and Maier, A. and Marquard, P. and Sch\"afer,
	G.},
	title = "{The fifth-order post-Newtonian Hamiltonian dynamics of
	two-body systems from an effective field theory approach}",
	eprint = "2110.13822",
	archivePrefix = "arXiv",
	primaryClass = "gr-qc",
	reportNumber = "DESY 21--151, DO--TH 21/27, SAGEX--21--30",
	month = "10",
	year = "2021"
}

@article{Blumlein:2021txj,
	author = {Bl\"umlein, J. and Maier, A. and Marquard, P. and Sch\"afer,
	G.},
	title = "{The 6th post-Newtonian potential terms at $O(G_N^4)$}",
	eprint = "2101.08630",
	archivePrefix = "arXiv",
	primaryClass = "gr-qc",
	reportNumber = "DESY 20--199, DESY-20-199, DO-TH 20/06, SAGEX-20-08",
	doi = "10.1016/j.physletb.2021.136260",
	journal = "Phys. Lett. B",
	volume = "816",
	pages = "136260",
	year = "2021"
}

@article{Blumlein:2020pyo,
	author = {Bl\"umlein, J. and Maier, A. and Marquard, P. and Sch\"afer,
	G.},
	title = "{The fifth-order post-Newtonian Hamiltonian dynamics of
	two-body systems from an effective field theory approach: potential
	contributions}",
	eprint = "2010.13672",
	archivePrefix = "arXiv",
	primaryClass = "gr-qc",
	reportNumber = "DESY 20--062, DESY-20-062, DO--TH 20/04,
	SAGEX--20--10",
	doi = "10.1016/j.nuclphysb.2021.115352",
	journal = "Nucl. Phys. B",
	volume = "965",
	pages = "115352",
	year = "2021"
}

@article{Blumlein:2020znm,
	author = {Bl\"umlein, J. and Maier, A. and Marquard, P. and Sch\"afer,
	G.},
	title = "{Testing binary dynamics in gravity at the sixth
	post-Newtonian level}",
	eprint = "2003.07145",
	archivePrefix = "arXiv",
	primaryClass = "gr-qc",
	reportNumber = "DESY-20-044, DESY 20--044, DO--TH 20/02,
	SAGEX--20--06, DO-TH 20/02, SAGEX-20-06",
	doi = "10.1016/j.physletb.2020.135496",
	journal = "Phys. Lett. B",
	volume = "807",
	pages = "135496",
	year = "2020"
}

@article{Blumlein:2020pog,
	author = {Bl\"umlein, J. and Maier, A. and Marquard, P. and Sch\"afer,
	G.},
	title = "{Fourth post-Newtonian Hamiltonian dynamics of two-body
	systems from an effective field theory approach}",
	eprint = "2003.01692",
	archivePrefix = "arXiv",
	primaryClass = "gr-qc",
	reportNumber = "DESY 20--025, DO--TH 20/01, SAGEX--20--03,
	DESY-20-025, DO-TH 20/01, SAGEX-20-03",
	doi = "10.1016/j.nuclphysb.2020.115041",
	journal = "Nucl. Phys. B",
	volume = "955",
	pages = "115041",
	year = "2020"
}

@article{Blumlein:2019bqq,
	author = {Bl\"umlein, J. and Maier, A. and Marquard, P. and Sch\"afer,
	G. and Schneider, C.},
	title = "{From Momentum Expansions to Post-Minkowskian Hamiltonians by
	Computer Algebra Algorithms}",
	eprint = "1911.04411",
	archivePrefix = "arXiv",
	primaryClass = "gr-qc",
	reportNumber = "DESY-19-185, DO-TH 19/21, SAGEX-19-25",
	doi = "10.1016/j.physletb.2019.135157",
	journal = "Phys. Lett. B",
	volume = "801",
	pages = "135157",
	year = "2020"
}

@article{Blumlein:2019zku,
	author = {Bl\"umlein, J. and Maier, A. and Marquard, P.},
	title = "{Five-Loop Static Contribution to the Gravitational
	Interaction Potential of Two Point Masses}",
	eprint = "1902.11180",
	archivePrefix = "arXiv",
	primaryClass = "gr-qc",
	reportNumber = "DESY 19-029, DO-TH 19/01, DESY-19-029, DO-TH-19/01",
	doi = "10.1016/j.physletb.2019.135100",
	journal = "Phys. Lett. B",
	volume = "800",
	pages = "135100",
	year = "2020"
}

\appendix
\chapter{Solution of the wave equation} \label{app:wave equation sol}
    This derivation can be found in most textbooks on field theory, e.g. \textcite{Kachelriess:2017cfe}, or \textcite{NoNonsenseQFT}.
    
    Indices will be ignored in this appendix, as the spatio-temporal dependence of the solution is \emph{assumed} to be independent of indices. Thus, $h\ind{_{\mu\nu}}(x^\alpha)=\epsilon\ind{_{\mu\nu}} h(x^\alpha)$ and $ \dalembertian h\ind{_{\mu\nu}}(x^\alpha)=\epsilon\ind{_{\mu\nu}} \dalembertian h(x^\alpha)$.

    Assuming the solution to be a superposition of plane waves $e^{-ik_\sigma x^\sigma}$, the most general form the wave can take is
    \begin{align} \label{eq:Fourier of h}
        h(x^\alpha) = \int \frac{ \dd[4]{k} }{(2\pi)^4} \left\{ a(k_\mu) e^{-ik\ind{_\sigma}x\ind{^\sigma}} + b(k_\mu) e^{ik\ind{_\sigma}x\ind{^\sigma}} \right\}.
    \end{align}

    For \eqref{eq:Fourier of h} to be a solution of the wave equation \eqref{eq:wave equation} the following must hold
    \begin{align}
        \nonumber \dalembertian h(x\ind{^{\alpha}}) = 0 = \int \frac{\dd[4]{k}}{(2\pi)^4 } \dalembertian \left\{ a(k_\mu) e^{-i k\ind{_\sigma} x\ind{^{\sigma}} } + b(k_\mu) e^{i k\ind{_\sigma} x\ind{^{\sigma}} } \right\}& \\
        \nonumber = \int \frac{\dd[4]{k}}{(2\pi)^4 } \eta\ind{^{\mu\nu}}\pdv{}{x^\mu}\pdv{}{x^\nu} \left\{ a(k_\mu) e^{-i k\ind{_\sigma} x\ind{^{\sigma}} } + b(k_\mu) e^{i k\ind{_\sigma} x\ind{^{\sigma}} } \right\}& \\
        \nonumber = \int \frac{\dd[4]{k}}{(2\pi)^4 } \eta\ind{^{\mu\nu}} \left( i^2 k_\mu k_\nu \right) \left\{ a(k_\mu) e^{-i k\ind{_\sigma} x\ind{^{\sigma}} } + b(k_\mu) e^{i k\ind{_\sigma} x\ind{^{\sigma}} } \right\}& \\
        = \int \frac{\dd[4]{k}}{(2\pi)^4 } \left( k_0^2 - \abs{\tvec{k}}^2 \right) \left\{ a(k_\mu) e^{-i k\ind{_\sigma} x\ind{^{\sigma}} } + b(k\ind{_\sigma}) e^{i k\ind{_\sigma} x\ind{^{\sigma}} } \right\}&,
    \end{align}
    which is a solution as long as $k_0^2=\abs{\tvec{k}}^2$. Since $k_0$ is identified as the temporal frequency it is required to be positive \ifthenelse{\boolean{NaturalUnits}}{$k_0 = \omega\geq0$}{$k_0 = \omega/c\geq0$} in order to be physical. Both these conditions can be imposed by $\dirac{}{k_0^2 - \omega_k^2} \cdot \heaviside{k_0}$, where $\dirac{}{x}$ is the \emph{Dirac delta function}, $\heaviside{x}$ is the \emph{Heaviside step function}, and $\omega_k$ is determined by the dispersion relation and is $\omega_k = \abs{\tvec{k}}$ for massless fields.\footnote{Massive fields must satisfie the \emph{Klein–Gordon equation} $(\dalembertian - m^2)\phi=0$. The solution is the same as that of massless fields shown here, but with $\omega_k^2=\abs{\tvec{k}}^2+m^2$. That is why $\abs{\tvec{k}}$ is renamed $\omega_k$ here, to make the result more easily transferable.}
    
    Implementing these restrictions \eqref{eq:Fourier of h} becomes
    \begin{align}\label{eq:C3}
        h(x^\alpha) = \int \frac{ \dd[4]{k} }{(2\pi)^4} \dirac{}{k_0^2 - \omega_k^2} \heaviside{k_0} \left\{ a(k_\mu) e^{-ik\ind{_\sigma}x\ind{^\sigma}} + b(k_\mu) e^{ik\ind{_\sigma}x\ind{^\sigma}} \right\}.
    \end{align}
    
    Performing the $k_0$ integral results with
    \begin{subequations}\label{eq:k0 integral}
    \begin{align}
        \int \frac{\dd{k_0}}{2\pi} & \dirac{}{k_0^2 - \omega_k^2} \heaviside{k_0} f(k_0) \label{eq:k0 integral 1} \\
        & = \int \frac{\dd{k_0}}{2\pi} \dirac{}{(k_0 - \omega_k)(k_0 + \omega_k)} \heaviside{k_0} f(k_0) \label{eq:k0 integral 2}\\
        & = \int \frac{\dd{k_0}}{2\pi} \frac{1}{2k_0} \left[ \dirac{}{k_0 - \omega_k} + \dirac{}{k_0 + \omega_k} \right] \heaviside{k_0} f(k_0) \label{eq:k0 integral 3}\\
        & = \frac{1}{2\omega_k} f(\omega_k). \label{eq:k0 integral 4}
    \end{align}
    \end{subequations}
    Step-by-step the above calculation first, \eqref{eq:k0 integral 1}, collapse all dependence on $k_0$ into a function $f(k_0)$, other than the Dirac delta function and Heaviside step function. In line \eqref{eq:k0 integral 2} the argument of the Dirac delta was expanded, and in line \eqref{eq:k0 integral 3} the Dirac delta was \emph{itself} expanded according to the relation
    \begin{align}
        \dirac{}{f(x)} = \sum_i \frac{ \dirac{}{x-a_i} }{ \dv{f}{x} }, \quad \forall a_i: f(a_i)=0.
    \end{align}
    Lastly, in line \eqref{eq:k0 integral 4}, the $k_0\geq0$ term was singled out by $\heaviside{k_0}$.
    
    All the steps of \eqref{eq:k0 integral} can be performed for \eqref{eq:C3}. Also requiring $h(x^\alpha)$ to be a real function can easily be done by demanding $h^\dagger(x^\alpha) = h(x^\alpha)$, which is obtained most generally by having $b(k_\mu)=a^\dagger(k_\mu)$.
    
    Thus, the most general solution of the wave equation for a real scalar field is
    \begin{align}
        \nonumber & \dalembertian h(x^\alpha) = 0, \\
        \Rightarrow \quad h(x^\alpha) = \int \frac{ \dd[3]{k} }{(2\pi)^3 \cdot 2\omega_k} & \left\{ a(\tvec{k}) e^{-ik\ind{_\sigma}x\ind{^\sigma}} + a^\dagger(\tvec{k}) e^{ik\ind{_\sigma}x\ind{^\sigma}} \right\}, \label{eq:scalar sol. wave eq.}
    \end{align}
    with $k_0=\abs{\tvec{k}}=\omega_k$. This is the solution presented in equation \eqref{eq:wave:general}.
    
    This is still a large class of solutions, but it is restricted to travel in the $\tvec{k}$-direction through space, with a velocity of
    \begin{align}
        v_g = \ifthenelse{\boolean{NaturalUnits}}{\pdv{\omega}{\abs{\tvec{k}}}=\pdv{k_0}{\abs{\tvec{k}}} = \pdv{\abs{\tvec{k}}}{\abs{\tvec{k}}} = 1}{\pdv{\omega
        }{
        \abs{\tvec{k}}} = \pdv{ck_0 }{\abs{\tvec{k}}} = c\pdv{\abs{\tvec{k}}}{\abs{\tvec{k}}} = c}.
    \end{align}
    Notice that both the group velocity $v_g\equiv \pdv{\omega}{\abs{\tvec{k}}}$ and the phase velocity $v_p\equiv \frac{\omega}{\abs{\tvec{k}}}$ are both equal to \ifthenelse{\boolean{NaturalUnits}}{unity}{$c$}.
\chapter{Equivalent one body problem and mass term manipulation} \label{app:equivOneBodyAndMassTerms}
\section{Rewriting to the equivalent one body problem}
    \begin{figure}[h!]
        \centering
        \includegraphics[width=0.6\textwidth]{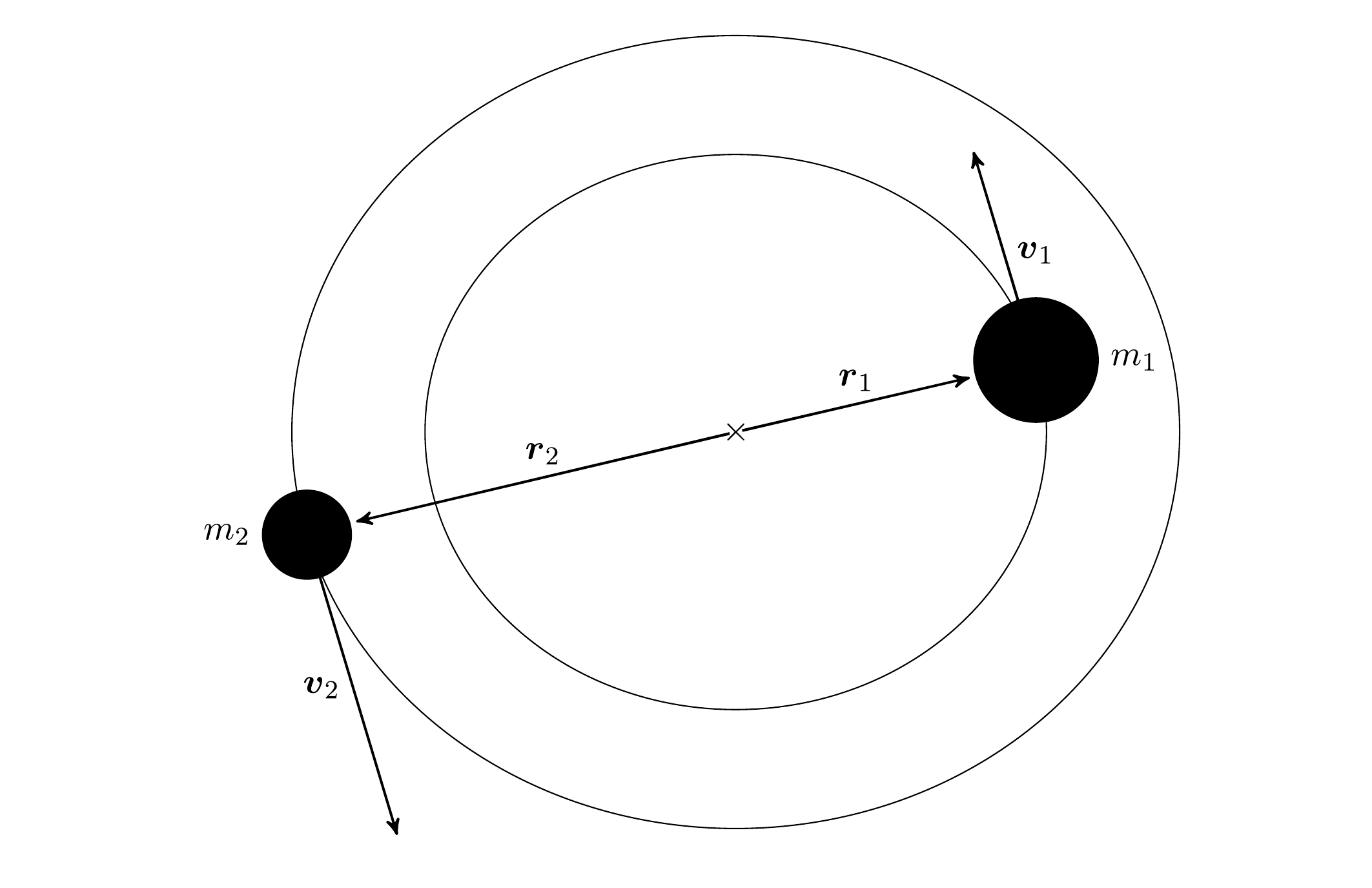}
        \caption{Diagram of a binary system.}
        \label{fig:BinarySystem:A}
    \end{figure}
    
    To solve the equation of motion of the \gls{2bodyprob} it is useful to rewrite the equations in terms of relative quantities, like the spatial separation $\tvec{r}$, and relative velocity $\tvec{v}\equiv \Dot{\tvec{r}}$. This will be done for the centre of mass frame in this appendix.

    Letting $\tvec{r}_i$, $i\in \{1,2\}$ be the position of object $i$ relative the centre of mass, which is placed in the origin, as in \cref{fig:BinarySystem:A} the following identity holds
    \begin{align} \label{eq:relativdissplacement:def}
        \tvec{r} \equiv \tvec{r}_2 - \tvec{r}_1,
    \end{align}
    \begin{align}
        \begin{split}\label{eq:centreOfMass}
            \tvec{0} \equiv \hspace{3pt} & (m_1+m_2)\tvec{R}_\text{CM} = m_1 \tvec{r}_1 + m_2 \tvec{r}_2, \\
            \Rightarrow \quad & m_1 \tvec{r}_1 = - m_2 \tvec{r}_2 = -m_2 \left( \tvec{r} + \tvec{r}_1 \right).
        \end{split}
    \end{align}
    In the last line of \eqref{eq:centreOfMass} equation \eqref{eq:relativdissplacement:def} was used to eliminate $\tvec{r}_2$. Similarly $\tvec{r}_1$ can be eliminated in favour of $\tvec{r}_2$. Thus $\tvec{r}_i$ can be expressed as
    \begin{subequations}\label{eq:relDisplacement:exchange}
        \begin{empheq}[box=\widefbox]{align} 
            \tvec{r}_1 = - & \frac{m_2}{m_1+m_2}\tvec{r} = -\frac{m_2}{M}\tvec{r}, \\
            \tvec{r}_2 = \hspace{3pt} & \frac{m_1}{m_1+m_2}\tvec{r} = \frac{m_1}{M} \tvec{r}.
        \end{empheq}
    \end{subequations}
    Here $M$ is the total mass of the binary. Since the velocity of each object in the centre of mass frame is $\tvec{v}_i = \Dot{\tvec{r}}_i$ it directly follows
    \begin{subequations}\label{eq:relVelocity:exchange}
        \begin{empheq}[box=\widefbox]{align} 
            \tvec{v}_1 = & -  \frac{m_2}{M}\tvec{v}, \\
            \tvec{v}_2 = & \hspace{3pt} \frac{m_1}{M} \tvec{v},
        \end{empheq}
    \end{subequations}
    where again $\tvec{v}=\Dot{\tvec{r}}$.
    
    Substituting $\tvec{r}_i$ and $\tvec{v}_i$ for the expressions of equations \eqref{eq:relDisplacement:exchange}-\eqref{eq:relVelocity:exchange} the Lagrangian, and thus the \acrshort{eom}, becomes a function of \emph{just} $\tvec{r}$ and $\tvec{v}$. Thus the \gls{2bodyprob} is reduced to solving for just the relative motion of one object, an equivalent one body problem.
    
    Explicitly the Newtonian Lagrangian becomes
    \begin{align*}
        L\ind{_{\text{Newt}}} & = \frac{1}{2}\left( m_1 v_1^2 + m_2 v_2^2 \right) + \frac{ Gm_1 m_2 }{\abs{ \tvec{r}_2 - \tvec{r}_1 }} = \frac{1}{2} \left( m_1 \frac{m_2^2}{M^2} + m_2 \frac{m_1^2}{M^2} \right)v^2 + \frac{Gm_1 m_2}{r} \\
        & = \frac{1}{2}\frac{m_1m_2}{M}\left( \frac{m_2+m_1}{M} \right)v^2 + \frac{G M\frac{m_1m_2}{M} }{r} = \frac{1}{2}\frac{m_1m_2}{M}v^2 + \frac{G M\frac{m_1m_2}{M} }{r} \\
        & \equiv \frac{1}{2}\mu v^2 + \frac{GM\mu}{r}.
    \end{align*}
    
    Bottom line is that preforming the substitution to $r$ and $v$ reduces the problem to describing the motion of \emph{one} particle with an effective mass of $\mu=\frac{m_1m_2}{M}$ in a gravitational potential produced by an effective mass of $M=m_1+m_2$, which is static and located at the position of the other particle.
    
\section{Mass term manipulation}
    Following the previous section it is hopefully clear what motivates the introduction of the total and reduced mass $M$ and $\mu$. The name reduced mass follows from the observation that in the extreme mass ratio, $m_1 \gg m_2$, $M =m_1+m_2 \simeq m_1$ and $\mu=\frac{m_1m_2}{M} \simeq \frac{m_1m_2}{m_1}=m_2$. I.e. in the test mass regime $M$ is the gravitational source and $\mu$ is the test mass exactly. Of course $M>\mu$ in all cases, with the largest value of $\mu_\text{max}=\frac{1}{4}M$ when $m_1=m_2$. 
    
    Moving beyond the Newtonian approximation there will appear other mass terms that are common in the literature. These are listen for convenience in equation \eqref{eq:masses:def}.
    
    In this thesis there will appear terms of the form $m_1 v_1^{n+1} + m_2 v_2^{n+1}$ and $m_1(-r_1)^{n+1} + m_2r_2^{n+1}$. In this section there will be tips for strategies to convert these expressions into \eqref{eq:masses:def} type mass terms. The result can be read of equations \eqref{eq:masses:powers}-\eqref{eq:masses:mixedpowers}.
    
    \begin{subequations} \label{eq:masses:def}
        \begin{empheq}[box=\widefbox]{align} 
            M & \equiv m_1 + m_2  & & \text{Total mass} \label{eq:totalMass:def} \\
            \mu & \equiv \frac{m_1 m_2}{m_1 + m_2} = \frac{m_1 m_2}{M} & & \text{Reduced mass} \label{eq:reducedMass:def} \\
            \eta & \equiv \frac{m_1 m_2}{\left(m_1 + m_2\right)^2} = \frac{\mu}{M} & & \text{Symmetric mass ratio} \label{eq:symmetricMassRatio:def} \\
            \M & \equiv \frac{\left( m_1 m_2 \right)^{3/5} }{ \left(m_1+m_2\right)^{1/5} } = \left( \mu^3 M^2 \right)^{1/5} = M\eta^{3/5} & & \text{Chirp mass} \label{eq:chirpMass:def}
        \end{empheq}
    \end{subequations}
    
    Here is a stepwise approach to deal with $m_1 v_1^{n+1} + m_2 v_2^{n+1}$ and $m_1(-r_1)^{n+1} + m_2r_2^{n+1}$ type expressions.
    \begin{enumerate}
        \item The product $m_1m_2=M\mu$. Identify and extract all common factors of $\mu$ from the expression.
        \item This will leave something like $\left(m_2^{n} \pm m_1^{n}\right)/M^n$.
        \begin{enumerate}
            \item If it is a sum, calculate $M^{n}=m_1^{n} + m_2^n + \dots$. Thus $m_2^n + m_1^n = M^n - \dots$ Then again look out for reduced masses and remember that $\mu/M=\eta$.\label{enu:massmanipulation:ifPluss}
            \item If it is a difference this will usually imply that $n$ is even. Let $n=2k$ and then $ m_2^{2k} - m_1^{2k}= (m_2^k - m_1^k)(m_2^k + m_1^k) $. The $(m_2^k + m_1^k)$ term can be expanded as in step \ref{enu:massmanipulation:ifPluss}, and hopefully this will be enough. The difference can be further expanded using $m_1^k-m_2^k=\sqrt{(m_1^k-m_2^k)^2}$ to get mixed terms which can be factored as $\mu$.
        \end{enumerate}
    \end{enumerate}
    
    To get some concrete examples, lets consider $(m_1m_2^4 + m_2 m_1^4)/M^4$.
    \begin{align*}
         \frac{m_1m_2^4 + m_2 m_1^4}{M^4} & = \frac{m_1m_2}{M}\frac{m_2^3+m_1^3}{M^3} = \mu \frac{M^3-3m_1m_2(m_1+m_2)}{M^3} \\ 
         & = \mu \left( 1-3\mu/M\right) = \mu \left(1-3\eta\right).
    \end{align*}
    $m_2^3+m_1^3$ was rewritten using equation \eqref{eq:m1m2ThirdPower}.
    
    Most of the expressions encountered at 1\acrshort{pn} will follow the same approach as the example above, with the exception of a term $(m_2m_1^3 - m_1 m_2^3)/M^3$ which appear in the octupole moment of the flux (see \cref{chap:flux}). It goes like this
    \begin{align*}
        \frac{m_2m_1^3 - m_1 m_2^3}{M^3} & = \mu \frac{m_1^2-m_2^2}{M^2} = \mu \frac{ (m_1-m_2)(m_1+m_2) }{M^2} = \mu \frac{m_1-m_2}{M} \\
        & = \mu \sqrt{ \frac{(m_1-m_2)^2}{M^2} } = \mu \sqrt{\frac{m_1^2 + m_2^2 - 2m_1m_2}{M^2}} = \mu \sqrt{\frac{M^2-4M\mu}{M^2}} \\
        & = \mu \sqrt{1-4\eta}.
    \end{align*}
    
    In equation \eqref{eq:masses:powers} the different sum of powers are listed, and in \eqref{eq:masses:mixedpowers} the mixed products are listed. These expressions are useful for mass term manipulations like those that appear in this thesis.
    
    \begin{subequations}\label{eq:masses:powers}
        \begin{empheq}[box=\widefbox]{align}
            m_1 + m_2 & = M, \\
            m_1^2 + m_2^2 & = M^2 - 2m_1m_2 = M^2 - 2M\mu, \\
            m_1^3 + m_2^3 & = M^3 - 3m_1^2m_2 - 3 m_1m_2^2 = M^3 - 3M^2\mu, \label{eq:m1m2ThirdPower}\\
        \begin{split}
            m_1^4 + m_2^4 & = M^4 - 4m_1^3m_2 - 6 m_1^2m_2^2 - 4m_1m_2^3 \\
            & = M^4 + 8M\mu^2 -4M^2\mu -6M^2\mu^2 .
        \end{split}
        \end{empheq}
    \end{subequations}
    \begin{subequations}\label{eq:masses:mixedpowers}
        \begin{empheq}[box=\widefbox]{align} 
            m_1 m_2 & = M\mu, \label{eq:masses:m1m2eqMmu} \\
           m_1 m_2^2 + m_1^2 m_2 & = M^2\mu, \\
           m_1 m_2^3 - m_1^3 m_2 & = M^3\mu\sqrt{1-4\eta}, \\
           m_1 m_2^4 + m_1^4 m_2 & = M^4\mu\left(1-3\eta \right).
        \end{empheq}
    \end{subequations}
    A handy trick is to combine equation \eqref{eq:masses:m1m2eqMmu} with expressions from \eqref{eq:masses:powers} to obtain \eqref{eq:masses:mixedpowers}-type expressions.
\chapter{Trigonometric identities}
\label{app:trig}
	The trigonometric identities that are used in this thesis can all be derived using Euler's formula
	\begin{align}\label{eq:app:EulersFormula}
		e^{i\theta} = \cos(\theta) + i \sin(\theta).
	\end{align}
	For the squared trigonometric functions one only needs to square this formula
	\begin{align*}
		\left(e^{i\theta}\right)^2 & = e^{i2\theta} = \cos(2\theta) + i \sin(2\theta) \\
		& = \left(\cos(\theta) + i \sin(\theta) \right)^2 = \cos[2](\theta) - \sin[2](\theta) + i2\sin(\theta)\cos(\theta) \\
		& = 2\cos[2](\theta) - 1 + i2\sin(\theta)\cos(\theta) = 1 - 2\sin[2](\theta) + i2\sin(\theta)\cos(\theta),
	\end{align*}
	where in the last line the Pythagorean identity $\sin[2](\theta)+\cos[2](\theta)=1$ was used to write the expression only by second powers in cosine or sine respectively. Comparing the real parts and imaginary parts of the first and third line produces the useful identities
	\begin{subequations}\label{eq:App:Trig:Square}
		\begin{empheq}[box=\widefbox]{align}
			\cos[2](\theta) & = \frac{1}{2} \left( 1+\cos(2\theta) \right), \\
			\sin[2](\theta) & = \frac{1}{2} \left( 1-\cos(2\theta) \right), \\
			\sin(\theta)&\cos(\theta) = \frac{1}{2}\sin(2\theta).
		\end{empheq}
	\end{subequations}
	
	Likewise the identities for the third power trigonometric functions can be obtained
	\begin{align*}
		\left(e^{i\theta}\right)^3 & = e^{i3\theta} = \cos(3\theta) + i \sin(3\theta) = \left( \cos(\theta) + i \sin(\theta) \right)^3\\
		& = \cos[3](\theta) - 3\cos(\theta)\sin[2](\theta) + i 3\cos[2](\theta)\sin(\theta) - i \sin[3](\theta) \\
		& = \left[4\cos[3](\theta) - 3\cos(\theta)\right] + i\left[ 3\sin(\theta) - 4\sin[3](\theta) \right],
	\end{align*}
	and then once again comparing the imaginary part of the first and third line the following identities are obtained.
	\begin{subequations}\label{eq:App:Trigg:Cube}
		\begin{empheq}[box=\widefbox]{align}
			\cos[3](\theta) & = \frac{1}{4} \left( 3\cos(\theta) + \cos(3\theta) \right), \\
			\sin[3](\theta) & = \frac{1}{4} \left( 3\sin(\theta) - \sin(3\theta) \right).
		\end{empheq}
	\end{subequations}

\end{document}